\documentclass[10pt,journal,compsoc]{IEEEtran}
\usepackage{cite}
\usepackage[T1]{fontenc}
\usepackage{bm}
\usepackage{amsmath}
\usepackage[table,xcdraw]{xcolor}

\usepackage{amsthm}
\usepackage{float}
\usepackage{setspace}
\usepackage{amssymb}
\usepackage{stfloats}
\usepackage{ragged2e}
\usepackage[ruled,vlined,linesnumbered]{algorithm2e}
\usepackage{amsfonts}
\usepackage{mathrsfs}
\usepackage{array,booktabs}
\usepackage{subfigure}
\usepackage{multirow}
\usepackage{cuted}
\usepackage{multicol}
\usepackage{graphicx}
\usepackage{hyperref}
\usepackage{threeparttable}
\usepackage{dcolumn}
\usepackage{makecell}
\usepackage{lipsum}
\usepackage{enumerate}
\usepackage{bbm}

\usepackage[protrusion=true,expansion=true]{microtype}
\pdfoutput=1

\newtheorem{Defn}{Definition}
\newtheorem{lem}{Lemma}

\newtheorem{Prop}{Proposition}
\newtheorem{thm}{Theorem}

\usepackage{xfrac}

\graphicspath{{figures/}}
\def\BibTeX{{\rm B\kern-.05em{\sc i\kern-.025em b}\kern-.08em
		T\kern-.1667em\lower.7ex\hbox{E}\kern-.125emX}}
\usepackage{balance}

\begin{document}
	\title{PAST: Pilot and Adaptive Orchestration for Timely and Resilient Service Delivery in Edge-Assisted UAV Networks under Spatio-Temporal Dynamics}
	\author{Houyi Qi, Minghui Liwang, \IEEEmembership{Senior Member, IEEE}, Liqun Fu, \IEEEmembership{Senior Member, IEEE}, Sai Zou, \IEEEmembership{Senior Member, IEEE}, Xinlei Yi, \IEEEmembership{Senior Member, IEEE}, Wei Ni, \IEEEmembership{Fellow, IEEE}, and Huaiyu Dai, \IEEEmembership{Fellow, IEEE}
		\thanks{This work was supported in part by National Natural Science Foundation of China under Grant nos. 62271424, 62503365, U23A20281; Shanghai Municipal Science and Technology Major Project under Grant no. 2021SHZDZX0100; Shanghai Pujiang Programme under Grant no. 24PJD117; Guizhou Provincial Science and Technology Plan Project ([2026]209); Aeronautical Science Foundation of China under Grant no. 2023Z066038001; Chinese Academy of Engineering, Strategic Research and Consulting Program under Grant no. 2023-XZ-65; and U.S. National Science Foundation (NSF) under Grant no. ECCS-2203214.
			H. Qi (houyiqi@tongji.edu.cn), M. Liwang (minghuiliwang@tongji.edu.cn), and X. Yi (xinleiyi@tongji.edu.cn) are with the Shanghai Research Institute for Intelligent Autonomous Systems, the State Key Laboratory of Autonomous Intelligent Unmanned Systems, Shanghai Key Laboratory of Intelligent Autonomous Systems, Department of Control Science and Engineering, Tongji University, Shanghai, China. L. Fu (liqun@xmu.edu.cn) is with the School of Informatics, Xiamen University, Fujian, China. S. Zou (dr-zousai@foxmail.com) is with College of Big Data and Information Engineering, Guizhou University. W. Ni (Wei.Ni@ieee.org) is with Data61, CSIRO. H. Dai (hdai@ncsu.edu) is with the Department of Electrical and Computer Engineering, North Carolina State University, NC, USA.
			
			Corresponding author: Minghui Liwang
	}}

	\IEEEtitleabstractindextext{
		\begin{abstract}\justifying
			Incentive-driven resource trading is essential for uncrewed aerial vehicle (UAV) applications with intensive, time-sensitive computing demands. Traditional spot trading suffers from negotiation delays and high energy costs, while conventional futures trading struggles to adapt to the dynamic, uncertain UAV--edge environment. To address these challenges, we propose \textit{PAST} (\textbf{p}ilot-and-\textbf{a}daptive \textbf{s}table \textbf{t}rading), a novel framework for edge-assisted UAV networks with spatio-temporal dynamism. PAST integrates two complementary mechanisms: \textit{PilotAO} (\textbf{pilot} trading \textbf{a}greements with \textbf{o}verbooking), a risk-aware, overbooking-enabled early-stage decision-making module that establishes long-term, mutually beneficial agreements and boosts resource utilization; and \textit{AdaptAO} (\textbf{adapt}ive trading \textbf{a}greements with \textbf{o}verbooking rate update), an intelligent adaptation module that dynamically updates agreements and overbooking rates based on UAV mobility, supply-demand variations, and agreement performance. {Together, these mechanisms integrate stable advance agreement construction with execution-feedback-driven adaptive renewal. Under the stated modeling assumptions, PilotAO satisfies the adopted individual-rationality and strong-stability criteria, as well as the competitive equilibrium and weak Pareto properties established for the considered trading model. Comprehensive experiments further demonstrate that, by coordinating advance planning with selective agreement renewal, PAST effectively balances stability, adaptability, and decision-making efficiency under dynamic UAV--edge conditions.}
			
		\end{abstract}
	}
	\maketitle

	\section{Introduction}
	\IEEEPARstart{W}{ith} their flexible deployment and relatively low operating cost, uncrewed aerial vehicles (UAVs) have become important platforms for urban monitoring, environmental surveillance, disaster response, and intelligent transportation~\cite{survey1,survry2,related work futures1}. Advances in sensing, communication, and computing technologies are further expanding their role in the emerging low-altitude economy~\cite{survey1,survry2}. Existing market analyses project that the global UAV market will reach \$48.5 billion by 2029~\cite{survey3}. As sensing and analytics applications increasingly adopt machine learning, UAVs face substantial computing and energy demands despite their limited onboard processing capabilities and power reserves~\cite{related work futures1}. These constraints hinder real-time data processing and efficient communications. Edge computing offers a promising solution by provisioning computing resources at the network edge, giving rise to edge-assisted UAV networks (EUNs), where on-ground edge servers (ESs) provide computing services to airborne UAVs. By transferring computation-intensive workloads from UAVs to terrestrial infrastructure, EUNs alleviate onboard service constraints and facilitate next-generation aerial applications~\cite{related work futures1,survey4}.
	
	\subsection{Challenges and Motivations}
	{Market-oriented\footnote{UAVs purchase on-demand computing services from ESs under pay-as-you-go pricing.} STD-EUNs enable incentive-driven airborne-edge service trading. Yet, three-dimensional (3D) UAV mobility dynamically alters ES accessibility and air-ground links, while tight mission windows and coupled task–resource dynamics make delayed, infeasible, or stale decisions costly. These challenges motivate three research questions (RQs) that guide our research.}
	
	\noindent\textit{\textbullet\ RQ1: How Can We Reduce the Overhead and Failure Risk of Online Trading in STD-EUNs?} {Online/spot trading establishes agreements in real time based on current supply, demand, and network conditions~\cite{related work futures1,jsac1}. 
		Although flexible, repeatedly negotiating associations and prices in each STU incurs substantial signaling, latency, and energy overhead in dynamic EUNs~\cite{related work futures1, related work futures2, matching}. Meanwhile, UAV mobility and time-varying ES capacities may invalidate agreements before execution, e.g., UAV may leave its selected ES's coverage~\cite{related work futures2}. Such failures are costly under tight mission deadlines and energy budgets and may discourage market participation.}

	\noindent\textit{\textbullet\ RQ2: How Can Pre-Signed Agreements and Overbooking Improve Market Efficiency?} {To mitigate limitations of online trading, futures trading leverages historical data to establish agreements in advance of transactions~\cite{related work futures2, matching}, thereby reducing real-time negotiation, enabling prompt service execution upon agreement confirmation. More importantly, inspired by overbooking practices in aviation~\cite{air}, hospitality management~\cite{hotel}, and telecommunications~\cite{tele}, \emph{overbooking} allows service providers to make advance commitments beyond their nominal execution capacity when not all commitments are expected to be executed simultaneously, improving resource utilization without expanding physical capacity. Similar mechanisms have been implemented across multiple layers of cloud data centers, including the kernel~\cite{kernel}, virtualization~\cite{virtual layers}, and container cluster schedulers such as OpenShift~\cite{openshift} and Mesos~\cite{MESOS}. Thus, combining futures agreements with risk-controlled overbooking offers a promising hybrid market for STD-EUNs. However, its effectiveness is challenged by time-varying air--ground channels, dynamic UAV trajectories and task arrivals, fluctuating ES workloads.} {Consequently, an agreement feasible and beneficial when established may become inaccessible, underutilized, or infeasible to fulfill at execution.}

	\noindent\textit{\textbullet\ RQ3: How Can Futures Agreements and Overbooking Rates Be Adaptively Updated under Market Uncertainty and Dynamics?} {Advance agreements with overbooking reduce trading overhead but must adapt to evolving mobility, task demand, channel quality, and ES availability. Fixed prices and static overbooking ratios\footnote{Effective overbooking schemes tailored for cloud environments have been proposed in~\cite{overbook_cloud1,overbook_cloud2}; however, these approaches typically employ static overbooking rates.} may cause idle reservations, inaccessible UAV--ES associations, resource shortages, and agreement non-fulfillment, degrading market efficiency and service reliability~\cite{overbooking2,overbooking3}.} {However, reconstructing the entire market upon every change would largely negate the latency and signaling benefits of pre-signed agreements.} {Deep reinforcement learning (DRL)-based policies can instead jointly adapt agreement retention, renewal, and overbooking based on task arrivals, resource availability, and execution outcomes, balancing agreement reuse and timely updates.}

	{The above RQs motivate designing \textit{Pilot-and-Adaptive Stable Trading} (PAST) framework, enabling reliable and responsive resource exchange between UAV buyers and capacity-constrained ES sellers.} {PAST comprises two stages: \textit{(i)} futures-trading stage, where \textbf{pilot} trading \textbf{a}greements\footnote{\emph{Pilot trading (PT)} refers to a preliminary trading mode that establishes tentative, risk-aware agreements before actual transactions.} with \textbf{o}verbooking (\textit{PilotAO}) are established via stable matching for risk-aware and flexible advance resource coordination; and \textit{(ii)} adaptive-adjustment stage, where \textbf{adapt}ive trading \textbf{a}greements with \textbf{o}verbooking rate update (\textit{AdaptAO}) use DRL-based techniques to selectively retain or renew pilot agreements and, upon renewal, determine the corresponding overbooking rate based on evolving execution states.} {Together, PilotAO reduces repeated negotiation through advance agreements, AdaptAO selectively renews them to track evolving UAV--edge conditions, balancing trading efficiency and responsiveness under spatio-temporal dynamism.}
	
	\subsection{Literature Investigation}
	Hereafter, we offer an in-depth analysis of existing literature, with a particular emphasis on resource sharing.
	
	\subsubsection{Status quo and limitations of online decision-making for resource sharing}
	{Existing studies on online resource sharing in UAV and edge networks have explored consensus-assisted coordination, auction-based trading, game-theoretic pricing, and matching-based resource allocation~\cite{related work spot1,related work spot2,related work spot3,related work spot4,related work spot5,wang2025lightweight}.} {\textit{Wang et al.}~\cite{wang2025lightweight} developed a lightweight consensus protocol that clusters UAVs by mobility similarity to improve scalability and reduce coordination overhead.} \textit{Xu et al.}~\cite{related work spot1} developed a blockchain-enabled double auction for joint spectrum and computing-resource trading in edge-assisted multi-UAV systems. \textit{Cheng et al.}~\cite{related work spot2} designed a reverse auction for federated-learning services among buyers, data providers, and UAV providers. \textit{Liu et al.}~\cite{related work spot3} proposed a QoE-oriented reverse auction for resource-constrained UAV-assisted crowdsensing. \textit{Wang et al.}~\cite{related work spot4} modeled resource pricing and allocation using a two-stage Stackelberg differential game. \textit{Raveendran et al.}~\cite{related work spot5} investigated many-to-many (M2M) matching for demand-aware fog-resource allocation over internet of things (IoT).
	
	{Despite their merits, existing market-oriented schemes largely rely on real-time negotiation, incurring substantial computation, signaling, energy overhead, and service-setup latency. Under volatile resource availability, such negotiations may fail to secure timely service, degrading efficiency and reliability. Consensus mechanisms ensure transaction integrity but do not address pricing, matching, or allocation. We therefore introduce pre-signed pilot agreements with overbooking to shift negotiation before execution, improve resource utilization, and enable futures-style trading.}

	\subsubsection{Potential and limitations of futures trading}
	Futures/offline trading enables participants to establish long-term agreements for anticipated needs, reducing decision-making overhead during actual transactions~\cite{related work futures1,related work futures2,related work futures3,related work futures4}. \textit{Liwang et al.}~\cite{related work futures1} designed futures trading for timely resource provision for multi-task UAVs. \textit{Qi et al.}~\cite{related work futures2} developed cross-layer pre-matching mechanisms for stable and cost-effective resource trading in dynamic cloud-assisted edge networks. \textit{Sheng et al.}~\cite{related work futures3} studied futures-based spectrum trading in wireless communication environments. \textit{Sexton et al.}~\cite{related work futures4} introduced futures-enabled resource slicing for wireless edge networks. {Recent studies further combine advance contracting with execution-stage adaptation. \textit{Wu et al.} integrated an overbooking-enabled pre-double auction with a real-time backup auction while ensuring truthfulness, individual rationality, and budget balance~\cite{twosauction}. \textit{Qi et al.} combined risk-aware long-term contracting with spot trading and reputation-assisted contract renewal to improve adaptation in dynamic edge markets~\cite{ohtrust}.}
	
	\begin{figure}[t]
		\setlength{\abovecaptionskip}{-1 mm}
		\centering
		\includegraphics[width=1\columnwidth]{images/SYSTEM_MODEL1.pdf}
		\caption{Diagram air-ground service trading over STD-EUNs.}
		\label{Scenario}
	\end{figure}
	
	{{These studies demonstrate the potential of combining advance contracting with execution-stage adaptation. Nevertheless, agreements based on historical or predicted information may become misaligned with evolving UAV mobility, task demand, channel conditions, and ES resource availability, causing inefficient allocation, resource shortages, and agreement non-fulfillment. Reconstructing the entire market upon every change restores responsiveness but largely sacrifices the latency and signaling benefits of advance agreements. To balance this trade-off, PAST combines stable advance agreements with state-dependent adaptation: PilotAO establishes risk-aware PT agreements with overbooking, while AdaptAO selectively retains or renews them and adjusts the overbooking rate based on the evolving execution state.}}

	\subsection{Bright Spots and Contributions}
	To the best of our knowledge, PAST is among the first to jointly address advance agreement construction, risk-aware overbooking, and feedback-driven renewal for UAV--ES service trading under multidimensional spatio-temporal uncertainties, with key contributions as follows:
	
	\noindent $\bullet$ \textit{A novel paradigm, PAST: Facilitating pilot and adaptive stable trading for STD-EUNs.} To address the uncertainty induced by UAV mobility, demand fluctuations, and volatile edge resources, we propose PAST, a unified framework that couples advance resource planning with state-dependent adaptation. By coordinating reusable PT agreements, risk-aware overbooking, and selective agreement renewal, PAST provides an integrated approach to efficient and adaptive UAV--edge service trading under spatio-temporal dynamics.

	\noindent $\bullet$ \textit{Function of PilotAO: Facilitating effective early-stage decisions.} Our PilotAO constitutes the futures-trading stage of PAST, establishing UAV-ES agreements that jointly specify task allocation, service pricing, and default provisions to hedge execution uncertainty. Formulated as a stable M2M matching problem, it balances the expected utilities and risks of UAV buyers and ES sellers. Its proactive overbooking mechanism permits ESs to admit commitments beyond nominal capacity based on anticipated service realization. Under the stated modeling assumptions, PilotAO further guarantees strong stability and individual rationality, and establishes competitive equilibrium and weak Pareto properties for the considered trading model, providing a principled basis for resilient UAV--edge service provision.
	
	\noindent $\bullet$ \textit{Function of AdaptAO: Enabling state-dependent agreement renewal.} We further design AdaptAO, which complements PilotAO by mitigating the performance degradation caused by stale or mismatched advance agreements. Using DRL, AdaptAO selectively retains or renews the current agreements and determines the overbooking rate upon renewal according to UAV mobility, demand fluctuations, resource availability, and execution feedback, thereby balancing agreement reuse with responsiveness to spatio-temporal dynamics.
	
	\noindent $\bullet$ \textit{Comprehensive experiments and evaluations on realistic data.} We conduct extensive trace-driven and model-based experiments using real-world mobility data and diverse dynamic settings. The results demonstrate the effectiveness of PAST in balancing social welfare, agreement fulfillment, resource utilization, and decision-making overhead, while further validating its adaptability under spatio-temporal dynamics.

	\section{Overview and Basic Modeling}
	This section first provides an overview of the proposed framework in Sec.~\ref{chap 2.1} and then presents the detailed modeling of UAVs and ESs in Sec.~\ref{chap 2.2}.
	
	\subsection{Overview}\label{chap 2.1}
	Consider the STD-EUN architecture illustrated in Fig.~\ref{Scenario}, where computing services are traded between two parties: \textit{(i)} UAVs in \(\bm{\mathcal U}=\{u_1,\ldots,u_i,\ldots,u_{|\bm{\mathcal U}|}\}\), each carrying out a mission that may require edge computing services; and \textit{(ii)} ground ESs in \(\bm{\mathcal S}=\{s_1,\ldots,s_j,\ldots,s_{|\bm{\mathcal S}|}\}\), which provide computing services to UAVs in return for payment. To characterize the spatio-temporal evolution of service demand and resource availability, we define two hierarchical trading time units as follows.
	
	\begin{Defn}[\textbf{Broader Trading Unit (BTU)}]
		A BTU is a mission-level trading horizon during which a UAV performs sensing, communication, and computing activities and may repeatedly request computing services from ground ESs.
	\end{Defn}
	
	\begin{Defn}[\textbf{Small Trading Unit (STU)}]
		An STU is a non-overlapping execution interval within a BTU during which task generation, PT-agreement execution, and the corresponding service outcomes are realized. Multiple task-level service transactions may occur within one STU.
	\end{Defn}
	
	\begin{figure}[t]
		\setlength{\abovecaptionskip}{-1 mm}
		\centering
		\includegraphics[width=0.8\columnwidth]{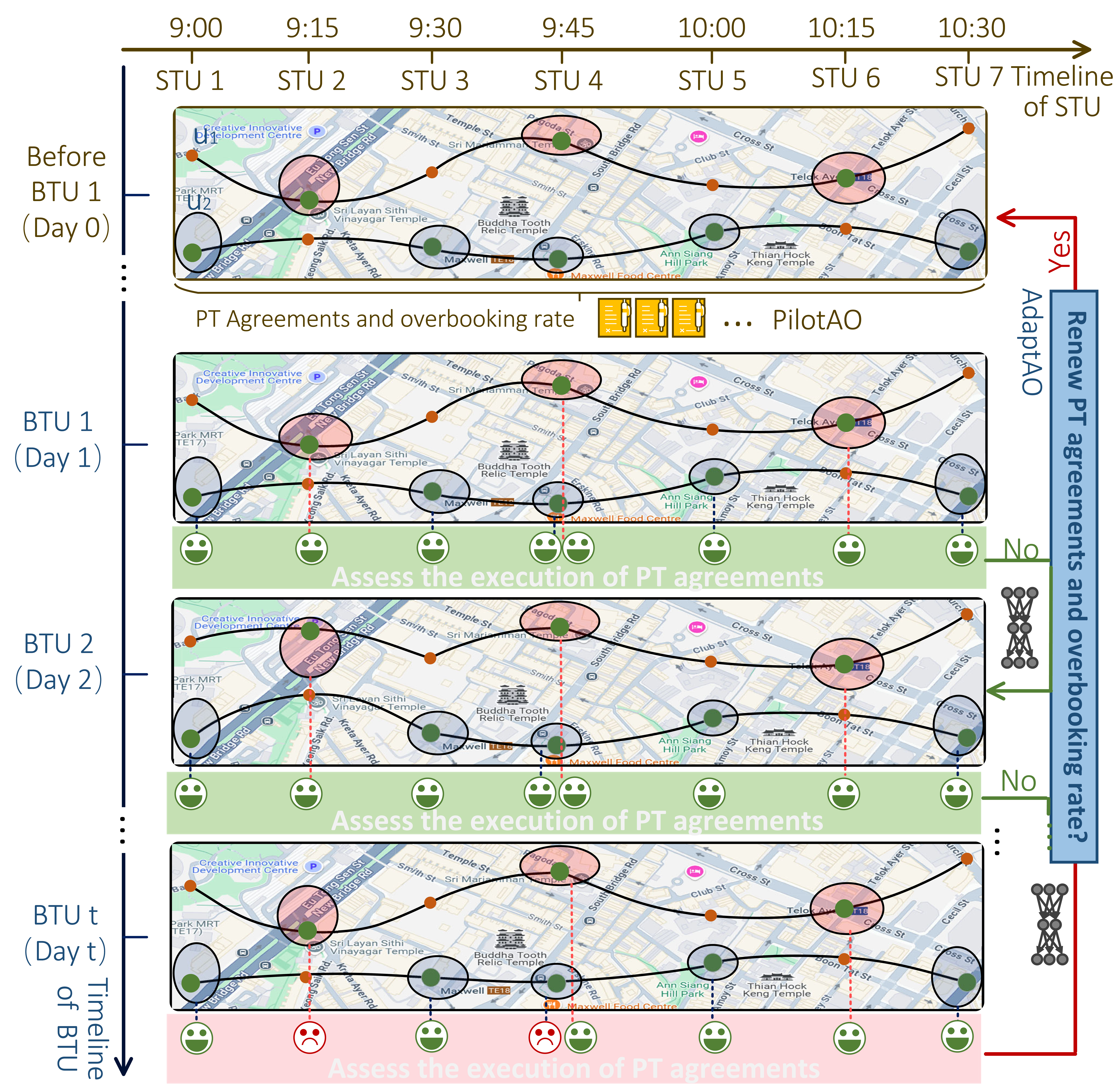}
		\caption{A timeline example of our proposed PAST.}
		\label{FSATO_BTU}
	\end{figure}
	
	To facilitate analysis, we partition the entire time horizon into multiple BTUs indexed by \(t\in\{1,2,\ldots,T\}\), each of which is further divided into \(N\) STUs indexed by \(n\in\{1,2,\ldots,N\}\). For example, a BTU spanning 9:00am--10:00am with a 15-minute STU duration contains four non-overlapping STUs, i.e., [9:00am, 9:15am), [9:15am, 9:30am), [9:30am, 9:45am), and [9:45am, 10:00am), as illustrated in Fig.~\ref{FSATO_BTU}. Before practical execution, UAV \(u_i\) is associated with a candidate task set for STU \(n\), denoted by \(\bm D_i^{\mathsf{(n)}}=\{d_{i,1}^{\mathsf{(n)}},\ldots,d_{i,m}^{\mathsf{(n)}},\ldots,d_{i,|\bm D_i^{\mathsf{(n)}}|}^{\mathsf{(n)}}\}\). Due to uncertain task generation, only a subset of these candidate tasks may actually arise during execution, denoted by \(\bm D_i^{\mathsf{act,(n)}}\subseteq\bm D_i^{\mathsf{(n)}}\).
	
	Each UAV mission may require traversal across multiple regions. For UAV \(u_i\in\bm{\mathcal U}\), its mission-region set is denoted by \(\bm C_i=\{C_{i,1},\ldots,C_{i,k},\ldots,C_{i,|\bm C_i|}\}\), where \(C_{i,k}\) specifies the coordinates of the \(k\)-th region. UAV \(u_i\) is required to visit all regions in \(\bm C_i\), while the travel time between regions depends on its flight speed and the corresponding inter-region distance. To characterize UAV mobility, we adopt a distance-biased state-transition process that accounts for previously visited regions. Equivalently, the mobility process is Markovian with respect to an augmented state consisting of the current region and the set of visited regions. We next illustrate this mobility process within one BTU.
	
	At the beginning of each BTU, UAV $u_i$ departs from its initial position and selects its first mission region from $\bm C_i$. The corresponding initial-region distribution is denoted by $\bm{\pi}_{i,0}=[\pi_1,\ldots,\pi_k,\ldots,\pi_{|\bm C_i|}]$, where $\pi_k$ represents the probability that $C_{i,k}$ is selected as the first mission region.
	As UAV $u_i$ traverses its mission regions, its location evolves through a sequence of state transitions. Owing to heterogeneous flight speeds, inter-region travel times, and task-generation patterns, $u_i$ may not participate in every STU within a BTU. To distinguish the mobility-transition order from the global STU index $n$, we introduce $\hat n_i\in\{1,2,\ldots,|\bm C_i|\}$ to index the sequence of actual participations of UAV $u_i$ within a BTU. Each participation corresponds to visiting a new mission region. The first participation corresponds to the initially selected region, while each participation with $\hat n_i\geq2$ is reached through one inter-region transition.\footnote{Due to differences in flight speed and task-generation patterns, UAVs may not generate resource demands in every STU. The index $\hat n_i$ therefore describes the participation and mobility-transition order of UAV $u_i$, rather than the global STU index $n$. If the initially visited region is counted as the first participation, visiting all $|\bm C_i|$ mission regions results in $|\bm C_i|$ participations and $|\bm C_i|-1$ inter-region transitions.}
	
	Let $X_{i,\hat n_i}\in\bm C_i$ denote the mission region visited by UAV $u_i$ at its $\hat n_i$-th participation, and let $\mathcal V_{i,\hat n_i}\subseteq\bm C_i$ denote the set of regions visited up to that participation. Conditional on the augmented state $(X_{i,\hat n_i},\mathcal V_{i,\hat n_i})$, the next region is selected only from the unvisited mission regions, with a probability inversely proportional to its distance from the current region. Specifically, for $\hat n_i<|\bm C_i|$, if $X_{i,\hat n_i}=C_{i,a}$, the transition probability to an unvisited region $C_{i,b}$ is given by $\Pr\!\left(X_{i,\hat n_i+1}=C_{i,b}\mid X_{i,\hat n_i}=C_{i,a},\mathcal V_{i,\hat n_i}\right)=\frac{d_{a,b}^{-1}\mathbbm 1_{\{C_{i,b}\notin\mathcal V_{i,\hat n_i}\}}}{\sum_{C_{i,c}\notin\mathcal V_{i,\hat n_i}}d_{a,c}^{-1}}$, where $d_{a,b}$ denotes the distance between regions $C_{i,a}$ and $C_{i,b}$. After the next region is selected, the visited-region set is updated as $\mathcal V_{i,\hat n_i+1}=\mathcal V_{i,\hat n_i}\cup\{X_{i,\hat n_i+1}\}$. Besides, $\mathbbm 1_{\{\cdot\}}$ denotes the indicator function, which equals 1 if the enclosed condition holds and 0 otherwise. Hence, the mobility process is Markovian with respect to the augmented state $(X_{i,\hat n_i},\mathcal V_{i,\hat n_i})$ and prevents repeated visits to the same mission region.
	
	Given the limited coverage of ground ESs, let $\mathcal E_i^{\mathsf{(n)}}$ denote the set of ESs accessible to UAV $u_i$ in STU $n$. This set evolves with the realized location of $u_i$ and determines its feasible service options during practical execution. Since the realized accessibility set $\mathcal E_i^{\mathsf{(n)}}$ is unavailable before practical execution, PilotAO instead uses the signing-stage candidate set $\widehat{\mathcal E}_i^{\mathsf{(n)}}$, which is inferred from historical or predicted UAV mobility and ES coverage information. Note that the realized set $\mathcal E_i^{\mathsf{(n)}}$ becomes observable only during practical execution and is subsequently used, together with realized task generation, to determine the executable-arrival indicator $\alpha_{i,j,m}^{\mathsf{(n)}}$.
	
	In the dynamic environment, uncertainties can arise from: \textit{(i)} fluctuating UAV task loads (across STUs); \textit{(ii)} uncertain position of UAVs within every STU due to Markovian mobility; \textit{(iii)} evolving spatial boundaries of UAV mission regions (across BTUs); and \textit{(iv)} variable ES resource availability, driven primarily by fluctuating inherent requestors\footnote{Since we consider a general and practical scenario where ESs are not dedicated solely to UAV services, they may also serve other requestors such as local users or background applications\cite{related work futures2}.}. These uncertainties impose challenges to timely and smooth edge service delivery.
	
	{Under these settings, we propose PAST, a unified framework for stable and adaptive UAV--ES collaboration.} PAST supports timely and adaptive services across BTUs and STUs through the following key functions:
	
	\noindent
	$\bullet$ \textit{Establishing risk-aware PT agreements with overbooking for future STU service demand (PilotAO).} PilotAO establishes pilot trading (PT) agreements between UAVs and ESs in advance, specifying service relationships and trading terms for future STUs. {Specifically, the PT agreement between UAV $u_i$ and ES $s_j$ for STU $n$ is defined as
		$
		\mathbb C_{i,j}^{\mathsf{(n)}}
		\triangleq
		\left(
		\mathcal T_{i,j}^{\mathsf{(n)}},
		\left\{p_{i,j,m}^{\mathsf{(n)}}\right\}_{d_{i,m}^{\mathsf{(n)}}\in\mathcal T_{i,j}^{\mathsf{(n)}}},
		q^{\mathsf U},
		q^{\mathsf E}
		\right),
		$
		where $\mathcal T_{i,j}^{\mathsf{(n)}}$ is the assigned task set, $\{p_{i,j,m}^{\mathsf{(n)}}\}_{d_{i,m}^{\mathsf{(n)}}\in\mathcal T_{i,j}^{\mathsf{(n)}}}$ contains the corresponding task-level service payments, and $q^{\mathsf U}$ and $q^{\mathsf E}$ are the contractual transfers triggered by UAV-side non-fulfillment and ES-side capacity shortage, respectively. All active PT agreements in STU $n$ are collected as
		$
		\bm{\mathbb C}^{\mathsf{(n)}}
		\triangleq
		\left\{
		\mathbb C_{i,j}^{\mathsf{(n)}}
		\,\middle|\,
		u_i\in\bm{\mathcal U}^{\mathsf{(n)}},
		s_j\in\bm{\mathcal S},
		\mathcal T_{i,j}^{\mathsf{(n)}}\neq\varnothing
		\right\},
		$
		which denotes the active PT-agreement set for STU $n$. PilotAO also sets an initial overbooking rate ($\tau$), allowing each ES to admit task-level commitments beyond its nominal execution capacity. This controlled redundancy accommodates uncertainty in UAV tasks, mobility, and ES workloads, enhancing resource utilization without increasing physical ES capacity\footnote{PT-agreement construction and overbooking rely on historical or predicted UAV mobility, task-demand patterns, and ES resource availability, balancing resource-utilization opportunities against the risk of unmet commitments.}. {Signing-stage mobility and coverage information further filters out UAV--ES associations unlikely to remain available during execution.} This reduces real-time negotiation and enhances the applicability of advance agreements.}

	\noindent
	$\bullet$ \textit{{Enabling state-dependent PT-agreement renewal and overbooking adaptation (AdaptAO).}} {Under the spatio-temporal dynamics of STD-EUNs, pre-signed PT agreements may become misaligned with the realized service environment due to changes in UAV accessibility and prediction errors in task demand or resource availability. AdaptAO therefore monitors agreement applicability, fulfillment performance, resource utilization, default behavior, and other execution-state information. Based on the observed state, it decides whether to retain or renew agreements and, upon renewal, determines the corresponding overbooking rate.} {Selective retention and renewal avoid unnecessary market-wide reconstruction while maintaining responsiveness to spatio-temporal changes.}
	
	By integrating PilotAO and AdaptAO, PAST aims to \textit{facilitate efficient, adaptive, and stable computing service delivery from various on-ground ESs to multiple UAVs in dynamic and uncertain environments}\footnote{{PilotAO assumes truthful reporting of task valuations and demands, UAV mobility and accessibility, ES costs and resource availability. We thus focus on agreement construction and adaptive renewal under truthful reporting, leaving incentive-compatible elicitation under strategic misreporting beyond our scope.}}. {In the following, we provide a toy example for better illustration.}

	\begin{figure}[]
		\setlength{\abovecaptionskip}{-1 mm}
		\centering
		\includegraphics[width=1\columnwidth]{images/SYSTEM_MODEL2.pdf}
		\caption{Framework and procedure in terms of a timeline associated with one STU in PAST.}
		\label{fig:fasto_timeline}
	\end{figure}
	\noindent
	\textit{Toy example}: Fig.~\ref{FSATO_BTU} illustrates the timeline of our PAST framework, where each UAV sequentially visits multiple mission regions within a BTU and may engage in resource trading during the corresponding STUs. On Day~0, using historical trading data, PilotAO pre-signs PT agreements for future STUs, providing guidance for subsequent BTUs. 
	On Day~1, UAVs carry out their pre-signed agreements in real transactions within each STU. As shown in Fig.~\ref{fig:fasto_timeline}, a single STU involves four UAVs and two ESs. In Transaction~a, UAV $u_1$ does not participate; however, thanks to the overbooking, ES $s_1$'s resource is still fully utilized by the remaining UAVs with signed PT agreements. In Transaction~b, ESs $s_1$ and $s_2$ experience temporary resource shortages while serving inherent requestors, leading to breaches of PT agreements with $u_1$ and $u_4$, which are later resolved through compensatory measures. PT agreements can become unreliable due to UAV mobility and demand variations. To handle this, we integrate AdaptAO into PAST, which continuously monitors agreement performance. The deviations observed in Transaction~b are incorporated into the next decision state. Based on this most recently available state, AdaptAO may retain the current configuration or reconstruct the PT agreements under a newly selected overbooking rate before the ensuing execution interval.
	
	\subsection{Basic Modeling}\label{chap 2.2}
	We next show the detailed modeling of UAVs and ESs:
	
	\noindent $\bullet$ \textit{Modeling of UAVs as service buyers:} To characterize the UAVs eligible to establish PT agreements for STU $n$, we use $\bm{\mathcal{U}}^\mathsf{(n)}$ to denote the signing-stage candidate UAV set. Actual task participation during practical execution is captured by $\bm D_i^{\mathsf{act,(n)}}$ and the executable-arrival indicator $\alpha_{i,j,m}^{\mathsf{(n)}}$. Specifically, each UAV $u_i \in \bm{\mathcal{U}}^\mathsf{(n)}$ is characterized by a 5-tuple:
	$u_i = \{f_i, e_i^\mathsf{tran}, \bm{D}^{\mathsf{(n)}}_i, e_i^{\mathsf{loc}}, \bm{l}_i^{\mathsf{uav,(n)}}\}$,
	where $f_i$ denotes the on-board computing capability of $u_i$ (e.g., CPU cycles/s). $e_i^\mathsf{tran}$ and $e_i^{\mathsf{loc}}$ represent the transmission power and local computing power (Watts), respectively, which are assumed to be constants over the entire time horizon. $\bm D_i^{\mathsf{(n)}}$ denotes the candidate task set considered during agreement construction, while $\bm D_i^{\mathsf{act,(n)}}\subseteq\bm D_i^{\mathsf{(n)}}$ denotes the subset actually generated during practical execution. For a task $d^\mathsf{(n)}_{i,m} \in \bm{D}^\mathsf{(n)}_i$, we use $r^\mathsf{(n)}_{i,m}$ to specify its required computing resources (e.g., CPU cycles). Moreover, $\bm{l}_i^{\mathsf{uav,(n)}} = \left (x_i^\mathsf{uav,(n)}, y_i^\mathsf{uav,(n)}, H_i^\mathsf{uav,(n)} \right )$ denotes the three-dimensional (3D) coordinates of $u_i$ at STU $n$, where $H_i^\mathsf{uav,(n)}$ is its altitude. 
	
	\noindent $\bullet$ \textit{Modeling of ESs as service sellers:} 
	Each ES $s_j \in \bm{\mathcal{S}}$ is modeled by a 4-tuple: $s_j = \{f_j^\mathsf{E}, e_j^\mathsf{E}, G_j, \bm{l}_j^\mathsf{ES}\}$, where $f_j^\mathsf{E}$ denotes the computing capability of $s_j$ (e.g., CPU cycles/s); $e_j^\mathsf{E}$ (Watts) represents the local computing power; $G_j$ is the total number of available resources (e.g., quantized by virtual machines), with each resource assumed to process one task for simplicity. Besides, $\bm{l}_j^\mathsf{ES} = (x^\mathsf{ES}_j, y^\mathsf{ES}_j, H^\mathsf{ES}_0)$ represents the fixed 3D coordinates of $s_j$, with $H^\mathsf{ES}_0$ being the ground altitude. 
	
	We further consider a practical setting in which ESs are not dedicated exclusively to UAV services but also serve terrestrial users, such as smartphones and vehicles. Their background workloads reduce the computing resources available to UAVs and introduce additional resource uncertainty and service competition. Let $\varepsilon_j^\mathsf{(n)}$ denote the inherent demand at ES $s_j$ in STU $n$, modeled as $\varepsilon_j^\mathsf{(n)}\sim\mathbf{P}(\kappa_j^\mathsf{(n)})$, where $\mathbf{P}(\cdot)$ denotes the Poisson distribution and $\kappa_j^\mathsf{(n)}$ is its mean. Since a Poisson realization may exceed the physical capacity $G_j$, the residual capacity available to UAV services is defined as $\max\{0,G_j-\varepsilon_j^\mathsf{(n)}\}$. {For ease of analysis, key notations are summarized in Table~\ref{tab:key_notation}.}
	\begin{table}[t!]
		\centering
		\footnotesize
		
		\caption{Key Notations}
		\label{tab:key_notation}
		\renewcommand{\arraystretch}{1.08}
		\setlength{\tabcolsep}{2pt}
		\begin{tabular}{
				@{}
				>{\raggedright\arraybackslash}p{2.3cm}
				>{\raggedright\arraybackslash}p{5.9cm}
				@{}
			}
			\toprule
			\textbf{Notation} & \textbf{Definition} \\
			\midrule
			
			$\bm{\mathcal U}$, $\bm{\mathcal U}^{\mathsf{(n)}}$, $\bm{\mathcal S}$
			&
			Sets of all UAVs, signing-stage candidate UAVs for STU $n$, and ESs
			\\
			
			$t$, $n$; BTU, STU
			&
			BTU and STU indices; broader and small trading units
			\\
			
			$\bm D_i^{\mathsf{(n)}}$, $\bm D_i^{\mathsf{act,(n)}}$, $d_{i,m}^{\mathsf{(n)}}$, $r_{i,m}^{\mathsf{(n)}}$
			&
			Candidate task set, realized task subset, task $m$, and its computational workload in CPU cycles
			\\
			
			$f_i$, $f_j^{\mathsf E}$, $G_j$, $\varepsilon_j^{\mathsf{(n)}}$
			&
			Computing capacities of UAV $u_i$ and ES $s_j$, nominal ES resource units, and inherent ES workload
			\\
			
			$\bm l_i^{\mathsf{uav,(n)}}$, $\bm l_j^{\mathsf{ES}}$, $\widehat{\mathcal E}_i^{\mathsf{(n)}}$, $\mathcal E_i^{\mathsf{(n)}}$
			&
			UAV and ES locations, signing-stage candidate ES set, and realized accessible ES set
			\\
			
			$g_{i,j}^{\mathsf{(n)}}$, $R_{i,j}^{\mathsf{(n)}}$, $T_{i,j,m}^{\mathsf{edge,(n)}}$
			&
			Channel gain, transmission rate, and edge-execution delay
			\\
			
			$\varphi^{\mathsf{(n)}}(u_i)$, $\varphi^{\mathsf{(n)}}(s_j)$, $\mathcal T_{i,j}^{\mathsf{(n)}}$
			&
			Matching sets of UAV $u_i$ and ES $s_j$, and their assigned task set
			\\
			
			$\mathbb C_{i,j}^{\mathsf{(n)}}$, $\bm{\mathbb C}^{\mathsf{(n)}}$, $p_{i,j,m}^{\mathsf{(n)}}$
			&
			PT agreement, active PT-agreement set, and service payment for task $d_{i,m}^{\mathsf{(n)}}$
			\\
			
			$q^{\mathsf U}$, $q^{\mathsf E}$, $\tau$
			&
			UAV-side non-fulfillment transfer, ES-side shortage compensation, and overbooking rate
			\\
			
			$\alpha_{i,j,m}^{\mathsf{(n)}}$, $\lambda_{i,j,m}^{\mathsf{(n)}}$
			&
			Executable-arrival indicator and conditional ES-shortage indicator
			\\
			
			$p_{\alpha,i,j,m}^{\mathsf{(n)}}$,
			$p_{\lambda\mid\alpha,i,j,m}^{\mathsf{(n)}}$
			&
			Executable-arrival probability and conditional ES-shortage probability
			\\
			
			$U^{\mathsf U}(\cdot)$, $U^{\mathsf S}(\cdot)$,
			$R^{\mathsf U}$, $R_1^{\mathsf S}$, $R_2^{\mathsf S}$
			&
			UAV and ES utilities; UAV-side and ES-side utility-shortfall risks; contract-level conditional ES-shortage risk
			\\
			
			$\mathbbm{s}_t^{\mathsf{(n)}}$, $\mathbbm{a}_t^{\mathsf{(n)}}$,
			$\mathbbm{r}_t^{\mathsf{(n)}}$
			&
			State, action, and immediate reward of AdaptAO
			\\
			
			\bottomrule
		\end{tabular}
	\end{table}

	\section{PilotAO of PAST: Establishing PT Agreements Before Practical Execution}
	PAST comprises two complementary stages. This section presents PilotAO, which establishes advance PT agreements between UAVs and ESs through risk-aware overbooking.
	
	\subsection{Key Modeling}
	To characterize the M2M associations between UAVs and ESs for future STUs, we first introduce matching notations:
	
	\noindent
	$\bullet$ $\varphi^\mathsf{(n)}(u_i)$: The set of ESs that serve UAV $u_i$ in STU $n$, i.e., $\varphi^\mathsf{(n)}(u_i)\subseteq\bm{\mathcal S}$;
	
	\noindent
	$\bullet$ $\varphi^\mathsf{(n)}(s_j)$: The set of UAVs that receive computing services from ES $s_j$ in STU $n$, i.e., $\varphi^\mathsf{(n)}(s_j)\subseteq\bm{\mathcal U}^\mathsf{(n)}$;
	
	For each pair $(u_i,s_j)$, let $\mathcal T_{i,j}^{\mathsf{(n)}}\subseteq\bm D_i^{\mathsf{(n)}}$ denote the subset of UAV $u_i$'s candidate tasks assigned to ES $s_j$ in STU $n$.
	
	\subsubsection{Utility and risk analysis of ESs}
	Processing UAV tasks incurs service costs at ESs. For task \(d_{i,m}^{\mathsf{(n)}}\) assigned to ES \(s_j\), we define the corresponding monetary service cost as
	\begin{equation}
		c_{i,j,m}^{\mathsf{E,(n)}}
		=
		\omega_1
		\frac{r_{i,m}^{\mathsf{(n)}}e_j^{\mathsf E}}
		{f_j^{\mathsf E}}
		+
		c_j^{\mathsf{hard}},
		\quad
		\forall d_{i,m}^{\mathsf{(n)}}\in\mathcal T_{i,j}^{\mathsf{(n)}}.
	\end{equation}
	Here, \(\omega_1\) converts the energy consumed for task execution into a monetary cost, while \(c_j^{\mathsf{hard}}\) denotes the constant hardware-depreciation cost of ES \(s_j\)~\cite{related work futures2}.
	
	{
		Due to UAV mobility and uncertain task generation, a pre-signed task--ES agreement may not become executable in every practical STU. To distinguish executable arrival from the final service outcome, we define the binary indicator \(\alpha_{i,j,m}^{\mathsf{(n)}}\). Specifically, \(\alpha_{i,j,m}^{\mathsf{(n)}}=1\) indicates that task \(d_{i,m}^{\mathsf{(n)}}\) is actually generated and the contracted ES \(s_j\) remains accessible under the realized UAV location. Formally,
		\begin{equation}
			\label{eq:executable_arrival}
			\alpha_{i,j,m}^{\mathsf{(n)}}
			=
			\begin{cases}
				1,
				&
				\text{if }
				d_{i,m}^{\mathsf{(n)}}\in\bm D_i^{\mathsf{act,(n)}}
				\text{ and }
				s_j\in\mathcal E_i^{\mathsf{(n)}},
				\\[1mm]
				0,
				&
				\text{otherwise}.
			\end{cases}
		\end{equation}
		The executable-arrival event only captures task generation and ES accessibility before capacity contention, while actual service is determined by the realized residual ES capacity.
	}
	
	The utility of ES $s_j$ consists of three components: \textit{(i)} the service payments received from successfully served tasks minus the corresponding service costs, \textit{(ii)} compensation paid for ES-side capacity shortages, and \textit{(iii)} transfers received from UAV-side non-fulfillment events. It is given by
	{\begin{equation}\label{Utility}
			{\small\begin{aligned}
					&U^\mathsf{S}(s_j,\varphi^\mathsf{(n)}(s_j),\bm{\mathbb C}^{\mathsf{(n)}})=\\&\sum_{u_i\in\varphi^\mathsf{(n)}(s_j)}\sum_{d^\mathsf{(n)}_{i,m}\in\mathcal{T}_{i,j}^\mathsf{(n)}}\alpha_{i, j,m}^\mathsf{(n)}(1-\lambda^\mathsf{(n)}_{i,j,m})\left( p^\mathsf{(n)}_{i,j,m}-c^\mathsf{E,(n)}_{i,j,m}\right)\\&+\sum_{u_i\in\varphi^\mathsf{(n)}(s_j)}\sum_{d^\mathsf{(n)}_{i,m}\in\mathcal{T}_{i,j}^\mathsf{(n)}}\left((1-\alpha_{i,j,m}^{\mathsf{(n)}})q^{\mathsf U}-\alpha_{i,j,m}^{\mathsf{(n)}}\lambda_{i,j,m}^{\mathsf{(n)}}q^{\mathsf E}\right),
			\end{aligned}}
	\end{equation}}
	where $p_{i,j,m}^{\mathsf{(n)}}$ denotes the service payment from UAV $u_i$ to ES $s_j$ for task $d_{i,m}^{\mathsf{(n)}}$.
	{
		Conditional on $\alpha_{i,j,m}^{\mathsf{(n)}}=1$, we set $\lambda_{i,j,m}^{\mathsf{(n)}}=1$ if task $d_{i,m}^{\mathsf{(n)}}$ cannot be served because the realized residual capacity of ES $s_j$ is insufficient. Such a task is treated as a compensated volunteer under the predefined execution rule\footnote{{Each task consumes one indivisible ES resource unit. In each STU, an ES first serves its inherent demand, then allocates its realized residual capacity to executable contracted tasks. If demand exceeds capacity, tasks are ranked by realized net service margin, with ties broken by UAV service value and then a predetermined fixed order. The highest-ranked tasks are served, while the rest become compensated volunteers under their PT agreements.}}. For consistency, we set $\lambda_{i,j,m}^{\mathsf{(n)}}=0$ whenever $\alpha_{i,j,m}^{\mathsf{(n)}}=0$.
	}
	
	{
		For notational clarity, define $p_{\alpha,i,j,m}^{\mathsf{(n)}}=\Pr\!\left(\alpha_{i,j,m}^{\mathsf{(n)}}=1\right)$ and $p_{\lambda\mid\alpha,i,j,m}^{\mathsf{(n)}}=\Pr\!\left(\lambda_{i,j,m}^{\mathsf{(n)}}=1\mid\alpha_{i,j,m}^{\mathsf{(n)}}=1\right)$. Accordingly, the probability that a pre-signed task becomes executable and is successfully served by ES $s_j$ is $p_{\alpha,i,j,m}^{\mathsf{(n)}}\!\left(1-p_{\lambda\mid\alpha,i,j,m}^{\mathsf{(n)}}\right)$, whereas the probability that it becomes executable but encounters an ES-side capacity shortage is $p_{\alpha,i,j,m}^{\mathsf{(n)}}p_{\lambda\mid\alpha,i,j,m}^{\mathsf{(n)}}$\footnote{{The executable-arrival and conditional-shortage probabilities are estimated from historical observations and joint model realizations, then refined with execution feedback. During each PilotAO invocation, the conditional-shortage estimate is fixed before terminal matching and thus serves as an ex ante risk estimate rather than the exact probability under realized matching. Estimation, online refinement, calibration, and error analysis are detailed in Appx.~B.}}. Since PT agreements are constructed before these execution outcomes are realized, PilotAO evaluates the expected ES utility as
		\begin{equation}
			\label{eq:expected_es_utility}
			{\small
				\begin{aligned}
					&\overline{U^{\mathsf S}}\!\left(
					s_j,
					\varphi^{\mathsf{(n)}}(s_j),
					\bm{\mathbb C}^{\mathsf{(n)}}
					\right)
					=
					\mathbb{E}\!\left[
					U^{\mathsf S}\!\left(
					s_j,
					\varphi^{\mathsf{(n)}}(s_j),
					\bm{\mathbb C}^{\mathsf{(n)}}
					\right)
					\right]
					\\
					=&\hspace{-4mm}
					\sum_{u_i\in\varphi^{\mathsf{(n)}}(s_j)}
					\sum_{d_{i,m}^{\mathsf{(n)}}\in\mathcal T_{i,j}^{\mathsf{(n)}}}
					\hspace{-2mm}p_{\alpha,i,j,m}^{\mathsf{(n)}}
					\left(
					1-p_{\lambda\mid\alpha,i,j,m}^{\mathsf{(n)}}
					\right)\hspace{-1mm}
					\left(
					p_{i,j,m}^{\mathsf{(n)}}
					-
					c_{i,j,m}^{\mathsf{E,(n)}}
					\right)
					\\
					+&
					\hspace{-4mm}\sum_{u_i\in\varphi^{\mathsf{(n)}}(s_j)}
					\sum_{d_{i,m}^{\mathsf{(n)}}\in\mathcal T_{i,j}^{\mathsf{(n)}}}\hspace{-2mm}
					\left[
					\left(
					1-p_{\alpha,i,j,m}^{\mathsf{(n)}}
					\right)
					q^{\mathsf U}
					\hspace{-.5mm}-\hspace{-.5mm}
					p_{\alpha,i,j,m}^{\mathsf{(n)}}
					p_{\lambda\mid\alpha,i,j,m}^{\mathsf{(n)}}
					q^{\mathsf E}
					\right].
			\end{aligned}}
	\end{equation}}
	
	Although PT agreements rely on historical or predicted information, realized UAV mobility and ES resource availability may deviate from signing-stage estimates, exposing ESs to utility shortfalls and capacity shortages. Accordingly, we consider two complementary ES-side risks.
	
	\noindent $\bullet$ \textit{ES-side utility-shortfall risk:} For each matched UAV $u_i\in\varphi^\mathsf{(n)}(s_j)$, let $U^{\mathsf S}(s_j,u_i,\mathbb C_{i,j}^{\mathsf{(n)}})$ denote the realized utility contribution of PT agreement $\mathbb C_{i,j}^{\mathsf{(n)}}$ to ES $s_j$. We characterize the corresponding utility-shortfall risk as
	\begin{equation}
		{\small
			\begin{aligned}
				R_1^\mathsf{S}\left(s_j,u_i,\mathbb C_{i,j}^{\mathsf{(n)}}\right)
				=
				\Pr\!\left(
				U^\mathsf{S}\left(s_j,u_i,\mathbb C_{i,j}^{\mathsf{(n)}}\right)
				<
				u_{\min}
				\right),
		\end{aligned}}
	\end{equation}
	where $u_{\min}$ denotes the minimum acceptable ES-side utility.
	
	\noindent $\bullet$ \textit{ES-side shortage risk conditional on executable arrival:} Even when a contracted task becomes executable, it may remain unserved if the residual capacity of its contracted ES is insufficient. We therefore define
	\begin{equation}
		{\small\begin{aligned}
				R_2^{\mathsf S}\left(s_j,u_i,d_{i,m}^{\mathsf{(n)}}\right)
				\hspace{-.5mm}=\hspace{-.5mm}
				\Pr\!\left(
				\lambda_{i,j,m}^{\mathsf{(n)}}=1
				\mid
				\alpha_{i,j,m}^{\mathsf{(n)}}=1
				\right)\hspace{-.5mm}=\hspace{-.5mm}
				p_{\lambda\mid\alpha,i,j,m}^{\mathsf{(n)}}.
		\end{aligned}}
		\label{eq:es_conditional_shortage_risk}
	\end{equation}
	A larger $R_2^{\mathsf S}$ indicates that an executable task is more likely to encounter ES-side capacity shortage. Accordingly, candidate task--ES contracts with risks above the prescribed tolerance are excluded during agreement construction.
	
	\subsubsection{Utility and risk analysis of UAVs}
	The air-ground wireless communication channel between UAVs and ES \( s_j \) is assumed to be dominated by the probabilistic line of sight (LoS) channel. We denote the probability of a LoS channel between ES \( s_j \) and UAV \( u_i \) at STU \(n \) as \( \epsilon^{\mathsf{(n)}}_{i,j} \), determined by environment-related parameters and the elevation angle of UAV \( u_i \) \cite{UAV}:
	\begin{equation}
		\epsilon^{\mathsf{(n)}}_{i,j} = \frac{1}{1 + \omega_2 \exp(-\omega_3 (\beta^{\mathsf{(n)}}_{i,j} - \omega_2))},
	\end{equation}
	where \( \omega_2 \) and \( \omega_3 \) are environment-related parameters, and
	\begin{equation}
		\beta^{\mathsf{(n)}}_{i,j} = \frac{180}{\pi}\arctan\left(\frac{H_i^{\mathsf{uav,(n)}}-H^{\mathsf{ES}}_0}{\| (x^{\mathsf{uav,(n)}}_i,y^{\mathsf{uav,(n)}}_i)-(x^{\mathsf{ES}}_j,y^{\mathsf{ES}}_j) \|}\right).
	\end{equation}
	represents the elevation angle in degrees between UAV \(u_i\) and ES \(s_j\) at STU \(n\). The channel gain is thus given by
	\begin{equation}{\small
			\begin{aligned}
				g^{\mathsf{(n)}}_{i,j} = \frac{g_0 (\epsilon^{\mathsf{(n)}}_{i,j} + \zeta (1 - \epsilon^{\mathsf{(n)}}_{i,j}))}{((H_i^{\mathsf{uav,(n)}}-H_0^{\mathsf{ES}})^2+\|(x_i^{\mathsf{uav,(n)}},y_i^{\mathsf{uav,(n)}})-(x_j^{\mathsf{ES}},y_j^{\mathsf{ES}})\|^2)},
		\end{aligned}}
	\end{equation}
	where \(g_0\) is the linear-scale channel gain at the reference distance \(l_0=1\,\text{m}\), and \(\zeta\) is the non-line-of-sight (NLoS) attenuation factor.
	The data transmission rate (e.g., Mbits/s)\footnote{{We assume orthogonal time--frequency allocation for concurrent UAV--ES uplinks within each STU, thereby avoiding co-channel interference. Spectrum reuse or non-orthogonal access would require an SINR-based rate formulation and is left for future work.}} from UAV \(u_i\) to ES \(s_j\) at STU \(n\) is given by
	
	\begin{equation}
		R^{\mathsf{(n)}}_{i,j} = B^{\mathsf{(n)}}_{i,j} \log_2 \left( 1 + \frac{g^{\mathsf{(n)}}_{i,j} e^{\mathsf{tran}}_{i}}{(\sigma^{\mathsf{(n)}}_{i,j})^2} \right),
	\end{equation}
	{where \(B_{i,j}^{\mathsf{(n)}}\) denotes the orthogonal bandwidth allocated to the uplink from UAV \(u_i\) to ES \(s_j\), and \((\sigma_{i,j}^{\mathsf{(n)}})^2\) denotes the corresponding receiver-noise power.} The transmission delay for task \( d^\mathsf{(n)}_{i,m} \) is
	\begin{equation}
		T^\mathsf{trans,(n)}_{i,j,m} = \frac{r^{\mathsf{(n)}}_{i,m} d'}{R^{\mathsf{(n)}}_{i,j}},
	\end{equation}
	where \(d'\) denotes the input-data size per CPU cycle, measured in bits/cycle.
	The local task completion time for \( u_i \)'s task \( d^\mathsf{(n)}_{i,m} \) is defined as \( \frac{r^\mathsf{(n)}_{i,m}}{f_i} \), while the edge computing completion time (i.e., \( u_i \) offloads \( d^\mathsf{(n)}_{i,m} \) to ES \( s_j \)) is
	\begin{equation}
		T^\mathsf{edge,(n)}_{i,j,m} = \frac{r^\mathsf{(n)}_{i,m}}{f_j^\mathsf{E}} + T^\mathsf{trans,(n)}_{i,j,m}.
	\end{equation}
	Thus, the time saved by using edge services is given by
	\begin{equation}
		T^\mathsf{save,(n)}_{i,j,m} = \frac{r^\mathsf{(n)}_{i,m}}{f_i} - T^\mathsf{edge,(n)}_{i,j,m},
	\end{equation}
	and the energy consumption that can be saved is given by
	\begin{equation}
		c^\mathsf{save,(n)}_{i,j,m} = \frac{r^\mathsf{(n)}_{i,m}}{f_i} e^{\mathsf{loc}}_i - T^\mathsf{trans,(n)}_{i,j,m} e^{\mathsf{tran}}_i.
	\end{equation}
	We define the valuation (profit from saving time and energy) that ES \( s_j \) serving task \( d^\mathsf{(n)}_{i,m} \) brings to UAV \( u_i \) as
	\begin{equation}
		v^{\mathsf{(n)}}_{i,j,m} = \omega_4 T^\mathsf{save,(n)}_{i,j,m} + \omega_5 c^\mathsf{save,(n)}_{i,j,m},
	\end{equation}
	where \( \omega_4 \) and \( \omega_5 \) are two positive weighting coefficients.
	Accordingly, the utility of a UAV from trading with ESs involves three components: \textit{(i)} the valuation from matched ESs minus payments; \textit{(ii)} penalties for breaching agreements (e.g., when $\alpha_{i, j,m}=0$ in the corresponding STU); and \textit{(iii)} compensation received when the UAV's task is selected as a volunteer. This can be expressed as
	\begin{equation}\label{equ.utility of UAV}{\small
			\begin{aligned}
				&U^\mathsf{U}\left (u_i, \varphi^{\mathsf{(n)}}(u_i),\bm{\mathbb C}^{\mathsf{(n)}}\right ) = \\& \sum_{s_j \in \varphi^{\mathsf{(n)}}(u_i)} \sum_{d_{i,m}^{\mathsf{(n)}}\in\mathcal T_{i,j}^{\mathsf{(n)}}} \alpha_{i, j,m}^{\mathsf{(n)}} (1 - \lambda^\mathsf{(n)}_{i,j,m}) \left( v_{i,j,m}^{\mathsf{(n)}} - p^\mathsf{(n)}_{i,j,m} \right) \\
				& +\hspace{-4mm} \sum_{s_j \in \varphi^{\mathsf{(n)}}(u_i)} \sum_{d_{i,m}^{\mathsf{(n)}}\in\mathcal T_{i,j}^{\mathsf{(n)}}} \left( - (1 - \alpha_{i, j,m}^{\mathsf{(n)}}) q^\mathsf{U} + \alpha_{i, j,m}^{\mathsf{(n)}} \lambda^\mathsf{(n)}_{i,j,m} q^\mathsf{E} \right).
		\end{aligned}}
	\end{equation}
	
	{
		Since PT agreements are constructed before the actual execution outcomes are realized, PilotAO evaluates the expected UAV utility as
		\begin{equation}
			\label{eq:expected_uav_utility}
			{\small
				\begin{aligned}
					&\overline{U^\mathsf{U}}
					\left(
					u_i,\varphi^\mathsf{(n)}(u_i),\bm{\mathbb C}^{\mathsf{(n)}}
					\right)
					=
					\mathbbm{E}
					\left[
					U^\mathsf{U}
					\left(
					u_i,\varphi^\mathsf{(n)}(u_i),\bm{\mathbb C}^{\mathsf{(n)}}
					\right)
					\right]
					\\
					&=\hspace{-4mm}
					\sum_{s_j\in\varphi^\mathsf{(n)}(u_i)}
					\sum_{d_{i,m}^{\mathsf{(n)}}\in\mathcal T_{i,j}^{\mathsf{(n)}}}
					p_{\alpha,i,j,m}^{\mathsf{(n)}}
					\left(
					1-p_{\lambda\mid\alpha,i,j,m}^{\mathsf{(n)}}
					\right)
					\hspace{-1.5mm}\left(
					\mathbbm{E}
					\left[
					v_{i,j,m}^{\mathsf{(n)}}
					\right]
					\hspace{-.5mm}-\hspace{-.5mm}
					p_{i,j,m}^{\mathsf{(n)}}
					\right)
					\\
					&+\hspace{-4mm}
					\sum_{s_j\in\varphi^\mathsf{(n)}(u_i)}
					\sum_{d_{i,m}^{\mathsf{(n)}}\in\mathcal T_{i,j}^{\mathsf{(n)}}}
					\left[
					p_{\alpha,i,j,m}^{\mathsf{(n)}}
					p_{\lambda\mid\alpha,i,j,m}^{\mathsf{(n)}}
					q^\mathsf{E}
					-
					\left(
					1-p_{\alpha,i,j,m}^{\mathsf{(n)}}
					\right)
					q^\mathsf{U}
					\right].
			\end{aligned}}
		\end{equation}
		Here, $\mathbbm{E}[v_{i,j,m}^{\mathsf{(n)}}]$ denotes the expected service valuation conditioned on successful execution, i.e., $\alpha_{i,j,m}^{\mathsf{(n)}}=1$ and $\lambda_{i,j,m}^{\mathsf{(n)}}=0$. Its numerical evaluation is provided in Appx.~A.}
	
	UAVs may suffer utility degradation when execution conditions deviate from signing-stage expectations. Mobility may make a contracted ES inaccessible and trigger UAV-side transfer, while changes in channel quality, task demand, and ES resource availability may reduce service benefit or increase non-fulfillment risk. We define the UAV utility-shortfall risk as
	\begin{equation}
		{\small\begin{aligned}
				\hspace{-1.5mm}R^{\mathsf{U}}\hspace{-.5mm}
				\left(
				u_i,
				\varphi^{\mathsf{(n)}}(u_i),
				\bm{\mathbb C}^{\mathsf{(n)}}
				\right)\hspace{-.5mm}=\hspace{-.5mm}
				\Pr\!\left(
				U^\mathsf{U}
				\left(
				u_i,
				\varphi^{\mathsf{(n)}}(u_i),
				\bm{\mathbb C}^{\mathsf{(n)}}
				\right)
				\hspace{-.5mm}<\hspace{-.5mm}
				u_{\min}
				\right).
		\end{aligned}}
	\end{equation}
	
	\subsection{Core Definitions of Matching}
	We next formalize the M2M matching relationship adopted in PilotAO.
	
	\begin{Defn}[M2M Matching in PilotAO]
		An M2M matching $\varphi^\mathsf{(n)}$ defines the association between the candidate UAV set $\bm{\mathcal U}^\mathsf{(n)}$ and the ES set $\bm{\mathcal S}$ for STU $n$ and satisfies the following conditions:
		
		\noindent
		$\bullet$ For each UAV $u_i\in\bm{\mathcal U}^\mathsf{(n)}$, $\varphi^\mathsf{(n)}(u_i)\subseteq\bm{\mathcal S}$;
		
		\noindent
		$\bullet$ For each ES $s_j\in\bm{\mathcal S}$, $\varphi^\mathsf{(n)}(s_j)\subseteq\bm{\mathcal U}^\mathsf{(n)}$;
		
		\noindent
		$\bullet$ For each UAV $u_i\in\bm{\mathcal U}^\mathsf{(n)}$ and ES $s_j\in\bm{\mathcal S}$, $u_i\in\varphi^\mathsf{(n)}(s_j)$ if and only if $s_j\in\varphi^\mathsf{(n)}(u_i)$;
		
		\noindent$\bullet$ Each UAV task can be assigned to at most one ES, such that $\mathcal T_{i,j}^{\mathsf{(n)}}\cap\mathcal T_{i,j'}^{\mathsf{(n)}}=\varnothing$ for any $s_j\neq s_{j'}$, $\forall u_i\in\bm{\mathcal U}^{\mathsf{(n)}}$.
	\end{Defn}
	
	We next introduce the notion of a \textit{blocking coalition}, which captures situations where participants have incentives to depart from the current matching and is central to the subsequent stability analysis~\cite{related work futures2}.
	
	\begin{Defn}[Blocking Coalition]
		Under a given matching $\varphi^\mathsf{(n)}$, an ES $s_j$ and a UAV subset $\mathbb U\subseteq\bm{\mathcal U}^\mathsf{(n)}$ form a blocking coalition, denoted by $(s_j;\mathbb U)$, if either of the following two types of mutually beneficial deviation exists.
		
		\noindent \textbf{Type 1 blocking coalition:} This type represents a reassignment of existing associations and satisfies the following two conditions:
		
		\noindent
		$\bullet$ ES $ s_j $ prefers a UAV set $ \mathbb{U} \subseteq \bm{\mathcal{U}}^\mathsf{(n)} $ rather than its currently matched UAV set $ \varphi^\mathsf{(n)}(s_j) $, i.e., 
		\begin{equation}\label{equ. 19}
			\begin{aligned}
				\overline{U^\mathsf{S}}\left(s_j,\mathbb{U},\bm{\mathbb C}^{\mathsf{(n)}}\right)>\overline{U^\mathsf{S}}\left(s_j,\varphi^\mathsf{(n)}(s_j),\bm{\mathbb C}^{\mathsf{(n)}}\right). 
			\end{aligned} 
		\end{equation}

		\noindent
		$\bullet$ Every UAV in \(\mathbb U\) prefers the proposed reassignment involving ES \(s_j\) to its currently matched ES set. That is, for any UAV \(u_i\in\mathbb U\), there exists an evicted ES subset \(\varphi^{\mathsf{(n)}\prime}(u_i)\subseteq\varphi^\mathsf{(n)}(u_i)\) such that
		\begin{equation}\label{eq:type1_uav_blocking_preference}
			\begin{aligned}
				&\overline{U^{\mathsf U}}\left(
				u_i,
				\left(
				\varphi^\mathsf{(n)}(u_i)\setminus
				\varphi^{\mathsf{(n)}\prime}(u_i)
				\right)\cup\{s_j\},
				\bm{\mathbb C}^{\mathsf{(n)}}
				\right)\\
				&>
				\overline{U^{\mathsf U}}\left(
				u_i,
				\varphi^\mathsf{(n)}(u_i),
				\bm{\mathbb C}^{\mathsf{(n)}}
				\right).
			\end{aligned}
		\end{equation} 
		
		\noindent \textbf{Type 2 blocking coalition:} This type represents an expansion of the current matching and satisfies the following two conditions:
		
		\noindent
		$\bullet$ ES \(s_j\) strictly prefers adding UAV set \(\mathbb U\subseteq\bm{\mathcal U}^{\mathsf{(n)}}\) to its current matching, i.e.,
		\begin{equation}
			{\small\begin{aligned}
					&\overline{U^{\mathsf S}}\left(
					s_j,
					\varphi^\mathsf{(n)}(s_j)\cup\mathbb U,
					\bm{\mathbb C}^{\mathsf{(n)}}
					\right)>
					\overline{U^{\mathsf S}}\left(
					s_j,
					\varphi^\mathsf{(n)}(s_j),
					\bm{\mathbb C}^{\mathsf{(n)}}
					\right).
			\end{aligned}}
			\label{eq:type2_es_blocking_preference}
		\end{equation}
		
		\noindent
		$\bullet$ Every UAV \(u_i\in\mathbb U\) strictly prefers adding \(s_j\) to its current matching, i.e.,
		\begin{equation}
			{\small\begin{aligned}
					\overline{U^{\mathsf U}}\left(
					u_i,
					\varphi^\mathsf{(n)}(u_i)\cup\{s_j\},
					\bm{\mathbb C}^{\mathsf{(n)}}
					\right)>
					\overline{U^{\mathsf U}}\left(
					u_i,
					\varphi^\mathsf{(n)}(u_i),
					\bm{\mathbb C}^{\mathsf{(n)}}
					\right).
			\end{aligned}}
			\label{eq:type2_uav_blocking_preference}
		\end{equation}
	\end{Defn}
	Type~1 blocking replaces existing matches, whereas Type~2 blocking expands the current matching without replacement.
	
	\subsection{Problem Formulation}
	Before service execution, UAV--ES service relationships are established through M2M matching. For each STU $n$, UAVs and ESs form mutually acceptable agreements that maximize their respective expected utilities subject to price, capacity, and risk constraints.
	Each $u_i\in\bm{\mathcal U}^{\mathsf{(n)}}$ seeks to maximize its expected utility for STU $n$:
	\begin{subequations}
		\begin{align}
			\bm{\mathcal{F}^\mathsf{U}}:~&\underset{{\varphi^\mathsf{(n)}\left(u_i\right),\bm{\mathbb C}^{\mathsf{(n)}}}}{\max}~\overline{U^\mathsf{U}}\left(u_i,\varphi^\mathsf{(n)}\left(u_i\right),\bm{\mathbb C}^{\mathsf{(n)}}\right)\tag{23}\label{ProUAV}\\
			\text{s.t.}~~~
			&\varphi^\mathsf{(n)}\left(u_i\right)\subseteq\bm{\mathcal{S}} \tag{23a}\label{23a}\\
			&\mathbbm{E}[v_{i,j,m}^{\mathsf{(n)}}]\geq p_{i,j,m}^{\mathsf{(n)}},~\forall s_j\in\varphi^\mathsf{(n)}(u_i),~\forall d_{i,m}^{\mathsf{(n)}}\in\mathcal T_{i,j}^{\mathsf{(n)}}\tag{23b}\label{23b}\\
			&R^{\mathsf U}\left(u_i,\varphi^\mathsf{(n)}(u_i),\bm{\mathbb C}^{\mathsf{(n)}}\right)\leq\rho_1.\tag{23c}\label{23c}
		\end{align}
	\end{subequations}
	{Here, $0<\rho_1\leq1$ denotes the UAV-side risk threshold. Constraint~(\ref{23a}) restricts the matching of UAV $u_i$ to ESs in $\bm{\mathcal S}$. Constraint~(\ref{23b}) ensures that the expected service valuation of each assigned task covers its corresponding payment, while constraint~(\ref{23c}) bounds the UAV-side utility-shortfall risk. Further derivations associated with these constraints are provided in Appx.~A.}
	
	Similarly, each $s_j\in\bm{\mathcal S}$ maximizes its expected utility as:
	\begin{subequations}
		\begin{align}
			\bm{\mathcal{F}^\mathsf{S}}:~&\underset{{\varphi^\mathsf{(n)}\left(s_j\right),\bm{\mathbb C}^{\mathsf{(n)}}}}{\max}~\overline{U^\mathsf{S}}\left(s_j,\varphi^\mathsf{(n)}\left(s_j\right),\bm{\mathbb C}^{\mathsf{(n)}}\right)\tag{24}\label{ProES}\\
			\text{s.t.}~~~
			&\varphi^\mathsf{(n)}\left(s_j\right)\subseteq\bm{\mathcal{U}}^\mathsf{(n)} \tag{24a}\label{equ.25a}\\
			&c_{i,j,m}^{\mathsf{E,(n)}}\leq p_{i,j,m}^{\mathsf{(n)}},~\forall u_i\in\varphi^\mathsf{(n)}(s_j),~\forall d_{i,m}^{\mathsf{(n)}}\in\mathcal T_{i,j}^{\mathsf{(n)}} \tag{24b}\label{equ.25b}\\
			&\sum_{u_i\in\varphi^\mathsf{(n)}(s_j)}|\mathcal T_{i,j}^{\mathsf{(n)}}|+\kappa_j^{\mathsf{(n)}}\leq(1+\tau)G_j\tag{24c}\label{equ.25c}\\
			&R^\mathsf{S}_1\left( s_j,u_i,\mathbb{C}_{i,j}^{\mathsf{(n)}} \right)\leq \rho_2,~\forall u_i\in \varphi^\mathsf{(n)} \left( s_j \right)\tag{24d}\label{equ.25d}\\
			&R_2^{\mathsf S}\left(s_j,u_i,d_{i,m}^{\mathsf{(n)}}\right)\leq\rho_3,~
			\forall u_i\in\varphi^\mathsf{(n)}(s_j),~
			\forall d_{i,m}^{\mathsf{(n)}}\in\mathcal T_{i,j}^{\mathsf{(n)}}.
			\tag{24e}\label{equ.25e}
		\end{align}
	\end{subequations}
	Here, $\tau$ denotes the overbooking rate, while $\rho_2,\rho_3\in(0,1]$ denote the ES-side risk thresholds. Constraint~(\ref{equ.25a}) restricts the ES matching to candidate UAVs in $\bm{\mathcal U}^{\mathsf{(n)}}$, and constraint~(\ref{equ.25b}) ensures that each service payment covers the corresponding ES service cost. Constraint~(\ref{equ.25c}) limits the signed task demand and predicted mean inherent workload to the overbooked signing capacity $(1+\tau)G_j$. Constraints~(\ref{equ.25d}) and~(\ref{equ.25e}) respectively bound the ES-side utility-shortfall risk and the shortage risk of executable contracted tasks.
	
	Our goal is to address the multi-objective optimization (MOO) problem induced by the UAV- and ES-side objectives in (\ref{ProUAV}) and (\ref{ProES}). Conventional M2M matching methods are not readily suited to the spatio-temporal dynamics, stochastic uncertainties, and associated execution risks of STD-EUNs. To cope with these challenges, we next develop PilotAO.

	\subsection{Solution Design: Structure of PilotAO}
	For each candidate task--ES contract, let $p_{i,j,m}^{\min}$ denote its minimum admissible initial payment satisfying $p_{i,j,m}^{\min}\geq c_{i,j,m}^{\mathsf{E,(n)}}$, ensuring ES-side price feasibility. We next present PilotAO in Alg.~\ref{Matching}, which extends the Gale--Shapley matching framework~\cite{related work futures2} to incorporate task pricing, risk constraints, and overbooking. PilotAO constructs a stable, risk-aware M2M matching between UAVs and ESs while satisfying price, capacity, and risk feasibility.
	\begin{algorithm}[t]
		{\small
			\caption{PilotAO}
			\label{Matching}
			\LinesNumbered
			
			\textbf{Initialization:}
			$p_{i,j,m}^{\mathsf{(n)}}\langle1\rangle\leftarrow p_{i,j,m}^{\min}$,
			$\mathrm{flag}_i\leftarrow1$,
			$\mathbb Y(s_j)\leftarrow\varnothing$,
			$\mathbb Z_{i,m}\leftarrow\varnothing$,
			$k^\prime\leftarrow1$\;
			
			Calculate the fixed task-specific ES preference lists $\bm L_{i,m}$\;
			
			\While{$\sum_{u_i\in\bm{\mathcal U}^{\mathsf{(n)}}}\mathrm{flag}_i>0$}{
				
				$\mathrm{flag}_i\leftarrow0$, $\forall u_i\in\bm{\mathcal U}^{\mathsf{(n)}}$\;
				
				$\widetilde{\mathbb Y}(s_j)\leftarrow\mathbb Y(s_j)$,
				$\forall s_j\in\bm{\mathcal S}$\;
				
				\For{$\forall u_i\in\bm{\mathcal U}^{\mathsf{(n)}}$}{
					
					\For{$\forall$ currently unassigned $d_{i,m}^{\mathsf{(n)}}\in\bm D_i^{\mathsf{(n)}}$}{
						
						\If{there exists $s_j\in\bm L_{i,m}\setminus\mathbb Z_{i,m}$ satisfying
							(\ref{23b})--(\ref{23c}) and
							$s_j\in\widehat{\mathcal E}_i^{\mathsf{(n)}}$}{
							
							$s_j^\star\leftarrow$ highest-ranked such ES in $\bm L_{i,m}$\;
							
							$\widetilde{\mathbb Y}(s_j^\star)
							\leftarrow
							\widetilde{\mathbb Y}(s_j^\star)
							\cup
							\{d_{i,m}^{\mathsf{(n)}}\}$\;
						}
					}
				}
				
				\For{$\forall s_j\in\bm{\mathcal S}$}{
					
					Rank tasks in $\widetilde{\mathbb Y}(s_j)$ by
					$\bar u_{i,j,m}^{\mathsf S,(n)}\langle k^\prime\rangle$\;
					
					$\mathbb Y(s_j)\leftarrow$
					highest-ranked feasible tasks satisfying
					(\ref{equ.25b})--(\ref{equ.25e})\;
					
					$\mathbb R(s_j)
					\leftarrow
					\widetilde{\mathbb Y}(s_j)\setminus\mathbb Y(s_j)$\;
				}
				
				$p_{i,j,m}^{\mathsf{(n)}}\langle k^\prime+1\rangle
				\leftarrow
				p_{i,j,m}^{\mathsf{(n)}}\langle k^\prime\rangle$\;
				
				\For{$\forall d_{i,m}^{\mathsf{(n)}}\in\mathbb R(s_j),~
					\forall s_j\in\bm{\mathcal S}$}{
					
					$p^+\leftarrow
					\min\!\left\{
					p_{i,j,m}^{\mathsf{(n)}}\langle k^\prime\rangle+\Delta p,
					\mathbbm E[v_{i,j,m}^{\mathsf{(n)}}]
					\right\}$\;
					
					\eIf{$p^+>
						p_{i,j,m}^{\mathsf{(n)}}\langle k^\prime\rangle$
						and (\ref{23b})--(\ref{23c}) remain satisfied}{
						
						$p_{i,j,m}^{\mathsf{(n)}}\langle k^\prime+1\rangle
						\leftarrow p^+$\;
						
						$\mathrm{flag}_i\leftarrow1$\;
					}{
						$\mathbb Z_{i,m}
						\leftarrow
						\mathbb Z_{i,m}\cup\{s_j\}$\;
					}
				}
				
				\If{$\sum_{u_i}\mathrm{flag}_i>0$}{
					$k^\prime\leftarrow k^\prime+1$\;
				}
			}
			
			$p_{i,j,m}^{\star}
			\leftarrow
			p_{i,j,m}^{\mathsf{(n)}}\langle k^\prime\rangle$\;
			
			Update the task-side and ES-side preferences using $p_{i,j,m}^{\star}$, and run fixed-price task-proposing deferred acceptance to obtain $\mathbb Y^\star(s_j)$\;
			
			$\mathcal T_{i,j}^{\mathsf{(n)}}\leftarrow\mathbb Y^\star(s_j)\cap\bm D_i^{\mathsf{(n)}}$\;
			
			$\varphi^\mathsf{(n)}(u_i)\leftarrow\{s_j\mid\mathcal T_{i,j}^{\mathsf{(n)}}\neq\varnothing\}$,
			$\quad
			\varphi^\mathsf{(n)}(s_j)\leftarrow\{u_i\mid\mathcal T_{i,j}^{\mathsf{(n)}}\neq\varnothing\}$,
			and construct $\mathbb C_{i,j}^{\mathsf{(n)}}$\;
			
			\textbf{Return:}
			$\{\mathbb C_{i,j}^{\mathsf{(n)}}\}_{i,j}$,
			$\varphi^\mathsf{(n)}(u_i)$,
			$\varphi^\mathsf{(n)}(s_j)$
		}
	\end{algorithm}

	\noindent
	{\textbf{Step 1. Initialization (lines 1--2):} Each candidate task--ES payment is initialized as $p_{i,j,m}^{\mathsf{(n)}}\langle1\rangle=p_{i,j,m}^{\min}$, with $\mathbb Y(s_j)=\varnothing$, $\mathbb Z_{i,m}=\varnothing$, and $k^\prime=1$. Here, $\mathbb Y(s_j)$ denotes the set of tasks temporarily retained by ES $s_j$, while $\mathbb Z_{i,m}$ records the ESs that are no longer considered by task $d_{i,m}^{\mathsf{(n)}}$. Before the matching iterations, the expected service valuation $\mathbbm E[v_{i,j,m}^{\mathsf{(n)}}]$ is evaluated for each feasible UAV--task--ES candidate contract and reused throughout PilotAO\footnote{{The expected service valuation $\mathbbm E[v_{i,j,m}^{\mathsf{(n)}}]$ is evaluated from the signing-stage UAV-location and channel statistics. Since these statistics remain unchanged during a given invocation of PilotAO, the resulting valuation is consistently used throughout its price-adjustment rounds. The corresponding numerical evaluation and computational analysis are provided in Appx.~A.}}. The task-specific ES preference list is defined as follows.}
	
	\begin{Defn}[\textbf{ES preference list}]
		For each task $d_{i,m}^{\mathsf{(n)}}$, the initial expected UAV-side utility contribution associated with candidate ES $s_j$ is defined as
		\begin{equation}
			{\small	\begin{aligned}
					&\hspace{-2mm}\bar u_{i,j,m}^{\mathsf U,(n)}\langle1\rangle
					={}
					p_{\alpha,i,j,m}^{\mathsf{(n)}}
					\left(
					1-p_{\lambda\mid\alpha,i,j,m}^{\mathsf{(n)}}
					\right)
					\left(
					\mathbbm E[v_{i,j,m}^{\mathsf{(n)}}]
					-p_{i,j,m}^{\mathsf{(n)}}\langle1\rangle
					\right)\\
					&+
					p_{\alpha,i,j,m}^{\mathsf{(n)}}
					p_{\lambda\mid\alpha,i,j,m}^{\mathsf{(n)}}
					q^{\mathsf E}
					-
					\left(
					1-p_{\alpha,i,j,m}^{\mathsf{(n)}}
					\right)
					q^{\mathsf U}.
			\end{aligned}}
			\label{eq:task_es_expected_uav_utility}
		\end{equation}
		The preference list $\bm L_{i,m}$ ranks the initially admissible ESs in non-increasing order of $\bar u_{i,j,m}^{\mathsf U,(n)}\langle1\rangle$, with ties resolved by a fixed tie-breaking rule. This ranking remains unchanged throughout the price-adjustment phase.
	\end{Defn}
	For ES-side task selection, the expected utility contribution of candidate task $d_{i,m}^{\mathsf{(n)}}$ to ES $s_j$ in round $k^\prime$ is
	\begin{equation}
		{\small	\begin{aligned}
				\bar u_{i,j,m}^{\mathsf S,(n)}\langle k^\prime\rangle
				={}&
				p_{\alpha,i,j,m}^{\mathsf{(n)}}
				\left(
				1-p_{\lambda\mid\alpha,i,j,m}^{\mathsf{(n)}}
				\right)
				\left(
				p_{i,j,m}^{\mathsf{(n)}}\langle k^\prime\rangle
				-c_{i,j,m}^{\mathsf{E,(n)}}
				\right)\\
				&+
				\left(
				1-p_{\alpha,i,j,m}^{\mathsf{(n)}}
				\right)
				q^{\mathsf U}
				-
				p_{\alpha,i,j,m}^{\mathsf{(n)}}
				p_{\lambda\mid\alpha,i,j,m}^{\mathsf{(n)}}
				q^{\mathsf E}.
		\end{aligned}}
		\label{eq:task_es_expected_es_utility}
	\end{equation}
	
	\noindent
	\textbf{Step 2. Task proposal (lines 3--10):} In each round $k^\prime$, every unassigned task $d_{i,m}^{\mathsf{(n)}}$ identifies the admissible ESs in $\bm L_{i,m}\setminus\mathbb Z_{i,m}$ according to the accessibility and UAV-side feasibility conditions, and proposes to the highest-ranked available ES.
	
	\noindent
	\textbf{Step 3. Task selection at ESs (lines 11--14):} Each ES $s_j$ jointly considers its previously retained tasks and newly received proposals in $\widetilde{\mathbb Y}(s_j)$. The candidate tasks are ranked according to $\bar u_{i,j,m}^{\mathsf S,(n)}\langle k^\prime\rangle$, after which $s_j$ retains the highest-ranked feasible tasks subject to constraints~(\ref{equ.25b})--(\ref{equ.25e}). The retained and rejected tasks are recorded in $\mathbb Y(s_j)$ and $\mathbb R(s_j)$, respectively.
	
	\noindent
	\textbf{Step 4. Payment update (lines 15--22):} For each rejected task $d_{i,m}^{\mathsf{(n)}}\in\mathbb R(s_j)$, its payment to ES $s_j$ is increased by at most $\Delta p$ if the resulting offer remains feasible under the UAV-side price and risk constraints. Otherwise, $s_j$ is added to $\mathbb Z_{i,m}$ and excluded from subsequent proposals of task $d_{i,m}^{\mathsf{(n)}}$.
	
	\noindent
	\textbf{Step 5. Termination and final assignment (lines 23--29):} If no rejected task admits a further feasible payment increase, the price-adjustment phase terminates; otherwise, Steps~2--4 are repeated. PilotAO then updates the task-side and ES-side preferences using the resulting payments and performs fixed-price task-proposing deferred acceptance to obtain the final task assignments. These assignments determine $\mathcal T_{i,j}^{\mathsf{(n)}}$, $\varphi^\mathsf{(n)}(u_i)$, $\varphi^\mathsf{(n)}(s_j)$, and the corresponding agreements $\mathbb C_{i,j}^{\mathsf{(n)}}$. 
	
	Upon termination of PilotAO, each ES $s_j$ establishes PT agreements with its matched UAVs according to the resulting matching and payments. During practical execution, however, deviations in UAV mobility, task demand, and ES resource availability may gradually reduce the applicability of these agreements and make the current overbooking rate less suitable for the realized market state. Such mismatches can lead to agreement non-fulfillment and performance degradation. To maintain adaptability under evolving conditions, we introduce AdaptAO, whose design and operation are presented in Sec.~\ref{Chap 4}.

	\subsection{Matching Properties of PilotAO}
	We next analyze the key matching properties of PilotAO\footnote{{The matching-property analysis follows the utility-based preference and ES-selection rules defined in PilotAO and assumes truthful reporting of the information used for PT-agreement construction. The formal conditions used in the corresponding proofs are provided in Appx.~D.}}.
	
	\begin{Defn}[Individual rationality of PilotAO]
		\label{def:pilotao_individual_rationality}
		For both ESs and UAVs, a matching \( \varphi^\mathsf{(n)} \) is individually rational when the following conditions are satisfied:
		
		\noindent
		$\bullet$ For ESs:
		\textit{(i)} the payment for every selected task covers the corresponding service cost, i.e., constraint~(\ref{equ.25b}) is satisfied;
		\textit{(ii)} the signed UAV-task demand together with the predicted inherent load does not exceed the overbooked signing capacity, i.e., constraint~(\ref{equ.25c}) is satisfied;
		\textit{(iii)} the ES-side utility-shortfall risk is controlled within the prescribed tolerance, i.e., constraint~(\ref{equ.25d}) is satisfied; and
		\textit{(iv)} the conditional ES-shortage risk of every selected task--ES contract is controlled within the prescribed tolerance, i.e., constraint~(\ref{equ.25e}) is satisfied.
		
		\noindent
		$\bullet$ For UAVs:
		\textit{(i)} the expected valuation obtained by \(u_i\) from \(s_j\) covers its corresponding payment, i.e., constraint (\ref{23b}) is satisfied; and
		\textit{(ii)} the risk of each UAV obtaining an undesired utility is controlled within a reasonable range, i.e., meet constraint (\ref{23c}).
	\end{Defn}

	\begin{Defn}[Fairness of PilotAO]
		A matching \(\varphi^\mathsf{(n)}\) is fair if and only if it admits no Type~1 blocking coalition.
	\end{Defn}
	
	\begin{Defn}[Non-wastefulness of PilotAO]
		A matching \(\varphi^\mathsf{(n)}\) is non-wasteful if and only if it admits no Type~2 blocking coalition.
	\end{Defn}
	
	\begin{Defn}[Strong stability of PilotAO] A matching $\varphi^\mathsf{(n)}$ of PilotAO is strongly stable if it is individual rational, fair, and non-wasteful.
	\end{Defn}
	
	{	Competitive equilibrium characterizes whether UAVs and ESs can reach mutually acceptable trading outcomes under feasible prices and resource constraints. For the discrete M2M market considered in this work, we adopt a competitive equilibrium criterion based on bilateral price feasibility, utility-based choices over the feasible contract domain, and the absence of further feasible utility-improving admissions~\cite{related work futures2}.
		
		\begin{Defn}[Competitive-equilibrium criterion for UAV--ES trading]
			A UAV--ES trading outcome satisfies the competitive equilibrium criterion adopted in this work if the following conditions hold:
			
			\noindent
			$\bullet$ Every selected task--ES contract satisfies bilateral price feasibility, i.e.,
			$
			c_{i,j,m}^{\mathsf{E,(n)}}
			\leq
			p_{i,j,m}^{\mathsf{(n)}}
			\leq
			\mathbbm E[v_{i,j,m}^{\mathsf{(n)}}].
			$
			
			\noindent
			$\bullet$ Under the resulting payments, each UAV task selects its preferred feasible ES according to the corresponding expected-utility ranking subject to constraints~(\ref{23a})--(\ref{23c}).
			
			\noindent
			$\bullet$ Under the resulting payments, each ES selects its preferred feasible task set according to the corresponding expected-utility ranking subject to constraints~(\ref{equ.25a})--(\ref{equ.25e}).
			
			\noindent
			$\bullet$ No additional task--ES contract can be added such that the involved UAV task strictly prefers the additional service relationship, the corresponding ES obtains a strictly higher expected utility, and the applicable price, capacity, and risk constraints remain satisfied.
		\end{Defn}
		
		In the considered discrete M2M market, weak Pareto optimality is defined with respect to the reallocations permitted by the proposed matching model and the UAV- and ES-side constraints in~(\ref{ProUAV}) and~(\ref{ProES}). A feasible Pareto improvement refers to a reallocation that strictly increases the expected utility of every participant affected by the change while leaving the other participants unchanged. If no such improvement exists, the resulting trading outcome is regarded as weakly Pareto optimal. Reallocations or side payments not represented in the proposed model are not considered~\cite{related work futures2,parato}.
		\begin{Prop}[Weak Pareto optimality of PilotAO]
			\label{prop:pilotao_weak_pareto}
			PilotAO yields a weakly Pareto-optimal service-trading outcome.
	\end{Prop}}
	
	Detailed properties of PilotAO are found in Appx.~D.
	
	\section{AdaptAO of PAST: Enabling Adaptive Agreements and Overbooking Updates During Practical BTUs}\label{Chap 4}
	PT agreements are constructed from historical or predicted information before execution. However, evolving UAV mobility, task demand, channel conditions, and ES availability can make retained agreements and overbooking rates mismatched with actual conditions, causing inaccessible UAV--ES associations, idle commitments, capacity shortages, and non-fulfillment. We therefore develop AdaptAO to determine agreement retention or renewal from the observed state and, upon renewal, the overbooking rate for PilotAO to reconstruct the agreements. This retain-or-renew mechanism enables agreement reuse while adapting to dynamic UAV--edge conditions.

	We formulate the sequential decision process of AdaptAO as a Markov decision process (MDP). The trading horizon contains $T$ BTUs, each consisting of $N$ STUs, where $t$ and $n$ denote the BTU and STU indices, respectively. For each STU $n$, the MDP is represented by the 5-tuple $(\mathbb S^{\mathsf{(n)}},\mathbb A^{\mathsf{(n)}},\nu^{\mathsf{(n)}},\mathbb P^{\mathsf{(n)}},\mathbb R^{\mathsf{(n)}})$~\cite{RL,DQN}. Here, $\mathbbm s_t^{\mathsf{(n)}}\in\mathbb S^{\mathsf{(n)}}$ denotes the state observed at STU $n$ of practical BTU $t$, and $\mathbbm a_t^{\mathsf{(n)}}\in\mathbb A^{\mathsf{(n)}}$ denotes the corresponding action. Moreover, $\nu^{\mathsf{(n)}}\in[0,1)$ is the discount factor, $\mathbb P^{\mathsf{(n)}}$ characterizes the state transition probability $\mathbb P(\mathbbm s_{t+1}^{\mathsf{(n)}}\mid\mathbbm s_t^{\mathsf{(n)}},\mathbbm a_t^{\mathsf{(n)}})$, and $\mathbb R^{\mathsf{(n)}}$ characterizes the immediate reward associated with each state--action transition. The state, action, and reward design are detailed below.
	
	\noindent
	{$\bullet$ \textit{State:}
		AdaptAO represents the state $\mathbbm{s}_{t}^{\mathsf{(n)}}$ as a normalized feature vector comprising \textit{(i)} the current decision stage; \textit{(ii)} estimates of task demand, UAV mobility, and residual ES resources; \textit{(iii)} realized agreement fulfillment, resource utilization, default rate, and social welfare; \textit{(iv)} the active overbooking rate; \textit{(v)} a combined agreement-status-and-retention-duration feature; \textit{(vi)} agreement applicability; and \textit{(vii)} prediction error. Together, these features capture the current system state, the suitability of the existing PT agreements, and the mismatch between the predicted and realized UAV--edge conditions.}
	
	\noindent
	{$\bullet$ \textit{Action:}
		AdaptAO adopts a discrete action space that jointly determines whether the current PT agreements are retained or renewed and, upon renewal, selects the corresponding overbooking rate:
		$\mathbb A^{\mathsf{(n)}}=
		\{a_{\mathsf{keep}}\}
		\cup
		\left\{
		a_{\mathsf{renew}}(\tau_k)
		\,\middle|\,
		\tau_k\in\mathcal T_\tau
		\right\}$,
		where $\mathcal T_\tau=\{\tau_1,\ldots,\tau_K\}$ denotes the finite set of candidate overbooking rates. Accordingly, the action space consists of one retention action and $K$ renewal actions. Action $a_{\mathsf{keep}}$ retains the current PT agreements and active overbooking rate without invoking PilotAO, whereas $a_{\mathsf{renew}}(\tau_k)$ invokes PilotAO to reconstruct the PT agreements using the selected overbooking rate $\tau_k$. For subsequent formulation, define the renewal-action set as
		$\mathbb A_{\mathsf{renew}}^{\mathsf{(n)}}=
		\left\{
		a_{\mathsf{renew}}(\tau_k)
		\,\middle|\,
		\tau_k\in\mathcal T_\tau
		\right\}$}\footnote{{The candidate overbooking rate set adopted in the implementation and the corresponding action-grid analysis are detailed in Appx.~F.}}.
	
	{\noindent
		$\bullet$ \textit{Reward:}
		After applying action $\mathbbm{a}_{t}^{\mathsf{(n)}}$ and executing the resulting agreement configuration in the ensuing interval, AdaptAO evaluates the realized service performance, agreement applicability, prediction accuracy, and renewal overhead through the following per-STU reward:
		\begin{equation}
			\label{27}
			{\small
				\begin{aligned}
					\mathbbm{r}_{t}^{\mathsf{(n)}}
					={}&
					\lambda_{\mathsf{sw}}
					\widetilde{SW}_{t}^{\mathsf{(n)}}
					+
					\lambda_{\mathsf{ful}}
					\mathrm{TRLC}_{t}^{\mathsf{(n)}}
					+
					\lambda_{\mathsf{util}}
					\mathrm{UoR}_{t}^{\mathsf{(n)}}
					+
					\lambda_{\mathsf{app}}
					A_{t}^{\mathsf{(n)}}
					\\
					&
					-
					\lambda_{\mathsf{def}}
					D_{t}^{\mathsf{(n)}}
					-
					\lambda_{\mathsf{err}}
					E_{t}^{\mathsf{(n)}}
					-
					C_{t}^{\mathsf{upd,(n)}}
					-
					\lambda_{\mathsf{ni}}
					\mathrm{NI}_{t}^{\mathsf{(n)}}.
			\end{aligned}}
		\end{equation}
		Here, $\widetilde{SW}_{t}^{\mathsf{(n)}}=SW_{t}^{\mathsf{(n)}}/s_{\mathsf{sw}}$ denotes the normalized realized social welfare, where $SW_{t}^{\mathsf{(n)}}$ is the sum of the realized UAV and ES utilities defined in~(\ref{Utility}) and~(\ref{equ.utility of UAV}). Moreover, $\mathrm{TRLC}_{t}^{\mathsf{(n)}}$, $\mathrm{UoR}_{t}^{\mathsf{(n)}}$, $A_{t}^{\mathsf{(n)}}$, and $D_{t}^{\mathsf{(n)}}$ denote the agreement-fulfillment ratio, residual-resource utilization ratio, agreement-applicability ratio, and default rate, respectively. $E_{t}^{\mathsf{(n)}}$ denotes the prediction error between the estimated and realized task-demand, UAV-mobility, and residual-resource states. The positive coefficients $\lambda_{\mathsf{sw}}$, $\lambda_{\mathsf{ful}}$, $\lambda_{\mathsf{util}}$, $\lambda_{\mathsf{app}}$, $\lambda_{\mathsf{def}}$, $\lambda_{\mathsf{err}}$, and $\lambda_{\mathsf{ni}}$ balance the corresponding reward components.
	}
	{
		The agreement-renewal cost is
		\begin{equation}
			\label{eq:adaptao_update_cost}
			{\small
				\begin{aligned}
					C_{t}^{\mathsf{upd,(n)}}
					\hspace{-.5mm}=\hspace{-.5mm}
					\mathbbm{1}_
					{\left\{
						\mathbbm{a}_{t}^{\mathsf{(n)}}
						\in
						\mathbb A_{\mathsf{renew}}^{\mathsf{(n)}}
						\right\}}
					\left[
					c_{\mathsf{fix}}
					+
					c_{\mathsf{early}}
					\left(
					H_{\mathsf{age}}
					-
					\mathrm{Age}_{t}^{-,\mathsf{(n)}}
					\right)^{+}
					\right],
			\end{aligned}}
		\end{equation}
		where $(x)^{+}=\max\{0,x\}$, $c_{\mathsf{fix}}$ denotes the fixed cost of reconstructing the PT agreements, while $c_{\mathsf{early}}$ controls the additional renewal cost when the current PT agreements have been retained for less than the reference retention horizon $H_{\mathsf{age}}$. Besides, $\mathrm{Age}_{t}^{-,\mathsf{(n)}}$ denotes the elapsed retention duration of the active PT-agreement set since its most recent construction or renewal, measured immediately before the current decision. Moreover, $\mathrm{NI}_{t}^{\mathsf{(n)}}$ denotes the protocol interactions generated by agreement renewal and equals zero when the current agreements are retained}\footnote{{Appx.~G provides the detailed definitions and normalization of the reward components, together with sensitivity analyses of $c_{\mathsf{fix}}$ and $c_{\mathsf{early}}$.}}.
	
	To learn the state-dependent retain-or-renew policy and the overbooking rate selection upon renewal, we employ a double deep Q-network (DDQN)~\cite{DDQN}\footnote{{Appx.~E compares the DDQN-based AdaptAO with alternative controllers and further examines the effects of the learned policy and an external rule-based safeguard.}}. The corresponding offline training procedure is summarized in Alg.~\ref{alg:adaptao_training}. During practical evaluation, the trained network parameters remain fixed, and AdaptAO selects the action with the highest estimated Q-value without further parameter updates.
	
	\begin{algorithm}[t!]
		{\small
			\setstretch{0.4}
			\caption{Offline DDQN Training for AdaptAO}
			\label{alg:adaptao_training}
			\LinesNumbered
			
			\textbf{Initialization:}
			initialize the primary-network parameters $\theta$,
			set $\hat{\theta}\leftarrow\theta$,
			and initialize replay buffer $\mathcal D\leftarrow\varnothing$\;
			
			\For{episode $=1$ \KwTo $E_{\max}$}{
				
				$\{\mathbb C_{i,j}^{\mathsf{(n)}}\}_{n=1}^{N}
				\leftarrow
				\mathrm{PilotAO}(\tau_0)$;
				set the active overbooking rate as $\tau\leftarrow\tau_0$\;
				
				\For{$t=1$ \KwTo $T$}{
					
					\For{$n=1$ \KwTo $N$}{
						
						Observe the most recently available state
						$\mathbbm{s}_{t}^{\mathsf{(n)}}$\;
						
						Select action
						$\mathbbm{a}_{t}^{\mathsf{(n)}}$
						using the $\varepsilon$-greedy policy\;
						
						\eIf{$\mathbbm{a}_{t}^{\mathsf{(n)}}=
							a_{\mathsf{renew}}(\tau_k)$}{
							$\bm{\mathbb C}^{\mathsf{(n)}}
							\leftarrow
							\mathrm{PilotAO}(\tau_k)$;
							$\tau\leftarrow\tau_k$\;
						}{
							Retain the current PT agreements and active overbooking rate\;
						}
						
						Execute the resulting PT agreements in the ensuing execution interval\;
						
						Observe reward $\mathbbm{r}_{t}^{\mathsf{(n)}}$
						and next state
						$\mathbbm{s}_{t+1}^{\mathsf{(n)}}$\;
						
						Store
						$(\mathbbm{s}_{t}^{\mathsf{(n)}},
						\mathbbm{a}_{t}^{\mathsf{(n)}},
						\mathbbm{r}_{t}^{\mathsf{(n)}},
						\mathbbm{s}_{t+1}^{\mathsf{(n)}})$
						in $\mathcal D$\;
						
						Sample a random minibatch $\mathcal B$ from $\mathcal D$\;
						
						\ForEach{$(\mathbbm{s}_{b}^{\mathsf{(n)}},
							\mathbbm{a}_{b}^{\mathsf{(n)}},
							\mathbbm{r}_{b}^{\mathsf{(n)}},
							\mathbbm{s}_{b}^{\mathsf{(n)}\prime})
							\in\mathcal B$}{
							
							$\mathbbm{a}_{b}^{\star}
							\leftarrow
							\arg\max_{\mathbbm a^\prime}
							Q\!\left(
							\mathbbm{s}_{b}^{\mathsf{(n)}\prime},
							\mathbbm a^\prime;
							\theta
							\right)$\;
							
							$y_b
							\leftarrow
							\mathbbm{r}_{b}^{\mathsf{(n)}}
							+
							\nu^{\mathsf{(n)}}
							\hat Q\!\left(
							\mathbbm{s}_{b}^{\mathsf{(n)}\prime},
							\mathbbm{a}_{b}^{\star};
							\hat{\theta}
							\right)$\;
						}
						
						Update $\theta$ by minimizing
						$
						L(\theta)
						=
						\frac{1}{|\mathcal B|}
						\sum_{b\in\mathcal B}
						\left[
						y_b
						-
						Q\!\left(
						\mathbbm{s}_{b}^{\mathsf{(n)}},
						\mathbbm{a}_{b}^{\mathsf{(n)}};
						\theta
						\right)
						\right]^2
						$\;
						
						Update the target network:
						$\hat{\theta}
						\leftarrow
						\mu\theta
						+
						(1-\mu)\hat{\theta}$\;
					}
				}
			}
		}
	\end{algorithm}
	
	\noindent
	\textbf{Step 1: Initialization (lines 1--3)}: The primary-network parameters $\theta$ are initialized, the target-network parameters are set as $\hat{\theta}\leftarrow\theta$, and the replay buffer $\mathcal D$ is initialized as empty. At the beginning of each training episode, PilotAO establishes the initial PT agreements for the considered STUs using the initial overbooking rate $\tau_0$.
	
	{\noindent 
		\textbf{Step 2: State observation and action selection (lines 4--11)}: At each decision stage, AdaptAO observes the available execution state $\mathbbm{s}_{t}^{\mathsf{(n)}}$, which summarizes the predicted resource and demand conditions, realized agreement performance, agreement applicability, retention duration, and prediction error. During offline training, an action is selected from $\mathbb A^{\mathsf{(n)}}$ using the $\varepsilon$-greedy policy~\cite{DQN,DDQN}. If $a_{\mathsf{keep}}$ is selected, the current PT agreements and active overbooking rate are retained. If $a_{\mathsf{renew}}(\tau_k)$ is selected, PilotAO reconstructs the PT agreements using the selected overbooking rate $\tau_k$ and the most recently available information.
		
		\noindent
		\textbf{Step 3: Agreement execution and transition collection (lines 12--14)}: The resulting PT agreements are applied in the ensuing execution interval. After execution, AdaptAO observes the realized system outcome, evaluates the reward $\mathbbm{r}_{t}^{\mathsf{(n)}}$ according to~(\ref{27}), and constructs the next state $\mathbbm{s}_{t+1}^{\mathsf{(n)}}$. The resulting transition $(\mathbbm{s}_{t}^{\mathsf{(n)}},\mathbbm{a}_{t}^{\mathsf{(n)}},\mathbbm{r}_{t}^{\mathsf{(n)}},\mathbbm{s}_{t+1}^{\mathsf{(n)}})$ is then stored in the replay buffer $\mathcal D$.
		
		\noindent
		\textbf{Step 4: DDQN model training (lines 15--20)}: A random minibatch is sampled from $\mathcal D$ for network training. For each sampled transition, the primary network selects the action with the largest estimated Q-value in the next state, while the target network evaluates the corresponding target value~\cite{DDQN}. The primary-network parameters $\theta$ are updated by minimizing the minibatch temporal-difference loss. The target-network parameters $\hat{\theta}$ are then softly updated toward $\theta$, where $\mu\in(0,1]$ denotes the soft-update coefficient.}
	
	\noindent
	\textbf{Step 5: Training completion and practical deployment (lines 2--20)}: Steps~2--4 are repeated throughout the prescribed training episodes. After offline training, the learned network parameters are fixed. During practical execution, AdaptAO directly selects the action with the largest estimated Q-value for the observed state without further network updates.
	
	Through this procedure, AdaptAO learns when to retain or reconstruct PT agreements as UAV--edge conditions evolve. Upon renewal, it determines the overbooking rate for PilotAO, balancing agreement reuse, adaptability, and renewal overhead.

	\section{Evaluation}
	We conduct comprehensive experiments to evaluate the performance of PAST. All experiments are implemented in Python~3.9 on a platform equipped with a 12th Gen Intel Core i9-12900H processor and an NVIDIA GeForce RTX~3060 GPU. The evaluation addresses the following questions:
	
	\noindent
	$\bullet$ \textit{Q1:} How does PAST compare with the benchmark methods in terms of running time, interaction overhead, and practical task-completion time? (Sec.~\ref{chap 5.3.1})
	
	\noindent
	$\bullet$ \textit{Q2:} How does PAST compare with the benchmark trading and matching methods in terms of social welfare and UAV/ES utilities? (Sec.~\ref{chap 5.3.2})
	
	\noindent
	$\bullet$ \textit{Q3:} How does AdaptAO affect agreement fulfillment and residual ES-resource utilization compared with the benchmark methods? (Sec.~\ref{chap 5.3.3})
	
	\noindent
	$\bullet$ \textit{Q4:} Do the resulting payments satisfy the bilateral price-feasibility conditions supporting the adopted individual-rationality criterion? (Sec.~\ref{chap 5.3.4})
	
	{	\subsection{Core Experimental Settings}
		\noindent
		$\bullet$ \textit{Map and scenario:}
		We adopt the 77th Chicago community area represented in the Chicago taxi trip dataset~\cite{dataset} as the primary geographical scenario. UAV mission regions and ES locations are distributed within the considered area to capture spatial heterogeneity in UAV--edge service availability.
		
		\noindent
		$\bullet$ \textit{UAV initialization and task-load modeling:}
		For each UAV, the initial and terminal locations are mapped to the first pick-up and last drop-off locations, respectively, of a selected taxi trajectory within a day, while intermediate mission regions are generated within the considered area and visited sequentially. The candidate task load $|\bm D_i^{\mathsf{(n)}}|$ in STU $n$ is mapped from the corresponding passenger demand. Historical passenger-demand observations over 30 days are used to estimate future UAV task loads for PilotAO, thereby introducing temporal demand variations into the signing-stage agreement construction.
		
		\noindent
		$\bullet$ \textit{UAV trajectory generation and dynamic enhancement:}
		Within each BTU, UAV movement among mission regions follows the Markovian transition model described in Sec.~\ref{chap 2.1}, with transition probabilities determined by the corresponding inter-region distances. To introduce additional spatial dynamics across BTUs, the mission-region coordinates are randomly perturbed with a probability of $5\%$ per BTU. This setting allows us to evaluate the ability of PAST to adapt to changes in UAV mobility and the resulting service-accessibility conditions.

		Other key parameters are configured with reference to the relevant literature. Specifically, the historical task-generation propensity is set to $p_{i,m}^{\mathsf{arr},(n)}\in[0.64,0.96]$~\cite{matching}, characterizing the likelihood that candidate task $d_{i,m}^{\mathsf{(n)}}$ is generated in STU $n$. PilotAO combines this task-generation propensity with the mobility-dependent accessibility of the contracted ES to estimate the executable-arrival probability $p_{\alpha,i,j,m}^{\mathsf{(n)}}$. The corresponding wireless conditions are used to evaluate the conditional expected service valuation, while the shortage probability $p_{\lambda\mid\alpha,i,j,m}^{\mathsf{(n)}}$ is estimated from sampled inherent ES workloads and the competition among executable contracted tasks\footnote{Detailed estimation procedure and its numerical settings are in Appx.~B.}. The remaining parameters are set as follows:}
	$f_i\in[1,1.5]\times10^9$ CPU cycles/s~\cite{related work futures1},
	$f_j^\mathsf{E}\in[1,3]\times10^{12}$ CPU cycles/s~\cite{related work futures2},
	$e_i^{\mathsf{loc}},e_j^{\mathsf E}\in[450,500]$ mW~\cite{related work futures2},
	$d^\prime=1/600$ bits/cycle~\cite{related work futures2},
	$G_j\in[10,15]$,
	$r_{i,m}^{\mathsf{(n)}}\in[6,9]\times10^{8}$ CPU cycles~\cite{related work futures1,related work futures2},
	$e_i^\mathsf{tran}\in[500,550]$ mW,
	$B_{i,j}^\mathsf{(n)}=20$ MHz~\cite{simulation1,related work futures1},
	$g_0=10^{-5}$ (i.e., $-50$ dB)~\cite{simulation1},
	$\varepsilon_j^{\mathsf{(n)}}\sim\mathbf{P}(\kappa_j^{\mathsf{(n)}})$,
	$\zeta=0.2$~\cite{UAV},
	$\kappa_j^{\mathsf{(n)}}\in[5,6]$,
	{
		$q^{\mathsf U}=0.65$,
		$q^{\mathsf E}=0.85$,
		$c_j^{\mathsf{hard}}=0.01$,
		$\rho_1=\rho_2=\rho_3=0.3$~\cite{matching},
		and $\Delta p=0.08$\footnote{{Appx.~C presents one-factor-at-a-time sensitivity analyses of the three risk thresholds, the UAV- and ES-side compensation parameters, and the PilotAO price increment, illustrating their local effects and associated performance tradeoffs over the tested ranges.}},}
	$\omega_2=15$~\cite{simulation1},
	$\omega_3=0.5$~\cite{simulation1},
	$H_i^{\mathsf{uav,(n)}}\in[30,100]$ m,
	$H_0^{\mathsf{ES}}\in[0,2]$ m,
	and $|\bm C_i|\in[3,8]$.
	The results reported in the main text are averaged over 30 independent Monte Carlo realizations, with error bars in bar plots and shaded regions in line plots indicating two-sided 95\% confidence intervals.
	
	\subsection{Benchmark Methods and Evaluation Metrics}
	To provide a comprehensive comparison, we consider the following benchmark methods.
	
	\noindent
	$\bullet$ \textbf{Conventional spot trading-based M2M matching (ConSM)}~\cite{parato}, which performs M2M matching online in each STU according to the current UAV--edge conditions.
	
	\noindent
	$\bullet$ \textbf{Conventional futures trading-based M2M matching without adaptation (ConFM\_NoA)}~\cite{matching,related work futures2}, which establishes PT agreements in advance through conventional futures-based M2M matching but does not adaptively retain or renew them during practical execution.
	
	\noindent
	$\bullet$ \textbf{Conventional futures trading-based M2M matching without overbooking (ConFM\_NoO)}~\cite{related work futures1}, which follows ConFM\_NoA with the overbooking mechanism disabled during advance PT-agreement construction.
	
	\noindent
	$\bullet$ \textbf{Greedy-oriented matching (GrdM)}~\cite{random}, in which ESs greedily prioritize UAV tasks that yield higher ES-side profits subject to their available resources.
	
	\noindent
	$\bullet$ \textbf{Random-based matching (RandM)}~\cite{random}, in which task proposals and ES acceptances are determined randomly.
	
	{To ensure a consistent basis for comparison, GrdM and RandM reuse PAST's action, overbooking rate, and environmental traces, with only the agreement-selection strategy varied. This setting isolates the effect of different agreement-selection strategies under identical adaptation and environmental conditions. Besides, we evaluate the following performance metrics from multiple dimensions:
		
		\noindent
		$\bullet$ \textbf{Realized utilities of ESs and UAVs:} The realized utilities obtained by ESs and UAVs according to (\ref{Utility}) and (\ref{equ.utility of UAV}).
		
		\noindent
		$\bullet$ \textbf{Social welfare (SW):} The sum of the realized utilities of all participating ESs and UAVs.
		
		\noindent
		$\bullet$ \textbf{Running time (RT, ms/STU):} RT measures the average wall-clock computation time required to obtain the trading decision for each STU in our implementation environment.
		
		\noindent
		$\bullet$ \textbf{Number of interactions (NI/STU):} NI measures the average number of proposal, acceptance, rejection, price-update, and agreement-confirmation interactions per STU.
		
		\noindent
		$\bullet$ \textbf{Practical task-completion time (PTCT, ms):} PTCT combines trading-decision computation, protocol-message, task-transmission, and edge-processing delays, capturing task-completion time with trading overhead.
		
		\noindent
		$\bullet$ \textbf{Task completion rate under PT agreements (TRLC):} In each STU, TRLC is defined as the ratio of successfully served executable contracted tasks to all executable contracted tasks.
		
		\noindent
		$\bullet$ \textbf{Utilization of residual ES resources (UoR):} UoR measures the fraction of realized residual ES capacity, after accounting for inherent ES workloads, that is utilized by successfully served UAV tasks.}
	
	{	\subsection{Performance Evaluations}
		\subsubsection{RT, NI, and PTCT}\label{chap 5.3.1}
		\begin{figure}[t]
			\centering
			\setlength{\abovecaptionskip}{-1 mm}
			\includegraphics[width=1\columnwidth]{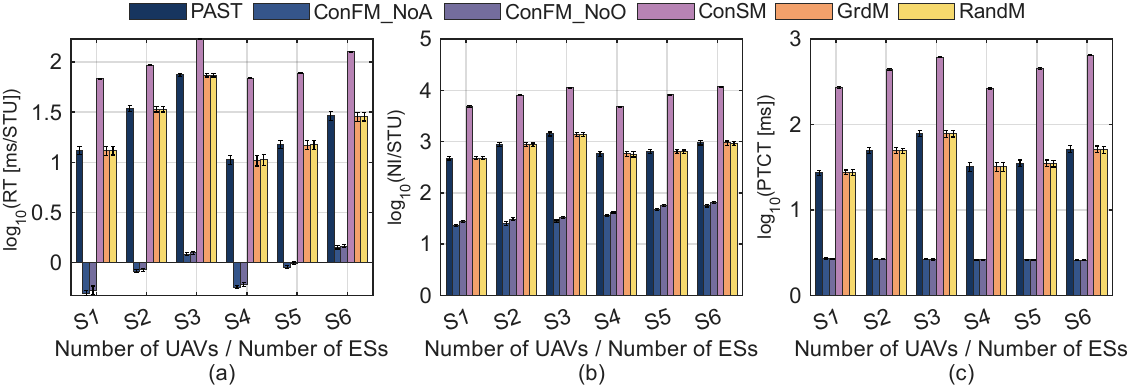}
			\caption{Performance comparisons in terms of (a) RT, (b) NI, and (c) PTCT. S1--S6 correspond to 40/6, 60/6, 80/6, 40/10, 60/10, and 80/10 UAVs/ESs, respectively. }
			\label{NI}
		\end{figure}
		
		We first compare the decision overhead and practical task-completion time of the considered methods under different market scales. As shown in Fig.~\ref{NI}, RT, NI, and PTCT denote the computation time for trading decisions, the resulting interaction overhead, and the task-completion time including trading-decision overhead, respectively. Logarithmic scales are used to facilitate comparisons across methods with different magnitudes.
		
		In Fig.~\ref{NI}(a), ConSM consistently incurs the largest RT since it reconstructs trading relationships according to the current UAV--edge conditions in each STU. In contrast, ConFM\_NoA and ConFM\_NoO achieve the lowest RT by directly reusing pre-established PT agreements without adaptive reconstruction. PAST also substantially reduces RT relative to ConSM by retaining current PT agreements whenever renewal is unnecessary, although its selective renewal incurs additional computation compared with the two static futures-trading baselines. GrdM and RandM exhibit RT values close to PAST, as their simpler agreement-selection strategies reduce reconstruction overhead. Fig.~\ref{NI}(b) shows a similar pattern for protocol-interaction overhead. ConSM incurs the largest NI because trading relationships are repeatedly established during execution, whereas ConFM\_NoA and ConFM\_NoO require the fewest interactions by retaining their pre-established PT agreements without adaptive renewal. PAST incurs more interactions than these static baselines because AdaptAO selectively invokes PilotAO to reconstruct PT agreements as execution conditions evolve, while maintaining substantially lower NI than ConSM. GrdM and RandM exhibit interaction overhead comparable to PAST under the same adaptation process, but use simpler agreement-selection strategies. Fig.~\ref{NI}(c) further compares PTCT. Since task transmission/edge processing follow the resulting decision, the corresponding decision time is included in the task-completion time. ConFM\_NoA and ConFM\_NoO therefore achieve the lowest PTCT by avoiding repeated decision overhead during execution. PAST incurs additional completion time due to selective agreement renewal, but remains substantially below ConSM across all considered market scales. Its PTCT is also close to that of GrdM and RandM, reflecting their comparable decision overhead during execution.
		
		Overall, the results reveal a clear trade-off between adaptive agreement management and decision overhead. The static futures-trading baselines achieve the lowest RT, NI, and PTCT by avoiding agreement reconstruction, whereas ConSM incurs the highest overhead by repeatedly determining trading relationships in each STU. PAST strikes a balance between these extremes by reusing existing PT agreements whenever appropriate and selectively renewing them to accommodate evolving execution conditions. The resulting benefits in trading performance are examined in the following experiments.

		\subsubsection{Social Welfare and UAV/ES Utilities}\label{chap 5.3.2}
		\begin{figure*}[t!]
			
			\centering
			\setlength{\abovecaptionskip}{-1 mm}
			\includegraphics[width=2\columnwidth]{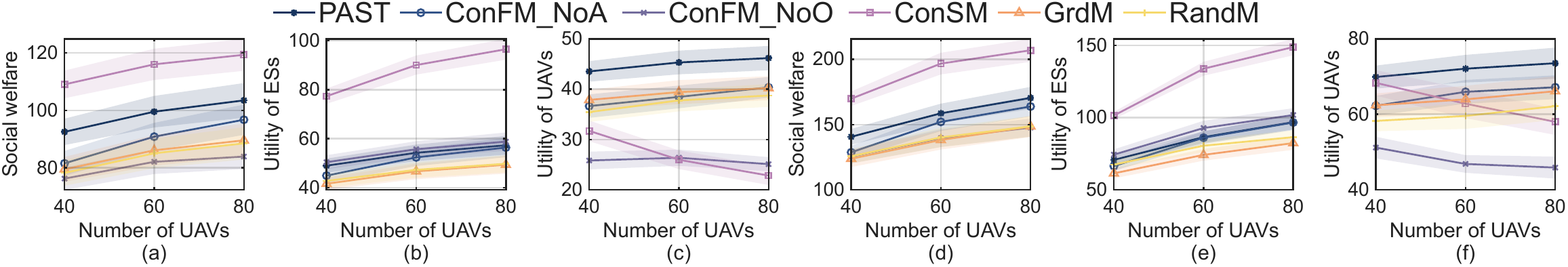}
			\caption{Performance comparisons in terms of SW, ES utility, and UAV utility, where (a)--(c) correspond to 6 ESs, (d)--(f) correspond to 10 ESs.}
			\label{SW}
		\end{figure*}
		
		We next evaluate trading outcomes in terms of SW and realized utilities obtained by ESs and UAVs. Fig.~\ref{SW} illustrates different numbers of UAVs under two ES-resource configurations: 6 ESs in Figs.~\ref{SW}(a)--\ref{SW}(c), 10 ESs in Figs.~\ref{SW}(d)--\ref{SW}(f).
		
		As shown in Figs.~\ref{SW}(a) and~\ref{SW}(d), ConSM achieves the highest SW across market scales by reconstructing trading relationships according to the realized UAV--edge conditions in each STU. PAST consistently ranks second and substantially outperforms the static futures-trading baselines, benefiting from selective renewal of PT agreements and overbooking rates when retained configurations become less suitable for evolving conditions. GrdM and RandM achieve lower SW because their simplified agreement-selection strategies do not jointly account for both trading sides. Figs.~\ref{SW}(b) and~\ref{SW}(e) show the realized ES utility. ConSM again achieves the highest ES utility through online reconstruction under the current market state. ConFM\_NoO also yields relatively high ES utility, slightly exceeding PAST in the considered settings, but at the cost of substantially lower UAV utility. PAST, in contrast, maintains competitive ES utility while preserving much higher UAV-side utility. ES utility generally increases with the number of UAVs as more tasks participate in the service market. The corresponding UAV-side results in Figs.~\ref{SW}(c) and~\ref{SW}(f) show that PAST consistently achieves the highest UAV utility. ConFM\_NoO yields the lowest UAV utility, indicating that disabling overbooking limits advance service opportunities. ConSM also exhibits a clear decline in UAV utility as the market grows, despite achieving the highest SW and ES utility. PAST maintains relatively high and stable UAV utility, demonstrating the benefit of combining overbooking with adaptive agreement renewal.
		
		Overall, ConSM maximizes aggregate SW primarily through higher ES-side utility, but incurs substantially greater repeated-decision overhead, as shown in Fig.~\ref{NI}. PAST achieves slightly lower SW while providing the highest UAV utility, competitive ES utility, and substantially lower decision overhead. These results demonstrate that PAST offers a favorable balance among aggregate trading performance, participant utilities, and repeated decision overhead.

		\subsubsection{TRLC and UoR}\label{chap 5.3.3}
		\begin{figure}[t]
			\centering
			\setlength{\abovecaptionskip}{-1 mm}
			\includegraphics[width=1\columnwidth]{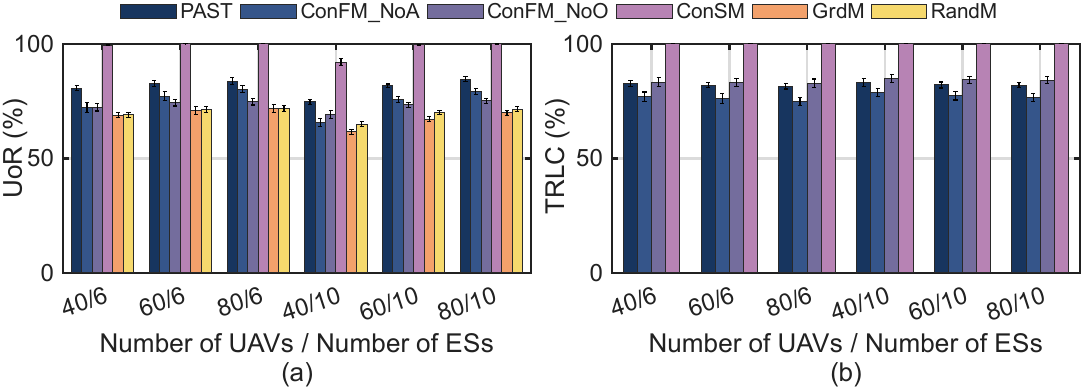}
			\caption{Performance comparisons in terms of (a) UoR and (b) TRLC. }
			\label{RoU}
		\end{figure}

		Fig.~\ref{RoU}(a) compares residual ES resource utilization under different market scales. ConSM achieves the highest UoR by determining trading relationships from the realized task demand and residual ES resources in each STU. PAST achieves lower UoR than ConSM because its PT agreements are established in advance and selectively renewed rather than reconstructed in every STU. Nevertheless, PAST consistently outperforms ConFM\_NoA and ConFM\_NoO by combining risk-aware overbooking with adaptive agreement renewal, allowing retained agreements to better accommodate variations in task arrivals, UAV--ES accessibility, and residual ES resources. GrdM and RandM achieve lower UoR because their simplified agreement-selection strategies use available residual resources less effectively. Fig.~\ref{RoU}(b) shows that ConSM also achieves the highest TRLC by making trading decisions from the realized task and resource states. Among the futures-trading methods, ConFM\_NoO achieves the highest TRLC. Compared with ConFM\_NoA, disabling overbooking reduces the number of advance task commitments relative to the nominal ES capacity and consequently lowers the exposure of executable contracted tasks to ES-side capacity shortage. This more conservative agreement construction improves fulfillment reliability, but results in lower residual-resource utilization and social welfare, as shown in Figs.~\ref{RoU}(a) and~\ref{SW}. PAST achieves higher TRLC than ConFM\_NoA because AdaptAO selectively renews PT agreements and overbooking rates as execution conditions evolve, while maintaining higher UoR than both static futures-trading baselines.
		
		Overall, ConSM achieves the highest resource utilization/agreement fulfillment at the cost of substantially higher repeated-decision overhead (see Fig.~\ref{NI}). PAST strikes a better balance through risk-aware overbooking and selective agreement renewal.}~{Appx.~F.2 examines the impact of overbooking rates on agreement construction, resource utilization, SW, and fulfillment, while Appx.~I evaluates PAST under varying UAV mobility, task demand, ES workload, and wireless conditions.}

	\subsubsection{Price Feasibility Supporting Individual Rationality}\label{chap 5.3.4}
	\begin{figure}[t]
		\centering
		\setlength{\abovecaptionskip}{-1 mm}
		\includegraphics[width=0.9\columnwidth,height=0.31\columnwidth]{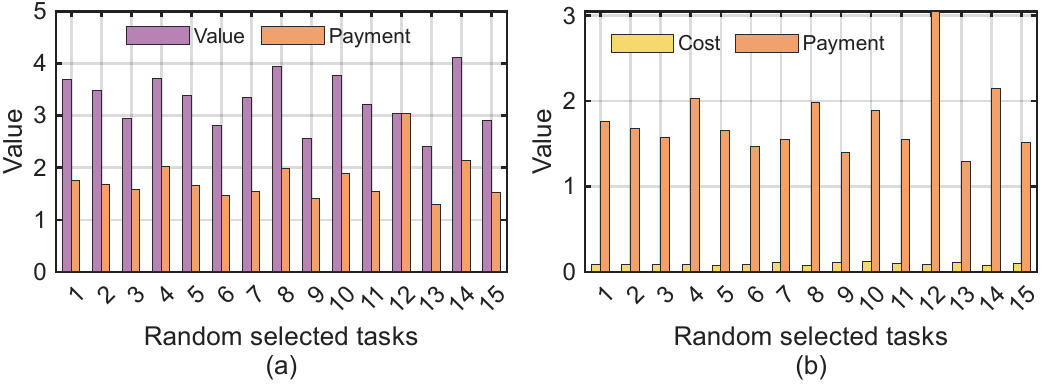}
		\caption{Bilateral price-feasibility results for 15 randomly selected signed task--ES agreements.}
		\label{IR}
	\end{figure}
	
	To examine the bilateral price-feasibility conditions associated with individual rationality, we randomly select 15 signed task--ES agreements. As shown in Fig.~\ref{IR}(a), the payment for each sampled task does not exceed its conditional expected service valuation, satisfying constraint~(\ref{23b}). Fig.~\ref{IR}(b) further shows that each payment exceeds the corresponding ES service cost, satisfying constraint~(\ref{equ.25b}). Hence, the sampled agreements satisfy the UAV- and ES-side price-feasibility conditions. The remaining feasibility conditions associated with individual rationality, including the corresponding capacity and risk constraints, are analyzed in Appx.~D.

	\section{Conclusion and Future Work}
	This paper presented PAST, a pilot-and-adaptive stable trading framework for STD-EUNs that combines PilotAO for risk-aware advance PT-agreement construction with AdaptAO for selective agreement retention and renewal. PilotAO satisfies individual rationality and strong stability, while also satisfying the competitive equilibrium and weak Pareto optimality established for the considered trading model. AdaptAO adjusts active PT agreements and the overbooking rate as UAV--edge conditions evolve. Experimental results demonstrate that PAST balances trading performance and decision efficiency while adapting to spatio-temporal dynamics.
	{Future work will evaluate PAST using measured UAV trajectories, air--ground channel traces, interference-aware SINR models, high-fidelity network simulations, and small-scale UAV--edge testbeds. We will also investigate privacy-preserving trading, heterogeneous task and service requirements, and broader air--ground systems with heterogeneous resource and mobility characteristics.}

	\newpage
	\clearpage
	\appendices

	\newpage
\clearpage
\appendices

\section{Detailed Derivations Associated With Mathematical Modeling}
\label{app:mathematical_derivations}

\noindent
\textbf{Conditional expected valuation of $v_{i,j,m}^{\mathsf{(n)}}$.}
{
	Based on (10)--(15), define}
\begin{equation}
	{
		\eta_{i,j}^{\mathsf{(n)}}
		=
		\frac{
			e_i^{\mathsf{tran}}
		}{
			\left(
			\sigma_{i,j}^{\mathsf{(n)}}
			\right)^2
		},
	}
	\label{eq:eta_definition}
\end{equation}
{
	and collect the valuation terms that are independent of the random channel gain as}
\begin{equation}
	{
		\Phi_{i,j,m}^{\mathsf{(n)}}
		=
		\omega_4
		\left(
		\frac{r_{i,m}^{\mathsf{(n)}}}{f_i}
		-
		\frac{r_{i,m}^{\mathsf{(n)}}}{f_j^{\mathsf E}}
		\right)
		+
		\omega_5
		\frac{r_{i,m}^{\mathsf{(n)}}}{f_i}
		e_i^{\mathsf{loc}},
	}
	\label{eq:phi_definition}
\end{equation}
\begin{equation}
	{
		\Psi_{i,m}^{\mathsf{(n)}}
		=
		r_{i,m}^{\mathsf{(n)}}d'
		\left(
		\omega_4+
		\omega_5e_i^{\mathsf{tran}}
		\right).
	}
	\label{eq:psi_definition}
\end{equation}

{
	Then, the valuation in (15) can be expressed as a function of the channel gain $g_{i,j}^{\mathsf{(n)}}$ as}
\begin{equation}
	{
		v_{i,j,m}^{\mathsf{(n)}}(g)
		=
		\Phi_{i,j,m}^{\mathsf{(n)}}
		-
		\frac{
			\Psi_{i,m}^{\mathsf{(n)}}
		}{
			B_{i,j}^{\mathsf{(n)}}
			\log_2
			\left(
			1+
			\eta_{i,j}^{\mathsf{(n)}}g
			\right)
		}.
	}
	\label{eq:valuation_as_gain}
\end{equation}

{
	To evaluate the conditional expected valuation used in the main text, let $f_{g_{i,j}^{\mathsf{(n)}}}(x)$ denote the probability density of the channel gain conditioned on $\alpha_{i,j,m}^{\mathsf{(n)}}=1$ and $\lambda_{i,j,m}^{\mathsf{(n)}}=0$. Its feasible support is denoted by}
\begin{equation}
	{
		\mathcal{G}_{i,j}^{\mathsf{(n)}}
		=
		\left[
		g_{i,j,\min}^{\mathsf{(n)}},
		g_{i,j,\max}^{\mathsf{(n)}}
		\right],
	}
\end{equation}
{
	where the bounds are induced by the UAV operating region and the feasible UAV--ES service range, with $g_{i,j,\min}^{\mathsf{(n)}}>0$. Taking the expectation of \eqref{eq:valuation_as_gain} over this conditional distribution gives}
{
	\begin{equation}
		{\small
			\begin{aligned}
				\mathbbm{E}
				\left[
				v_{i,j,m}^{\mathsf{(n)}}
				\right]
				={}&
				\Phi_{i,j,m}^{\mathsf{(n)}}
				-
				\Psi_{i,m}^{\mathsf{(n)}}
				\int_{\mathcal{G}_{i,j}^{\mathsf{(n)}}}
				\frac{
					f_{g_{i,j}^{\mathsf{(n)}}}(x)
				}{
					B_{i,j}^{\mathsf{(n)}}
					\log_2
					\left(
					1+
					\eta_{i,j}^{\mathsf{(n)}}x
					\right)
				}
				\,\mathrm{d}x.
		\end{aligned}}
		\label{eq:exact_expected_valuation}
\end{equation}}

{
	The expectation in \eqref{eq:exact_expected_valuation} involves the reciprocal of the transmission rate. Since the channel gain depends nonlinearly on UAV location, propagation distance, elevation angle, and LoS probability, this integral generally does not admit a simple closed-form expression. We therefore evaluate it using deterministic numerical quadrature with $Q$ quadrature points. For candidate contract $(u_i,d_{i,m}^{\mathsf{(n)}},s_j)$, let}
\begin{equation}
	{
		\left\{
		g_{i,j,q}^{\mathsf{(n)}},
		\varpi_{i,j,q}^{\mathsf{(n)}}
		\right\}_{q=1}^{Q}
	}
\end{equation}
{
	denote the quadrature nodes and the corresponding normalized weights over $\mathcal{G}_{i,j}^{\mathsf{(n)}}$, where}
\begin{equation}
	{
		\varpi_{i,j,q}^{\mathsf{(n)}}\geq0,
		\qquad
		\sum_{q=1}^{Q}
		\varpi_{i,j,q}^{\mathsf{(n)}}=1.
	}
\end{equation}
{
	The expected valuation used in constraint~(\ref{23b}) is then approximated as
	\begin{equation}
		{\small
			\begin{aligned}
				\mathbbm{E}
				\left[
				v_{i,j,m}^{\mathsf{(n)}}
				\right]
				\approx{}&
				\Phi_{i,j,m}^{\mathsf{(n)}}
				-
				\Psi_{i,m}^{\mathsf{(n)}}
				\sum_{q=1}^{Q}
				\frac{
					\varpi_{i,j,q}^{\mathsf{(n)}}
				}{
					B_{i,j}^{\mathsf{(n)}}
					\log_2
					\left(
					1+
					\eta_{i,j}^{\mathsf{(n)}}
					g_{i,j,q}^{\mathsf{(n)}}
					\right)
				}.
			\end{aligned}
			\label{appx1}}
\end{equation}}

{
	For each feasible UAV--task--ES candidate contract, (\ref{appx1}) is evaluated before the proposal and price-adjustment iterations of PilotAO. Since the signing-stage location and channel statistics remain unchanged during a given invocation of PilotAO, the resulting expected valuation is consistently reused for feasibility checking, preference construction, and payment adjustment.}

{
	The quadrature-based valuation also provides the model-based fallback when the finite joint-realization set contains no successful-service realization for estimating the conditional service value. In subsequent PilotAO invocations, once an online conditional service-value estimate becomes available from successful executions, it is combined with the model-based estimate as described in Appx.~\ref{app:joint_probability_estimation}.}

{
	\noindent\textbf{Computational complexity of the expectation evaluation.}
	Let
	\begin{equation}
		\mathcal{X}^{\mathsf{(n)}}
		=
		\left\{
		\left(
		u_i,
		d_{i,m}^{\mathsf{(n)}},
		s_j
		\right)
		\,\middle|\,
		u_i\in\bm{\mathcal{U}}^{\mathsf{(n)}},
		d_{i,m}^{\mathsf{(n)}}\in\bm{D}_{i}^{\mathsf{(n)}},
		s_j\in\widehat{\mathcal{E}}_{i}^{\mathsf{(n)}}
		\right\}
	\end{equation}
	denote the feasible UAV--task--ES candidate-contract set for STU $n$, and let
	\begin{equation}
		C^{\mathsf{(n)}}
		=
		\left|
		\mathcal{X}^{\mathsf{(n)}}
		\right|.
	\end{equation}
	
	Evaluating the $Q$-point deterministic quadrature requires $\mathcal O(Q)$ operations for each candidate contract. If the candidate-specific channel statistics are additionally estimated from $L_h$ historical UAV-location and channel observations, the corresponding preprocessing time and storage complexities are
	\begin{equation}
		\mathcal O\!\left((L_h+Q)C^{\mathsf{(n)}}\right)
		\quad\text{and}\quad
		\mathcal O\!\left(C^{\mathsf{(n)}}\right),
		\label{eq:app_valuation_preprocessing_complexity}
	\end{equation}
	respectively. If the conditional channel distributions and their quadrature nodes and weights are obtained in advance, the preprocessing time reduces to $\mathcal O\!\left(QC^{\mathsf{(n)}}\right)$.
	
	Before the proposal and price-adjustment iterations of PilotAO, the expected valuation in (\ref{appx1}) is evaluated for each candidate contract and stored for subsequent use. Hence, the valuation required by constraint~(\ref{23b}), preference construction, and payment adjustment can be directly retrieved during the matching iterations. The numerical expectation evaluation therefore introduces a preprocessing cost that is independent of the number of subsequent PilotAO iterations.
}

\noindent
\textbf{Expectation terms in the expected-utility expressions.}
From the definitions of $p_{\alpha,i,j,m}^{\mathsf{(n)}}$ and $p_{\lambda\mid\alpha,i,j,m}^{\mathsf{(n)}}$, together with the convention that $\lambda_{i,j,m}^{\mathsf{(n)}}=0$ whenever $\alpha_{i,j,m}^{\mathsf{(n)}}=0$, we have
$\mathbbm{E}[\alpha_{i,j,m}^{\mathsf{(n)}}]=p_{\alpha,i,j,m}^{\mathsf{(n)}}$,
$\mathbbm{E}[\alpha_{i,j,m}^{\mathsf{(n)}}\lambda_{i,j,m}^{\mathsf{(n)}}]
=p_{\alpha,i,j,m}^{\mathsf{(n)}}p_{\lambda\mid\alpha,i,j,m}^{\mathsf{(n)}}$,
and
$\mathbbm{E}[\alpha_{i,j,m}^{\mathsf{(n)}}(1-\lambda_{i,j,m}^{\mathsf{(n)}})]
=p_{\alpha,i,j,m}^{\mathsf{(n)}}(1-p_{\lambda\mid\alpha,i,j,m}^{\mathsf{(n)}})$.
These identities yield the execution-probability terms in~\eqref{eq:expected_es_utility} and~\eqref{eq:expected_uav_utility}, while the conditional expected service valuation $\mathbbm{E}[v_{i,j,m}^{\mathsf{(n)}}]$ is evaluated according to~\eqref{appx1}.

\noindent
\textbf{Derivation related to constraint~(\ref{23c}).}
Constraint~(\ref{23c}) bounds the probability that the realized UAV utility falls below the minimum acceptable level. To obtain a tractable sufficient condition, define
\begin{equation}
	\hat{U}^{\mathsf{U}}
	\left(
	u_i,
	\varphi^{\mathsf{(n)}}(u_i),
	\bm{\mathbb C}^{\mathsf{(n)}}
	\right)
	=
	U_{\max}^{\mathsf U,(n)}
	-
	U^{\mathsf{U}}
	\left(
	u_i,
	\varphi^{\mathsf{(n)}}(u_i),
	\bm{\mathbb C}^{\mathsf{(n)}}
	\right),
\end{equation}
where $U_{\max}^{\mathsf U,(n)}$ is a finite deterministic upper bound on the UAV utility over all feasible execution outcomes, $U_{\min}^{\mathsf U}\triangleq u_{\min}$, and $U_{\max}^{\mathsf U,(n)}>U_{\min}^{\mathsf U}$. By definition, $\hat U^{\mathsf U}\geq0$. The UAV-side utility-shortfall risk can therefore be rewritten as
\begin{align}
	&R^{\mathsf{U}}
	\left(
	u_i,
	\varphi^{\mathsf{(n)}}(u_i),
	\bm{\mathbb C}^{\mathsf{(n)}}
	\right)
	=
	\Pr\!\left(
	U^{\mathsf{U}}
	\left(
	u_i,
	\varphi^{\mathsf{(n)}}(u_i),
	\bm{\mathbb C}^{\mathsf{(n)}}
	\right)
	<
	U_{\min}^{\mathsf U}
	\right)
	\notag\\
	&=
	\Pr\!\left(
	\hat U^{\mathsf U}
	\left(
	u_i,
	\varphi^{\mathsf{(n)}}(u_i),
	\bm{\mathbb C}^{\mathsf{(n)}}
	\right)
	>
	U_{\max}^{\mathsf U,(n)}
	-
	U_{\min}^{\mathsf U}
	\right).
	\label{eq:23c_rewrite}
\end{align}

Since $\hat U^{\mathsf U}$ is nonnegative, Markov's inequality gives
\begin{equation}
	\begin{aligned}
		&\Pr\!\left(
		\hat U^{\mathsf U}
		>
		U_{\max}^{\mathsf U,(n)}
		-
		U_{\min}^{\mathsf U}
		\right)
		\leq
		\Pr\!\left(
		\hat U^{\mathsf U}
		\geq
		U_{\max}^{\mathsf U,(n)}
		-
		U_{\min}^{\mathsf U}
		\right)
		\\
		&\leq
		\frac{
			\mathbbm{E}
			\left[
			\hat U^{\mathsf U}
			\left(
			u_i,
			\varphi^{\mathsf{(n)}}(u_i),
			\bm{\mathbb C}^{\mathsf{(n)}}
			\right)
			\right]
		}{
			U_{\max}^{\mathsf U,(n)}
			-
			U_{\min}^{\mathsf U}
		}
		\\
		&=
		\frac{
			U_{\max}^{\mathsf U,(n)}
			-
			\mathbbm{E}
			\left[
			U^{\mathsf U}
			\left(
			u_i,
			\varphi^{\mathsf{(n)}}(u_i),
			\bm{\mathbb C}^{\mathsf{(n)}}
			\right)
			\right]
		}{
			U_{\max}^{\mathsf U,(n)}
			-
			U_{\min}^{\mathsf U}
		}.
	\end{aligned}
	\label{eq:23c_markov}
\end{equation}

Accordingly, the following conservative sufficient condition guarantees constraint~(\ref{23c}):
\begin{equation}
	\frac{
		U_{\max}^{\mathsf U,(n)}
		-
		\mathbbm{E}
		\left[
		U^{\mathsf U}
		\left(
		u_i,
		\varphi^{\mathsf{(n)}}(u_i),
		\bm{\mathbb C}^{\mathsf{(n)}}
		\right)
		\right]
	}{
		U_{\max}^{\mathsf U,(n)}
		-
		U_{\min}^{\mathsf U}
	}
	\leq
	\rho_1.
	\label{eq:23c_final}
\end{equation}

\noindent
\textbf{Derivation related to constraint~(\ref{equ.25d}).}
Constraint~(\ref{equ.25d}) bounds the probability that the realized ES-side utility contribution of a PT agreement falls below the minimum acceptable level. To obtain a tractable sufficient condition, define
\begin{equation}
	\hat{U}^{\mathsf{S}}
	\left(
	s_j,u_i,\mathbb{C}_{i,j}^{\mathsf{(n)}}
	\right)
	=
	U_{\max}^{\mathsf{S,(n)}}
	-
	U^{\mathsf{S}}
	\left(
	s_j,u_i,\mathbb{C}_{i,j}^{\mathsf{(n)}}
	\right),
\end{equation}
where $U_{\max}^{\mathsf S,(n)}$ is a finite deterministic upper bound on the corresponding ES-side utility over all feasible execution outcomes, $U_{\min}^{\mathsf S}\triangleq u_{\min}$, and $U_{\max}^{\mathsf S,(n)}>U_{\min}^{\mathsf S}$. By definition, $\hat U^{\mathsf S}\geq0$. The ES-side utility-shortfall risk can therefore be rewritten as
\begin{equation}
	\begin{aligned}
		&R_1^{\mathsf{S}}
		\left(
		s_j,u_i,\mathbb{C}_{i,j}^{\mathsf{(n)}}
		\right)
		\\
		&=
		\Pr\!\left(
		U^{\mathsf{S}}
		\left(
		s_j,u_i,\mathbb{C}_{i,j}^{\mathsf{(n)}}
		\right)
		<
		U_{\min}^{\mathsf{S}}
		\right)
		\\
		&=
		\Pr\!\left(
		\hat U^{\mathsf S}
		\left(
		s_j,u_i,\mathbb C_{i,j}^{\mathsf{(n)}}
		\right)
		>
		U_{\max}^{\mathsf S,(n)}
		-
		U_{\min}^{\mathsf S}
		\right).
	\end{aligned}
	\label{eq:23d_rewrite}
\end{equation}

Since $\hat U^{\mathsf S}$ is nonnegative, Markov's inequality gives
\begin{equation}
	\begin{aligned}
		&\Pr\!\left(
		\hat U^{\mathsf S}
		>
		U_{\max}^{\mathsf S,(n)}
		-
		U_{\min}^{\mathsf S}
		\right)
		\leq
		\Pr\!\left(
		\hat U^{\mathsf S}
		\geq
		U_{\max}^{\mathsf S,(n)}
		-
		U_{\min}^{\mathsf S}
		\right)
		\\
		&\leq
		\frac{
			\mathbbm{E}\!\left[
			\hat U^{\mathsf S}
			\left(
			s_j,u_i,\mathbb C_{i,j}^{\mathsf{(n)}}
			\right)
			\right]
		}{
			U_{\max}^{\mathsf S,(n)}
			-
			U_{\min}^{\mathsf S}
		}
		\\
		&=
		\frac{
			U_{\max}^{\mathsf S,(n)}
			-
			\mathbbm{E}\!\left[
			U^{\mathsf S}
			\left(
			s_j,u_i,\mathbb C_{i,j}^{\mathsf{(n)}}
			\right)
			\right]
		}{
			U_{\max}^{\mathsf S,(n)}
			-
			U_{\min}^{\mathsf S}
		}.
	\end{aligned}
	\label{eq:23d_markov}
\end{equation}

Accordingly, the following conservative sufficient condition guarantees constraint~(\ref{equ.25d}):
\begin{equation}
	\frac{
		U_{\max}^{\mathsf S,(n)}
		-
		\mathbbm{E}\!\left[
		U^{\mathsf S}
		\left(
		s_j,u_i,\mathbb C_{i,j}^{\mathsf{(n)}}
		\right)
		\right]
	}{
		U_{\max}^{\mathsf S,(n)}
		-
		U_{\min}^{\mathsf S}
	}
	\leq
	\rho_2,
	\quad
	\forall u_i\in\varphi^{\mathsf{(n)}}(s_j).
	\label{eq:23d_final}
\end{equation}

{
	\section{Probability Estimation and Error Sensitivity}
	\label{app:probability_estimation}
	
	This appendix details the estimation of the executable-arrival probability and the conditional ES-shortage probability used by PilotAO and further examines the sensitivity of agreement construction and adaptive renewal to estimation errors. For notational compactness, a candidate task--ES contract is denoted by $x=(u_i,d_{i,m}^{\mathsf{(n)}},s_j)$.
	
	\subsection{Joint Probability Estimation and Online Refinement}
	\label{app:joint_probability_estimation}
	
	For candidate contract $x$, $p_{\alpha,x}$ denotes the probability that task $d_{i,m}^{\mathsf{(n)}}$ is generated and the contracted ES $s_j$ remains accessible under the realized UAV location. Conditional on $\alpha_x=1$, $p_{\lambda\mid\alpha,x}$ denotes the probability that the task cannot be served because the realized residual capacity of $s_j$ is insufficient. Accordingly, the probabilities of successful service and ES-side capacity shortage are $p_{\alpha,x}\left(1-p_{\lambda\mid\alpha,x}\right)$ and $p_{\alpha,x}p_{\lambda\mid\alpha,x}$, respectively.
	
	Before each invocation of PilotAO, the estimator constructs $L$ joint model-based realizations, with $L=16$ used by default. Each realization captures a common signing-stage state, including UAV mobility and accessibility, wireless conditions, task attributes, inherent ES workloads, and the executable states of candidate agreements competing for the same ES resources. Jointly generating these quantities preserves the dependence induced by the common UAV--edge environment and ES-capacity contention.
	
	Let $\mathcal Z_{\ell}$ denote the mobility, accessibility, wireless, task, and resource state in realization $\ell$. The executable-arrival probability is estimated from the realization-specific conditional probabilities as
	\begin{equation}
		\widehat p_{\alpha,x}^{\mathsf{MC}}
		=
		\frac{1}{L}
		\sum_{\ell=1}^{L}
		\Pr\!\left(
		\alpha_x=1
		\mid
		\mathcal Z_{\ell}
		\right).
		\label{eq:app_mc_executable_arrival}
	\end{equation}
	This form uses the conditional probabilities associated with the sampled states rather than relying solely on binary executable-arrival realizations.
	
	For the conditional shortage probability, executable-arrival states are sampled jointly across candidate contracts according to their realization-specific probabilities. Let $\alpha_{x,\ell}$ and $\lambda_{x,\ell}$ denote the resulting executable-arrival and shortage indicators for contract $x$ in realization $\ell$. Define
	\begin{equation}
		N_{\alpha,x}
		=
		\sum_{\ell=1}^{L}\alpha_{x,\ell},
		\qquad
		N_{\lambda,x}
		=
		\sum_{\ell=1}^{L}
		\alpha_{x,\ell}\lambda_{x,\ell}.
		\label{eq:app_shortage_counts}
	\end{equation}
	Because some candidate contracts may have only a small number of executable-arrival realizations, the conditional-shortage estimate is regularized as
	\begin{equation}
		\widehat p_{\lambda\mid\alpha,x}^{\mathsf{MC}}
		=
		\frac{
			N_{\lambda,x}
			+
			2p_{\lambda}^{\mathsf{prior}}
		}{
			N_{\alpha,x}+2
		},
		\qquad
		p_{\lambda}^{\mathsf{prior}}=0.10,
		\label{eq:app_probability_smoothing}
	\end{equation}
	where the prior term reduces the sensitivity of the estimate when executable-arrival observations are sparse.
	
	Observed execution outcomes are subsequently used to refine the probability estimates through online updates. Before execution feedback becomes available, the online estimates are initialized by their corresponding model-based values. After observing contract $x$ at decision instant $t$, the executable-arrival estimate is updated as
	\begin{equation}
		p_{\alpha,x,t+1}^{\mathsf{online}}
		=
		(1-\beta_{\alpha})
		p_{\alpha,x,t}^{\mathsf{online}}
		+
		\beta_{\alpha}\alpha_{x,t},
		\qquad
		\beta_{\alpha}=0.18.
		\label{eq:app_online_executable_arrival}
	\end{equation}
	Because ES-side shortage is defined conditional on executable arrival, its online estimate is updated only when $\alpha_{x,t}=1$:
	\begin{equation}
		p_{\lambda\mid\alpha,x,t+1}^{\mathsf{online}}
		=
		(1-\beta_{\lambda})
		p_{\lambda\mid\alpha,x,t}^{\mathsf{online}}
		+
		\beta_{\lambda}\lambda_{x,t},
		~
		\beta_{\lambda}=0.18,
		~
		\text{if }\alpha_{x,t}=1.
		\label{eq:app_online_conditional_shortage}
	\end{equation}
	When $\alpha_{x,t}=0$, $p_{\lambda\mid\alpha,x,t}^{\mathsf{online}}$ remains unchanged because the observation provides no information about shortage conditional on executable arrival.
	
	For the conditional service value, the model-based estimate is obtained from successful-service realizations as
	\begin{equation}
		\widehat{\bar v}_{x}^{\mathsf{MC}}
		=
		\frac{
			\sum_{\ell=1}^{L}
			\alpha_{x,\ell}
			\left(1-\lambda_{x,\ell}\right)
			v_{x,\ell}
		}{
			\sum_{\ell=1}^{L}
			\alpha_{x,\ell}
			\left(1-\lambda_{x,\ell}\right)
		},
		\label{eq:app_mc_conditional_value}
	\end{equation}
	whenever at least one successful-service realization is available. Otherwise, the quadrature-based conditional valuation in Appx.~\ref{app:mathematical_derivations} is used as the model-based fallback. The online conditional service-value estimate is initialized upon the first successful execution of contract $x$. Thereafter, whenever $\alpha_{x,t}=1$ and $\lambda_{x,t}=0$, it is updated as
	\begin{equation}
		\bar v_{x,t+1}^{\mathsf{online}}
		=
		(1-\beta_v)
		\bar v_{x,t}^{\mathsf{online}}
		+
		\beta_v v_{x,t},
		\qquad
		\beta_v=0.18.
		\label{eq:app_online_conditional_value}
	\end{equation}
	When no successful service is observed, $\bar v_{x,t}^{\mathsf{online}}$ remains unchanged.
	
	At subsequent invocations of PilotAO, the model-based and online estimates are combined according to
	\begin{equation}
		\widehat{\xi}_{x}^{\mathsf{used}}
		=
		0.65\,
		\widehat{\xi}_{x}^{\mathsf{MC}}
		+
		0.35\,
		\widehat{\xi}_{x}^{\mathsf{online}},
		\qquad
		\xi_x
		\in
		\left\{
		p_{\alpha,x},
		p_{\lambda\mid\alpha,x},
		\bar v_x
		\right\}.
		\label{eq:app_probability_blending}
	\end{equation}
	For the conditional service value, the model-based estimate is used directly until its online counterpart has been initialized by a successful execution. The resulting estimates are then used in subsequent invocations of PilotAO.
	
	\subsection{Numerical Calibration}
	\label{app:probability_calibration}
	
	We evaluate $L\in\{8,16,32,64\}$ over ten independently generated model-based market states. For each state, an independent 1,000-draw reference is used as the high-sample benchmark, and 20 warm-up STUs are used to initialize the online estimates. Table~\ref{tab:app_probability_calibration} reports the resulting calibration metrics. The MAEs measure the estimation errors of $p_{\alpha}$ and $p_{\lambda\mid\alpha}$, and the Brier score evaluates the predicted successful-service probability $\widehat p_{\alpha,x}(1-\widehat p_{\lambda\mid\alpha,x})$ against the realized Bernoulli outcome $\alpha_x(1-\lambda_x)$. Value NMAE and Value P90 Error measure the mean and 90th-percentile normalized absolute errors of the conditional service-value estimate, respectively, while Value Support denotes the proportion of candidate instances with a well-defined conditional service-value reference.
	
	\begin{table}[t!]
		\centering 
		\caption{Calibration results under different numbers of joint realizations. Values are means over ten independent model-based market states, each using an independent 1,000-draw reference.}
		\label{tab:app_probability_calibration}
		\scriptsize
		\renewcommand{\arraystretch}{1.12}
		\setlength{\tabcolsep}{2.5pt}
		\resizebox{\columnwidth}{!}{%
			\begin{tabular}{c|cccccc}
				\toprule
				$L$
				&
				$p_{\alpha}$ MAE
				&
				$p_{\lambda\mid\alpha}$ MAE
				&
				Brier
				&
				Value NMAE (\%)
				&
				\shortstack{Value P90\\Error (\%)}
				&
				\shortstack{Value\\Support (\%)}
				\\
				\midrule
				8
				& 0.0386
				& 0.1343
				& 0.1443
				& 3.97
				& 8.54
				& 98.41
				\\
				\textbf{16 (default)}
				& \textbf{0.0281}
				& \textbf{0.0921}
				& \textbf{0.1353}
				& \textbf{3.36}
				& \textbf{7.34}
				& \textbf{98.41}
				\\
				32
				& 0.0218
				& 0.0655
				& 0.1313
				& 2.73
				& 6.30
				& 98.41
				\\
				64
				& 0.0178
				& 0.0453
				& 0.1291
				& 2.21
				& 5.03
				& 98.41
				\\
				\bottomrule
			\end{tabular}%
		}
	\end{table}
	
	Increasing $L$ consistently reduces the estimation errors over the tested range. In particular, the MAEs of both probability estimates and the two conditional-service-value error measures decrease as more joint realizations are used, while the Brier score decreases from $0.1443$ at $L=8$ to $0.1291$ at $L=64$. These results indicate that enlarging the joint-realization set improves the numerical accuracy of the signing-stage estimates, with progressively smaller gains at larger values of $L$.
	
	At the default setting $L=16$, the MAEs of $p_{\alpha}$ and $p_{\lambda\mid\alpha}$ are $0.0281$ and $0.0921$, with two-sided Student-$t$ 95\% confidence intervals of $0.0259$--$0.0304$ and $0.0795$--$0.1047$, respectively. The corresponding Value NMAE and Value P90 Error are $3.36\%$ and $7.34\%$. Since the computational cost of probability estimation increases with the number of joint realizations, we adopt $L=16$ as the default setting, providing a balance between estimation accuracy and preprocessing cost. Value Support remains $98.41\%$ for all tested values of $L$ because it is determined by the common high-draw reference rather than by the finite-$L$ estimator.

	\subsection{Sensitivity to Estimation Bias}
	\label{app:probability_bias}
	
	To examine the propagation of probability-estimation errors through PAST, we independently add biases to $p_{\alpha}$ and $p_{\lambda\mid\alpha}$ after combining the model-based and online estimates and before clipping them to $[0,1]$. Biases in $\{-0.15,-0.10,-0.05,0,0.05,0.10,0.15\}$ are considered, and Table~\ref{tab:app_probability_bias_summary} reports the two endpoints together with the zero-bias reference. All configurations use 60 UAVs, 10 ESs, 30 matched evaluation seeds, and 100 STUs per seed, with the trained DDQN parameters fixed throughout the evaluation. The resulting differences therefore originate from changes in PilotAO agreement construction induced by the biased estimates and their subsequent effects on AdaptAO.
	
	\begin{table}[t!]
		\centering 
		\caption{System responses to independently injected probability-estimation biases. Values are means over 30 matched evaluation seeds with 100 STUs per seed. UoR, TRLC, and ES shortage are reported as percentages.}
		\label{tab:app_probability_bias_summary}
		\scriptsize
		\renewcommand{\arraystretch}{1.12}
		\setlength{\tabcolsep}{2.8pt}
		
		\textbf{Panel A: Agreement volume and system outcomes}
		
		\vspace{1mm}
		{%
			\begin{tabular}{cc|rrrr}
				\toprule
				\textbf{Biased Estimate}
				&
				\textbf{Bias}
				&
				\textbf{Signed/STU}
				&
				\textbf{SW}
				&
				\textbf{UoR (\%)}
				&
				\textbf{TRLC (\%)}
				\\
				\midrule
				$p_{\alpha}$
				& $-0.15$
				& 45.96
				& 116.31
				& 60.61
				& 89.48
				\\
				$p_{\alpha}$
				& $0$
				& 76.28
				& 158.60
				& 81.69
				& 82.03
				\\
				$p_{\alpha}$
				& $+0.15$
				& 77.40
				& 159.42
				& 82.21
				& 81.87
				\\
				\midrule
				$p_{\lambda\mid\alpha}$
				& $-0.15$
				& 82.44
				& 164.97
				& 85.86
				& 80.78
				\\
				$p_{\lambda\mid\alpha}$
				& $0$
				& 76.28
				& 158.60
				& 81.69
				& 82.03
				\\
				$p_{\lambda\mid\alpha}$
				& $+0.15$
				& 59.45
				& 139.79
				& 70.00
				& 86.72
				\\
				\bottomrule
			\end{tabular}%
		}
		
		\vspace{2mm}
		\textbf{Panel B: Shortage exposure and agreement renewal}
		
		\vspace{1mm}
		{%
			\begin{tabular}{cc|rr}
				\toprule
				\textbf{Biased Estimate}
				&
				\textbf{Bias}
				&
				\textbf{ES Shortage (\%)}
				&
				~~\textbf{\shortstack{Renewals/100 STUs}}
				\\
				\midrule
				$p_{\alpha}$
				& $-0.15$
				& 10.52
				& 71.87
				\\
				$p_{\alpha}$
				& $0$
				& 17.97
				& 14.93
				\\
				$p_{\alpha}$
				& $+0.15$
				& 18.13
				& 14.73
				\\
				\midrule
				$p_{\lambda\mid\alpha}$
				& $-0.15$
				& 19.22
				& 11.40
				\\
				$p_{\lambda\mid\alpha}$
				& $0$
				& 17.97
				& 14.93
				\\
				$p_{\lambda\mid\alpha}$
				& $+0.15$
				& 13.28
				& 44.00
				\\
				\bottomrule
			\end{tabular}%
		}
	\end{table}
	
	The response to errors in $p_{\alpha}$ is markedly asymmetric. Underestimating the executable-arrival probability restricts agreement admission: at a bias of $-0.15$, Signed/STU decreases from $76.28$ to $45.96$, accompanied by substantial reductions in SW and UoR. The resulting increase in TRLC is therefore associated with a considerably smaller admitted agreement set rather than improved aggregate service delivery. By contrast, a positive bias produces only limited changes in the final agreement volume and execution outcomes. This saturation indicates that, once the executable-arrival estimate is sufficiently large, association, utility, capacity, and risk constraints become the dominant limits on additional agreement admission.
	
	Biases in $p_{\lambda\mid\alpha}$ primarily alter the balance between agreement participation and fulfillment reliability. Underestimation admits more agreements and increases residual-resource utilization, but also increases ES-side shortage and reduces TRLC. Overestimation has the opposite effect, yielding a smaller and more conservative agreement set with lower utilization but higher fulfillment. Thus, the conditional-shortage estimate directly affects how aggressively PilotAO uses residual ES capacity under uncertain executable demand.
	
	The estimation biases also propagate to AdaptAO through the agreement configurations generated by PilotAO. In particular, the renewal frequency rises from $14.93$ to $71.87$ renewals per 100 STUs under a $-0.15$ bias in $p_{\alpha}$, while a $+0.15$ bias produces little change. For $p_{\lambda\mid\alpha}$, underestimation reduces the renewal frequency to $11.40$, whereas overestimation increases it to $44.00$. Because the DDQN parameters remain fixed, these changes result from the altered execution states induced by biased agreement construction rather than from differences in policy training.
	
	Overall, the two probability estimates affect different aspects of PAST. Errors in $p_{\alpha}$ mainly influence the admission of potential service opportunities, whereas errors in $p_{\lambda\mid\alpha}$ regulate the tradeoff between agreement participation, residual-resource utilization, and fulfillment reliability. Their effects further propagate to AdaptAO by changing the agreement configurations and subsequent execution states.	}

{
	\section{Sensitivity Analysis of PilotAO Parameters}
	\label{app:pilot_parameter_sensitivity}
	
	This appendix examines the sensitivity of PilotAO to its principal operating parameters. We separately vary the three risk thresholds $\rho_1$--$\rho_3$, the compensation parameters $q^{\mathsf U}$ and $q^{\mathsf E}$, and the price increment $\Delta p$, while keeping the remaining mechanism components and environmental realizations unchanged. Each configuration is evaluated using 60 UAVs, 10 ESs, 30 matched evaluation seeds, and 100 STUs per seed. The trained DDQN parameters, task and mobility realizations, wireless conditions, probability estimator, and all non-varied parameters are shared across the corresponding comparisons. The default parameter setting is $\rho_1=\rho_2=\rho_3=0.3,~
	q^{\mathsf U}=0.65,~
	q^{\mathsf E}=0.85,~
	\Delta p=0.08. $
	The following results characterize the effects of these parameters over the tested ranges.
	
	\subsection{Sensitivity to the Risk Thresholds}
	\label{app:risk_threshold_sensitivity}
	
	As defined in the main text, the three thresholds bound distinct contractual risks:
	\begin{equation}
		R^{\mathsf U}\leq\rho_1,\qquad
		R_1^{\mathsf S}\leq\rho_2,\qquad
		R_2^{\mathsf S}\leq\rho_3.
		\label{eq:app_three_risk_constraints}
	\end{equation}
	Here, $R^{\mathsf U}$ denotes the UAV-side utility-shortfall risk, $R_1^{\mathsf S}$ denotes the ES-side utility-shortfall risk, and $R_2^{\mathsf S}$ denotes the conditional ES-shortage risk of an executable candidate contract. Smaller thresholds impose stricter risk screening and reduce the set of admissible agreements. Table~\ref{tab:app_risk_sensitivity} reports the corresponding one-factor-at-a-time sensitivity results. Risk-Feasible/Call denotes the mean number of candidate agreements satisfying the applicable PilotAO risk constraints in one agreement-construction invocation, while Signed/STU denotes the mean number of task-level PT agreements ultimately signed per STU.
	
	\begin{table}[t!]
		\centering
		
		\caption{One-factor-at-a-time sensitivity to the three PilotAO risk thresholds. Values are means over 30 matched evaluation seeds with 100 STUs per seed under the 60-UAV/10-ES setting.}
		\label{tab:app_risk_sensitivity}
		\footnotesize
		\renewcommand{\arraystretch}{1.10}
		\setlength{\tabcolsep}{2.6pt}
		\begin{tabular}{cc|ccccc}
			\toprule
			\textbf{\shortstack{Varied\\Threshold}}
			&
			\textbf{Value}
			&
			\textbf{\shortstack{Risk-Feasible\\/Call}}
			&
			\textbf{\shortstack{Signed\\/STU}}
			&
			\textbf{SW}
			&
			\textbf{\shortstack{UoR\\(\%)}}
			&
			\textbf{\shortstack{TRLC\\(\%)}}
			\\
			\midrule
			$\rho_1$
			& 0.1
			& 67.56
			& 29.74
			& 83.98
			& 43.76
			& 92.93
			\\
			$\rho_1$
			& \textbf{0.3}
			& \textbf{233.48}
			& \textbf{76.28}
			& \textbf{158.60}
			& \textbf{81.69}
			& \textbf{82.03}
			\\
			$\rho_1$
			& 0.5
			& 251.40
			& 76.80
			& 159.58
			& 82.13
			& 82.05
			\\
			\midrule
			$\rho_2$
			& 0.1
			& 113.62
			& 51.24
			& 126.43
			& 62.54
			& 88.81
			\\
			$\rho_2$
			& \textbf{0.3}
			& \textbf{233.48}
			& \textbf{76.28}
			& \textbf{158.60}
			& \textbf{81.69}
			& \textbf{82.03}
			\\
			$\rho_2$
			& 0.5
			& 242.61
			& 76.63
			& 159.35
			& 81.96
			& 81.91
			\\
			\midrule
			$\rho_3$
			& 0.1
			& 105.34
			& 48.70
			& 122.64
			& 60.47
			& 89.58
			\\
			$\rho_3$
			& \textbf{0.3}
			& \textbf{233.48}
			& \textbf{76.28}
			& \textbf{158.60}
			& \textbf{81.69}
			& \textbf{82.03}
			\\
			$\rho_3$
			& 0.5
			& 247.82
			& 77.14
			& 159.27
			& 82.24
			& 81.42
			\\
			\bottomrule
		\end{tabular}
	\end{table}
	
	Across all three thresholds, stricter risk screening reduces agreement participation while improving fulfillment reliability. This effect is most pronounced for $\rho_1$: reducing it from $0.3$ to $0.1$ sharply contracts both the risk-feasible candidate pool and the signed agreement set, accompanied by lower SW and UoR but higher TRLC. Tightening either ES-side threshold produces the same qualitative pattern. The responses of $\rho_2$ and $\rho_3$ are nevertheless distinct because they constrain different risks: $\rho_2$ limits ES-side utility shortfall, whereas $\rho_3$ directly limits the conditional shortage probability of executable candidate contracts. In particular, the stricter $\rho_3$ setting yields a slightly smaller signed agreement set than the corresponding $\rho_2$ setting.
	
	Relaxing the thresholds beyond the default value produces much smaller changes in the final outcomes. Although the number of risk-feasible candidates continues to increase, Signed/STU, SW, and UoR remain comparatively stable, indicating that agreement admission is increasingly determined by the remaining price, capacity, utility, and risk conditions once the varied constraint is sufficiently relaxed. The three thresholds therefore provide complementary controls over the tradeoff between agreement participation and fulfillment reliability. We retain $\rho_1=\rho_2=\rho_3=0.3$ as the default operating point within the tested range.
	
	\subsection{Sensitivity to the Compensation Parameters}
	\label{app:compensation_sensitivity}
	
	The parameters $q^{\mathsf U}$ and $q^{\mathsf E}$ specify the contractual transfers associated with two distinct agreement non-fulfillment events. If a signed task does not become executable, the corresponding UAV transfers $q^{\mathsf U}$ to the ES. By contrast, if an executable contracted task cannot be served because the realized residual ES capacity is insufficient, the ES transfers $q^{\mathsf E}$ to the affected UAV.
	
	For a candidate agreement $x$, define its successful-service probability as
	\begin{equation}
		s_x
		=
		p_{\alpha,x}
		\left(
		1-p_{\lambda\mid\alpha,x}
		\right).
		\label{eq:app_success_probability}
	\end{equation}
	Let $\bar v_x$ denote the conditional expected service valuation of agreement $x$, and let $p_x$ and $c_x$ denote the corresponding payment and ES service cost, respectively. The expected utility contributions to the UAV and ES are then
	\begin{equation}
		\overline U_x^{\mathsf U}
		=
		s_x(\bar v_x-p_x)
		-
		(1-p_{\alpha,x})q^{\mathsf U}
		+
		p_{\alpha,x}p_{\lambda\mid\alpha,x}q^{\mathsf E},
		\label{eq:app_uav_contract_utility}
	\end{equation}
	and
	\begin{equation}
		\overline U_x^{\mathsf S}
		=
		s_x(p_x-c_x)
		+
		(1-p_{\alpha,x})q^{\mathsf U}
		-
		p_{\alpha,x}p_{\lambda\mid\alpha,x}q^{\mathsf E}.
		\label{eq:app_es_contract_utility}
	\end{equation}
	For a fixed agreement and execution distribution, the transfer terms involving $q^{\mathsf U}$ and $q^{\mathsf E}$ enter the UAV and ES utilities with opposite signs and therefore cancel when the two utilities are aggregated. Their direct effect is therefore to redistribute utility between the two trading sides. Changing either parameter can nevertheless affect the resulting agreement set indirectly through the expected utilities, risk conditions, and preference rankings used by PilotAO.
	
	\begin{table*}[t!]
		\centering
		\caption{One-factor-at-a-time sensitivity to the two compensation parameters. Transfer/STU denotes the mean total compensation transferred between UAVs and ESs per STU.}
		\label{tab:app_compensation_sensitivity}
		\footnotesize
		\renewcommand{\arraystretch}{1.13}
		\setlength{\tabcolsep}{7pt}
		{%
			\begin{tabular}{cc|rrrrrrr}
				\toprule
				\textbf{Varied Parameter}
				&
				\textbf{Value}
				&
				\textbf{UAV Utility}
				&
				\textbf{ES Utility}
				&
				\textbf{SW}
				&
				\textbf{UoR (\%)}
				&
				\textbf{TRLC (\%)}
				&
				\textbf{Signed/STU}
				&
				\textbf{Transfer/STU}
				\\
				\midrule
				$q^{\mathsf U}$
				& 0.325
				& 76.15
				& 83.46
				& 159.61
				& 82.24
				& 81.73
				& 77.28
				& 13.34
				\\
				$q^{\mathsf U}$
				& \textbf{0.650}
				& \textbf{71.60}
				& \textbf{87.00}
				& \textbf{158.60}
				& \textbf{81.69}
				& \textbf{82.03}
				& \textbf{76.28}
				& \textbf{16.52}
				\\
				$q^{\mathsf U}$
				& 1.300
				& 63.60
				& 94.69
				& 158.29
				& 81.51
				& 82.25
				& 75.84
				& 23.34
				\\
				\midrule
				$q^{\mathsf E}$
				& 0.425
				& 66.32
				& 91.84
				& 158.16
				& 81.51
				& 81.64
				& 76.30
				& 11.83
				\\
				$q^{\mathsf E}$
				& \textbf{0.850}
				& \textbf{71.60}
				& \textbf{87.00}
				& \textbf{158.60}
				& \textbf{81.69}
				& \textbf{82.03}
				& \textbf{76.28}
				& \textbf{16.52}
				\\
				$q^{\mathsf E}$
				& 1.700
				& 85.79
				& 73.56
				& 159.35
				& 81.95
				& 82.29
				& 76.49
				& 25.84
				\\
				\bottomrule
			\end{tabular}%
		}
	\end{table*}
	
	The results reflect the opposite transfer directions of the two compensation parameters. Increasing $q^{\mathsf U}$ shifts utility from UAVs to ESs: over the tested range, UAV utility decreases from $76.15$ to $63.60$, while ES utility increases from $83.46$ to $94.69$. The corresponding increase in Transfer/STU is accompanied by only limited changes in SW, UoR, TRLC, and Signed/STU. Thus, $q^{\mathsf U}$ primarily affects the distribution of utility between the two trading sides rather than the aggregate service outcomes.
	
	Increasing $q^{\mathsf E}$ produces the opposite redistribution. UAV utility increases from $66.32$ to $85.79$, whereas ES utility decreases from $91.84$ to $73.56$, with Transfer/STU increasing accordingly. The principal execution metrics again remain comparatively stable. This pattern is consistent with $q^{\mathsf E}$ compensating the UAV when an executable contracted task cannot be served because of insufficient residual ES capacity.
	
	Overall, the compensation parameters primarily allocate the utility consequences of agreement non-fulfillment between UAVs and ESs. Their comparatively limited effects on SW, resource utilization, agreement fulfillment, and agreement volume are consistent with the compensation transfers canceling when UAV and ES utilities are aggregated. The remaining variations arise indirectly through their effects on expected utilities, risk feasibility, preference rankings, and agreement construction.
	
	\subsection{Sensitivity to the Price Increment}
	\label{app:price_increment_sensitivity}
	
	In PilotAO, a task rejected by an ES may increase its proposed payment by $\Delta p$, provided that the updated payment does not exceed its expected service valuation. For an admissible candidate contract $(i,j,m)$ satisfying
	$p_{i,j,m}^{\min}\leq\mathbbm E[v_{i,j,m}^{\mathsf{(n)}}]$,
	the number of strict payment increases is bounded by
	\begin{equation}
		K_{i,j,m}
		\leq
		\left\lceil
		\frac{
			\mathbbm E
			\left[
			v_{i,j,m}^{\mathsf{(n)}}
			\right]
			-
			p_{i,j,m}^{\min}
		}{
			\Delta p
		}
		\right\rceil.
		\label{eq:app_price_level_bound}
	\end{equation}
	Thus, $\Delta p$ determines the granularity of payment adjustment: a smaller increment permits more intermediate payment levels, whereas a larger increment reaches the valuation bound in fewer adjustments.
	
	Table~\ref{tab:app_price_increment_sensitivity} reports the resulting payment, agreement, and overhead measures. Active Calls denotes the fraction of PilotAO invocations containing at least one payment update, Rounds/Call denotes the mean number of payment-adjustment rounds among these invocations, and $p^\star$ denotes the mean terminal payment.
	
	\begin{table*}[t!]
		\centering
		
		\caption{Sensitivity to the PilotAO price increment under the 60-UAV/10-ES setting. Price-update messages, protocol interactions, and payment-stage computation time are reported per STU over the 100-STU evaluation horizon.}
		\label{tab:app_price_increment_sensitivity}
		\footnotesize
		\renewcommand{\arraystretch}{1.13}
		\setlength{\tabcolsep}{4pt}
		{%
			\begin{tabular}{c|rrrrrrrr}
				\toprule
				$\boldsymbol{\Delta p}$
				&
				\textbf{Active Calls (\%)}
				&
				\textbf{Rounds/Call}
				&
				$\boldsymbol{p^\star}$
				&
				\textbf{Signed/STU}
				&
				\textbf{SW}
				&
				\textbf{Price Msg./STU}
				&
				\textbf{NI/STU}
				&
				\textbf{Price Time (ms/STU)}
				\\
				\midrule
				0.01
				& 99.57
				& 125.54
				& 1.761
				& 77.36
				& 161.75
				& 390.66
				& 930.45
				& 2.718
				\\
				\textbf{0.08}
				& \textbf{100.00}
				& \textbf{35.38}
				& \textbf{1.806}
				& \textbf{76.28}
				& \textbf{158.60}
				& \textbf{84.58}
				& \textbf{640.51}
				& \textbf{1.047}
				\\
				0.20
				& 99.82
				& 10.69
				& 1.895
				& 76.25
				& 157.97
				& 54.83
				& 582.12
				& 0.876
				\\
				\bottomrule
			\end{tabular}%
		}
	\end{table*}
	
	The results show that the payment-adjustment overhead decreases markedly as $\Delta p$ increases. With $\Delta p=0.01$, the finer sequence of admissible payment levels requires substantially more adjustment rounds, price-update messages, protocol interactions, and payment-stage computation time. It also yields the lowest mean terminal payment and slightly higher Signed/STU and SW than the two coarser settings. Increasing the increment to $\Delta p=0.08$ substantially reduces this overhead while preserving similar agreement volume and welfare, and a further increase to $\Delta p=0.20$ shortens the adjustment process further but results in a higher mean terminal payment.
	
	These results are consistent with the bound in~\eqref{eq:app_price_level_bound}: finer payment increments provide greater adjustment resolution at the cost of additional communication and computation, whereas coarser increments reduce the number of intermediate payment levels. We therefore use $\Delta p=0.08$ as the default value within the tested range, providing an intermediate tradeoff between payment-adjustment resolution and agreement-management overhead.	}

\section{Property Analysis of PilotAO}
\label{app:pilotao_properties}

{
	We first establish the finite termination of PilotAO and then analyze the matching properties of its terminal outcome. After the price-adjustment phase terminates, the payments are fixed and PilotAO reconstructs the task-side and ES-side preferences over the remaining admissible task--ES contracts before performing the final deferred-acceptance procedure. The terminal contract domain is finite, ties are resolved deterministically, and the valuation, probability, and risk quantities used to construct the terminal preferences remain fixed throughout this procedure.
	
	The matching-property analysis relies on the following terminal-stage conditions. First, each task is unit-demand over the terminal-admissible contract domain and has a fixed strict ranking over its admissible ESs, with deterministic tie-breaking. Hence, its terminal choice is responsive in the standard unit-demand sense. Second, for each ES $s_j$, let $C_j(\mathcal X)$ denote the set of task contracts selected from an offered terminal-contract set $\mathcal X$. We require the induced terminal ES choice function to satisfy substitutability and irrelevance of rejected contracts: for offered sets $\mathcal X'\subseteq\mathcal X$, a contract chosen from $\mathcal X$ and still available in $\mathcal X'$ is not rejected merely because other contracts are removed, while contracts rejected from an offer set do not become chosen solely after other rejected contracts are removed. We further assume downward-closed terminal feasibility, i.e., every subset of a feasible selected contract set remains feasible. These are conditions on the terminal choice rule over the fixed admissible contract domain; they are not asserted for arbitrary set-dependent preferences or feasibility rules outside this terminal stage. Finally, the information used for PT-agreement construction is assumed to be truthfully reported.
	
	These conditions are sufficient for the deferred-acceptance arguments used below. In particular, responsiveness characterizes the unit-demand task-side choice, downward-closedness characterizes terminal feasibility, and substitutability together with irrelevance of rejected contracts characterizes the ES-side choice rule. No continuity assumption is required because the terminal contract domain and the admissible payment levels are finite. We do not invoke a separate gross-substitutes condition beyond the stated terminal choice-function assumptions.
	
	For the weak-Pareto result, the reallocation class referred to in the main text is understood as the feasible matching deviations represented by the Type~1 and Type~2 blocking-coalition notions defined there. Thus, it includes feasible reassignment of existing task--ES relationships and feasible expansion of the current matching, but excludes coordinated reallocations that cannot be represented by either blocking-coalition type, as well as side payments outside the proposed trading model. Throughout the following matching-property analysis, Type~1 and Type~2 blocking deviations are understood to be feasible deviations within the terminal-admissible contract domain and therefore satisfy the applicable UAV- and ES-side feasibility constraints. Under these terminal-stage conditions and this matching scope, the following results characterize the outcome returned by PilotAO.
	
	\begin{lem}[Convergence of PilotAO]
		Alg.~\ref{Matching} terminates after a finite number of rounds.
	\end{lem}
	
	\begin{proof}
		For every admissible UAV--task--ES option, the payment is initialized at $p_{i,j,m}^{\min}$ and can only increase by the strictly positive increment $\Delta p$. Constraint~(\ref{23b}) and the payment-update rule impose the finite upper bound $\mathbbm E[v_{i,j,m}^{\mathsf{(n)}}]$. Hence, for an admissible option satisfying $p_{i,j,m}^{\min}\leq\mathbbm E[v_{i,j,m}^{\mathsf{(n)}}]$, the number of strict payment increases is bounded by
		\begin{equation}
			K_{i,j,m}
			\leq
			\left\lceil
			\frac{
				\mathbbm E[v_{i,j,m}^{\mathsf{(n)}}]
				-
				p_{i,j,m}^{\min}
			}{
				\Delta p
			}
			\right\rceil.
		\end{equation}
		After a rejection, either the corresponding payment strictly increases or, if no further feasible increase is available, ES $s_j$ is added to $\mathbb Z_{i,m}$ and the corresponding task--ES option is removed from the candidate set. Since the numbers of UAVs, tasks, ESs, candidate options, and admissible payment levels are finite, only finitely many strict payment increases and candidate removals can occur. Therefore, the price-adjustment phase terminates after finitely many rounds.
		
		After the terminal payments are obtained, PilotAO performs fixed-price task-proposing deferred acceptance over the finite terminal-admissible contract domain. Each task proposes to any admissible ES at most once during this completion step. Since the numbers of tasks and admissible task--ES options are finite, the deferred-acceptance procedure also terminates after finitely many proposals. Hence, Alg.~\ref{Matching} terminates in finite time.
	\end{proof}

	\begin{lem}[Individual rationality of PilotAO]
		The matching returned by PilotAO is individually rational.
	\end{lem}
	
	\begin{proof}
		We verify the UAV-side and ES-side conditions separately.
		
		\noindent
		\textbf{(i) UAV side.}
		For each UAV $u_i\in\bm{\mathcal U}^{\mathsf{(n)}}$, every task--ES contract retained in the terminal matching satisfies constraint~(\ref{23b}), so that the conditional expected service valuation of the task is no smaller than the corresponding payment. The resulting UAV assignment also satisfies constraint~(\ref{23c}), which bounds the UAV-side utility-shortfall risk by $\rho_1$. Hence, the terminal UAV assignment satisfies the price- and risk-feasibility conditions specified in Definition~\ref{def:pilotao_individual_rationality}.
		
		\noindent
		\textbf{(ii) ES side.}
		For each ES $s_j\in\bm{\mathcal S}$, every selected task satisfies constraint~(\ref{equ.25b}), so that its payment covers the corresponding ES service cost. Constraint~(\ref{equ.25c}) ensures that the signed UAV-task demand together with the predicted inherent workload remains within the overbooked signing capacity. Constraints~(\ref{equ.25d}) and~(\ref{equ.25e}) further bound the ES-side utility-shortfall risk and the conditional shortage risk of the selected contracts by $\rho_2$ and $\rho_3$, respectively. Therefore, the terminal ES assignment satisfies the corresponding price, capacity, and risk-feasibility conditions.
		
		Since the terminal assignments of both UAVs and ESs satisfy the conditions in Definition~\ref{def:pilotao_individual_rationality}, the matching returned by PilotAO is individually rational.
	\end{proof}

	\begin{lem}[Fairness of PilotAO]
		Under the above choice-function conditions, the matching returned by PilotAO is fair.
	\end{lem}
	
	\begin{proof}
		According to the fairness definition, a matching is fair if it admits no Type~1 blocking coalition. Suppose, for contradiction, that after PilotAO terminates, an ES $s_j$ and a UAV subset $\mathbb U$ form a Type~1 blocking coalition.
		
		At the terminal payments, PilotAO reconstructs the task-side and ES-side preference orders from the terminal expected utility contributions and performs task-proposing deferred acceptance over the terminal-admissible contract domain. Under the responsive UAV-side choice rule, if the proposed reassignment in the assumed blocking coalition strictly improves the utility of an involved UAV, at least one corresponding task-level contract with $s_j$ must be preferred to that task's terminal alternative. Such a task therefore proposes to $s_j$ before proposing to any lower-ranked admissible ES or remaining unmatched.
		
		If this proposal is rejected, $s_j$ has selected a feasible set of higher-priority contracts according to its terminal expected-utility ranking. By substitutability and irrelevance of rejected contracts, removing other rejected proposals cannot make a previously rejected contract become part of a strictly preferred feasible choice solely because the surrounding offer set has changed. Together with responsiveness and deterministic tie-breaking, the terminal ES choice therefore cannot strictly prefer a feasible reassignment containing the rejected contracts to its retained terminal set.
		
		Hence, $s_j$ cannot strictly prefer the proposed UAV set $\mathbb U$ while all UAVs involved in the reassignment simultaneously prefer the corresponding deviation. This contradicts the definition of a Type~1 blocking coalition. Therefore, no Type~1 blocking coalition exists, and the matching returned by PilotAO is fair.
	\end{proof}

	\begin{lem}[Non-wastefulness of PilotAO]
		The matching returned by PilotAO is non-wasteful.
	\end{lem}
	
	\begin{proof}
		According to the definition of non-wastefulness, it is sufficient to show that no Type~2 blocking coalition exists. Suppose, for contradiction, that after PilotAO terminates, an ES $s_j$ and a UAV subset $\mathbb U$ can form a Type~2 blocking coalition.
		By definition, the corresponding additional task--ES relationships must be feasible for both sides. Specifically, the involved UAVs prefer adding $s_j$ to their current assignments, while $s_j$ can accept the corresponding tasks without violating constraints~(\ref{equ.25b})--(\ref{equ.25e}) and can thereby obtain a higher expected utility. Hence, these task--ES options are included among the feasible alternatives considered in the final fixed-price deferred-acceptance procedure.
		
		During this procedure, each task proposes to ESs according to its final preference order. Therefore, any task that prefers $s_j$ to its final assignment must have proposed to $s_j$ before moving to a less preferred ES or remaining unmatched. If the task is rejected, $s_j$ has already retained a feasible set of tasks that it prefers under the final ES-side ranking and feasibility constraints. Since the payments, preference rankings, and feasibility conditions remain fixed during this final procedure, a task rejected in favor of such a feasible set cannot later become a preferred additional choice merely because other rejected proposals are removed.
		
		Therefore, the additional tasks assumed in the Type~2 blocking coalition cannot remain outside the final matching while being preferred by the involved UAVs and simultaneously providing $s_j$ with a feasible utility improvement. This contradicts the definition of a Type~2 blocking coalition. Hence, no Type~2 blocking coalition exists, and the matching returned by PilotAO is non-wasteful.
	\end{proof}

	\begin{thm}[Strong stability of PilotAO]
		The matching returned by PilotAO is strongly stable.
	\end{thm}
	
	\begin{proof}
		By the preceding lemmas, the matching returned by PilotAO is individually rational, fair, and non-wasteful. It therefore satisfies the strong-stability definition in the main text.
	\end{proof}

	\begin{thm}[Competitive equilibrium of PilotAO]
		The trading outcome returned by PilotAO satisfies the competitive equilibrium criterion defined in the main text.
	\end{thm}
	
	\begin{proof}
		We verify the four conditions of the competitive equilibrium criterion.
		
		First, every selected task--ES contract satisfies bilateral price feasibility. Constraint~(\ref{23b}) ensures that the payment does not exceed the corresponding conditional expected service valuation, while constraint~(\ref{equ.25b}) ensures that the payment is no lower than the corresponding ES service cost.
		
		Second, after the final payments are determined, each task ranks the feasible ESs according to its expected UAV-side utility and proposes following this preference order. The final assignment is therefore consistent with the task-side preference among the ESs satisfying constraints~(\ref{23a})--(\ref{23c}).
		
		Third, each ES ranks the received task proposals according to their expected ES-side utility contributions and retains its preferred feasible task set subject to constraints~(\ref{equ.25a})--(\ref{equ.25e}). Hence, the resulting ES-side assignment is consistent with its utility ranking and feasibility requirements.
		
		Finally, suppose that an additional task--ES contract could be added to the final matching such that the involved task prefers this additional service relationship and the corresponding ES can accept it without violating the price, capacity, or risk constraints while strictly increasing its expected utility. The corresponding UAV and ES would then form a Type~2 blocking coalition, contradicting the non-wastefulness of the matching returned by PilotAO.
		
		Therefore, all four conditions of the competitive equilibrium criterion are satisfied, and the trading outcome returned by PilotAO satisfies the competitive equilibrium property.
	\end{proof}
	
	\setcounter{Prop}{0}
	\begin{Prop}[Weak Pareto optimality of PilotAO]
		The trading outcome returned by PilotAO is weakly Pareto optimal within the reallocation class considered in the main text.
	\end{Prop}
	
	\begin{proof}
		Suppose, for contradiction, that there exists a feasible Pareto improvement within the considered reallocation class. By definition, the expected utility of every participant affected by the corresponding matching change is strictly increased, while the utilities of all other participants remain unchanged.
		
		Under the reallocation scope specified above, such a matching change is represented by one of the two blocking-coalition types defined in the main text. If the improvement replaces existing task--ES relationships, the affected ES and UAVs constitute a Type~1 blocking coalition, contradicting the fairness of the matching returned by PilotAO. If the improvement adds feasible task--ES relationships without replacing existing ones, the affected ES and UAVs constitute a Type~2 blocking coalition, contradicting non-wastefulness.
		
		Hence, neither type of feasible Pareto improvement exists within the considered reallocation class. Therefore, the trading outcome returned by PilotAO is weakly Pareto optimal within this class.
\end{proof}}

{
	\section{AdaptAO Decision Rationale and Controller Comparison}
	\label{app:ddqn_controller_analysis}
	
	This appendix explains the decision rationale of AdaptAO and compares alternative controllers under the same action space and execution model. In particular, agreement retention or renewal affects not only the current reconstruction overhead but also the agreement configuration and overbooking rate used in subsequent execution. We therefore examine how different controllers balance agreement reuse and timely renewal under evolving UAV--edge conditions. We also distinguish the learned DDQN policy from the external rule-based safeguard in the subsequent analysis. The comparison focuses on social welfare, residual-resource utilization, agreement fulfillment, and renewal overhead.
	
	\subsection{Decision Coupling and Common Action Space}
	\label{app:sequential_adaptao}
	
	At each decision instant, AdaptAO observes the most recently available execution state and decides whether to retain or renew the current PT agreements. Retention avoids the computation and interaction overhead of rerunning PilotAO, but the existing agreements may become less suitable as task demand, UAV mobility, channel conditions, and residual ES resources evolve. Renewal incurs an immediate reconstruction cost, while updating both the PT agreements and the overbooking rate for the ensuing execution interval. Consequently, the current decision affects not only the immediate execution outcome but also the agreement configuration carried into subsequent decision stages.
	
	AdaptAO represents the execution state by a 12-dimensional normalized feature vector comprising the decision-stage index; estimates of task demand, UAV mobility, and residual ES resources; the realized TRLC, UoR, default rate, and social welfare; the active overbooking rate; a combined agreement-status and retention-duration feature; agreement applicability; and prediction error. These features characterize both the current execution condition and the suitability of the retained agreements. Thus, similar instantaneous utilization or shortage levels need not imply the same renewal decision when agreement applicability, retention duration, or the predicted execution conditions differ.
	
	To capture this temporal coupling, under policy $\pi$, the long-term value of taking action $\mathbbm{a}_t^{\mathsf{(n)}}$ in state $\mathbbm{s}_t^{\mathsf{(n)}}$ is defined as
	\begin{equation}
		Q^{\pi}
		\left(
		\mathbbm{s}_t^{\mathsf{(n)}},
		\mathbbm{a}_t^{\mathsf{(n)}}
		\right)
		=
		\mathbb{E}_{\pi}
		\left[
		\sum_{k=0}^{\infty}
		\left(
		\nu^{\mathsf{(n)}}
		\right)^k
		\mathbbm{r}_{t+k}^{\mathsf{(n)}}
		\,\middle|\,
		\mathbbm{s}_t^{\mathsf{(n)}},
		\mathbbm{a}_t^{\mathsf{(n)}}
		\right],
		\label{eq:app_adaptao_long_term_value}
	\end{equation}
	where $\nu^{\mathsf{(n)}}$ is the discount factor. This value accounts for the immediate renewal cost together with the subsequent effects of the selected agreement configuration on agreement applicability, social welfare, resource utilization, fulfillment, and later renewal decisions. AdaptAO therefore evaluates retention and renewal according to their cumulative effects rather than the current execution outcome alone.
	
	The continuously valued multidimensional state makes exhaustive tabular value learning impractical. AdaptAO therefore uses DDQN to approximate the state--action value function and map the observed execution state to the long-term values of the candidate agreement-management actions.
	
	For a consistent controller comparison, all methods operate over the common action space
	\begin{equation}
		\mathcal{A}
		=
		\left\{
		a_{\mathsf{keep}}
		\right\}
		\cup
		\left\{
		a_{\mathsf{renew}}(\tau_k)
		\,\middle|\,
		\tau_k\in\mathcal{T}_{\tau}
		\right\},
		\label{eq:app_common_action_space}
	\end{equation}
	where
	\begin{equation}
		\mathcal{T}_{\tau}
		=
		\left\{
		0,0.1,0.2,0.3,0.4
		\right\}.
		\label{eq:app_tau_grid}
	\end{equation}
	Action $a_{\mathsf{keep}}$ retains the current PT agreements and active overbooking rate without invoking PilotAO, whereas $a_{\mathsf{renew}}(\tau_k)$ reconstructs the PT agreements through PilotAO using the selected overbooking rate $\tau_k$. The resulting action space contains one retention action and five renewal actions.
	
	The following controllers are compared under this common action space:
	
	\begin{itemize}
		\item \textbf{PAST-DDQN:} The proposed state-dependent controller, which estimates the long-term value of each candidate action from the 12-dimensional execution state.
		
		\item \textbf{PAST-UCB:} An upper-confidence-bound controller that treats the six actions as independent arms and selects among them according to accumulated rewards and exploration bonuses, without conditioning explicitly on the current execution state.
		
		\item \textbf{PAST-SA:} A stochastic-approximation controller that updates a target overbooking rate according to observed resource idleness and ES-side shortage. The target is mapped to the candidate grid in~\eqref{eq:app_tau_grid}, with agreement renewal triggered when the selected grid value changes.
		
		\item \textbf{PAST-Threshold:} A deterministic controller that determines retention, renewal, and the corresponding overbooking rate from fixed thresholds on agreement applicability, retention duration, default rate, resource idleness, and ES-side shortage.
		
		\item \textbf{PAST-Random:} A state-independent reference that uniformly selects among the same six actions.
	\end{itemize}
	
	All controllers use the same PilotAO procedure, task and resource realizations, agreement-execution rules, candidate overbooking rates, and matched evaluation seeds. They differ only in the rule used to select retention or renewal and, when renewal is selected, the corresponding overbooking rate.

	\subsection{Comparison of AdaptAO Controllers}
	\label{app:controller_matched_evaluation}
	
	We evaluate the controllers using 60 UAVs, 10 ESs, 30 matched evaluation seeds, and 100 STUs per seed. PAST-DDQN uses the fixed default model trained with seed $7000$ for 220 episodes, each containing 100 STUs, under the 40-UAV/6-ES configuration, and is evaluated under the 60-UAV/10-ES setting without further parameter updates. The reported confidence intervals therefore characterize variation across the matched evaluation environments for the same trained DDQN model. For each seed, TRLC is calculated at the STU level as the fraction of executable contracted tasks that are successfully served and then averaged over the evaluation horizon, with zero assigned when no executable contracted task is present. NI counts proposal, acceptance, rejection, price-update, and final agreement-confirmation interactions, while decision time includes controller inference, probability preprocessing, matching computation, and modeled protocol-message delay.
	
	\begin{table}[t!]
		\centering
		\caption{Mean performance of the AdaptAO controllers over 30 matched evaluation seeds with 100 STUs per seed under the 60-UAV/10-ES setting.}
		\label{tab:app_controller_comparison}
		\scriptsize
		\renewcommand{\arraystretch}{1.12}
		\resizebox{\columnwidth}{!}{%
			\begin{tabular}{l|rrrrrr}
				\toprule
				\textbf{Controller}
				&
				\textbf{SW}
				&
				\textbf{UoR (\%)}
				&
				\textbf{TRLC (\%)}
				&
				\textbf{\shortstack{Renewals\\/100 STUs}}
				&
				\textbf{NI/STU}
				&
				\textbf{\shortstack{Decision Time\\(ms/STU)}}
				\\
				\midrule
				PAST-DDQN
				& 158.60
				& 81.69
				& 82.03
				& 14.93
				& 640.51
				& 57.33
				\\
				PAST-UCB
				& 159.44
				& 80.84
				& 83.56
				& 70.63
				& 2980.75
				& 263.34
				\\
				PAST-SA
				& 136.44
				& 69.75
				& 89.01
				& 2.70
				& 151.16
				& 13.28
				\\
				PAST-Threshold
				& 149.27
				& 74.55
				& 90.72
				& 76.80
				& 4062.54
				& 334.65
				\\
				PAST-Random
				& 161.83
				& 81.69
				& 83.69
				& 82.93
				& 3624.36
				& 315.20
				\\
				\bottomrule
			\end{tabular}%
		}
	\end{table}
	
	Table~\ref{tab:app_controller_paired_difference} reports the paired PAST-DDQN-minus-baseline differences over the same 30 evaluation seeds. Each entry gives the mean paired difference and its 95\% confidence interval, with UoR and TRLC differences expressed in percentage points.
	
	\begin{table*}[t!]
		\centering
		\caption{Paired PAST-DDQN-minus-baseline differences over 30 matched evaluation seeds. Each entry reports the paired mean and its 95\% confidence interval. UoR and TRLC differences are expressed in percentage points.}
		\label{tab:app_controller_paired_difference}
		\footnotesize
		\renewcommand{\arraystretch}{1.15}
		\setlength{\tabcolsep}{4.5pt}
		\resizebox{\textwidth}{!}{%
			\begin{tabular}{l|cccccc}
				\toprule
				\textbf{Baseline}
				&
				$\boldsymbol{\Delta}\mathrm{\textbf{SW}}$
				&
				$\boldsymbol{\Delta}\mathrm{\textbf{UoR}}$
				&
				$\boldsymbol{\Delta}\mathrm{\textbf{TRLC}}$
				&
				$\boldsymbol{\Delta}\mathrm{\textbf{Renewals}}$
				&
				$\boldsymbol{\Delta}\mathrm{\textbf{NI/STU}}$
				&
				$\boldsymbol{\Delta}\mathrm{\textbf{Time}}$
				\\
				\midrule
				UCB
				&
				$-0.85\,[-2.88,1.18]$
				&
				$+0.85\,[-0.20,1.91]$
				&
				$-1.53\,[-2.42,-0.65]$
				&
				$-55.70\,[-61.33,-50.07]$
				&
				$-2340.25\,[-2609.23,-2071.26]$
				&
				$-206.01\,[-228.48,-183.54]$
				\\
				SA
				&
				$+22.15\,[18.53,25.77]$
				&
				$+11.94\,[9.88,13.99]$
				&
				$-6.99\,[-7.93,-6.04]$
				&
				$+12.23\,[10.75,13.72]$
				&
				$+489.35\,[432.61,546.09]$
				&
				$+44.05\,[38.84,49.25]$
				\\
				Threshold
				&
				$+9.32\,[7.54,11.11]$
				&
				$+7.15\,[6.05,8.24]$
				&
				$-8.69\,[-9.31,-8.08]$
				&
				$-61.87\,[-64.31,-59.42]$
				&
				$-3422.03\,[-3596.79,-3247.27]$
				&
				$-277.32\,[-289.32,-265.33]$
				\\
				Random
				&
				$-3.23\,[-4.65,-1.81]$
				&
				$+0.001\,[-0.825,0.827]$
				&
				$-1.66\,[-2.19,-1.13]$
				&
				$-68.00\,[-69.83,-66.17]$
				&
				$-2983.85\,[-3141.77,-2825.94]$
				&
				$-257.87\,[-268.66,-247.08]$
				\\
				\bottomrule
			\end{tabular}%
		}
	\end{table*}
	
	The controller comparison reveals distinct operating points in the tradeoff between execution performance and agreement-management overhead. Relative to UCB, DDQN shows no statistically clear difference in SW or UoR, while its TRLC is modestly lower; however, it requires substantially fewer renewals, interactions, and decision time. A similar pattern is observed relative to Threshold and Random: both renew agreements much more frequently, whereas DDQN substantially reduces agreement-management overhead. Threshold attains higher TRLC but lower SW and UoR, while Random attains higher SW and TRLC with no statistically clear UoR difference.
	
	SA represents the opposite operating regime. Its infrequent renewals yield the lowest interaction and decision-time overhead and the highest TRLC among the compared controllers, but are accompanied by substantially lower SW and UoR than DDQN. The paired results therefore indicate that neither aggressive reconstruction nor highly conservative retention is uniformly preferable across the considered metrics. By conditioning retain-or-renew decisions on the observed execution state, DDQN maintains comparatively high welfare and residual-resource utilization while avoiding most of the renewal overhead incurred by the more frequently updating controllers.

	\subsection{Analysis of the Learned Policy and Rule-Based Safeguard}
	\label{app:safeguard_decomposition}
	
	We further distinguish the contribution of the learned DDQN policy from that of an external rule-based safeguard. The safeguard is applied after DDQN selects an action: when $a_{\mathsf{keep}}$ is selected while a predefined agreement-update condition is satisfied, the keep action is replaced by agreement renewal. We compare the following three variants:
	
	\begin{itemize}
		\item \textbf{Pure DDQN:} The action selected by the learned six-action policy is executed directly without external intervention.
		
		\item \textbf{DDQN+Guard:} A keep action selected by DDQN can be replaced by safeguard-triggered renewal.
		
		\item \textbf{Guard-only:} The DDQN policy is disabled, and agreement retention or renewal is determined solely by the rule-based safeguard. When renewal is triggered, PilotAO reconstructs the PT agreements using the fixed overbooking rate $\tau=0.2$.
	\end{itemize}
	
	\begin{table*}[b!]
		\centering
		\caption{Comparison of the learned DDQN policy and the rule-based safeguard under the 60-UAV/10-ES setting. Panel~A reports means over 30 matched evaluation seeds with 100 STUs per seed. Action counts and renewals are reported per 100 STUs. In Panel~B, each paired relative difference is calculated as $100(\mathrm{Pure~DDQN}-\mathrm{Comparator})/\mathrm{Comparator}$ and is followed by its 95\% confidence interval.}
		\label{tab:app_safeguard_decomposition}
		\scriptsize
		\renewcommand{\arraystretch}{1.15}
		\setlength{\tabcolsep}{4pt}
		
		\textbf{Panel A: Executed actions, system outcomes, and agreement-management overhead}
		
		\vspace{1mm}
		{%
			\begin{tabular}{l|rrrr|rrrr|r}
				\toprule
				\textbf{Variant}
				&
				\textbf{\shortstack{Learned\\Renew}}
				&
				\textbf{\shortstack{Guard\\Renew}}
				&
				\textbf{Keep}
				&
				\textbf{\shortstack{Total\\Renewals}}
				&
				\textbf{SW}
				&
				\textbf{UoR (\%)}
				&
				\textbf{TRLC (\%)}
				&
				\textbf{\shortstack{Edge-Served\\(\%)}}
				&
				\textbf{NI/STU}
				\\
				\midrule
				Pure DDQN
				& 14.93
				& 0.00
				& 85.07
				& 14.93
				& 158.60
				& 81.69
				& 82.03
				& 45.73
				& 640.51
				\\
				DDQN+Guard
				& 12.67
				& 26.67
				& 60.67
				& 39.33
				& 160.94
				& 81.96
				& 81.78
				& 46.10
				& 1610.13
				\\
				Guard-only
				& 0.00
				& 35.63
				& 64.37
				& 35.63
				& 161.69
				& 82.69
				& 83.47
				& 46.47
				& 1573.70
				\\
				\bottomrule
			\end{tabular}%
		}
		
		\vspace{2mm}
		\textbf{Panel B: Paired Pure-DDQN-minus-comparator relative differences}
		
		\vspace{1mm}
		{%
			\begin{tabular}{l|cccc}
				\toprule
				\textbf{Comparator}
				&
				$\boldsymbol{\Delta}\mathrm{\textbf{SW}}$ \textbf{(\%)}
				&
				$\boldsymbol{\Delta}\mathrm{\textbf{UoR}}$ \textbf{(\%)}
				&
				$\boldsymbol{\Delta}\mathrm{\textbf{TRLC}}$ \textbf{(\%)}
				&
				$\boldsymbol{\Delta}\mathrm{\textbf{Edge}}$ \textbf{(\%)}
				\\
				\midrule
				DDQN+Guard
				&
				$-1.457\,[-2.158,-0.757]$
				&
				$-0.332\,[-1.217,0.553]$
				&
				$+0.298\,[-0.413,1.008]$
				&
				$-0.811\,[-1.684,0.063]$
				\\
				Guard-only
				&
				$-1.914\,[-2.873,-0.956]$
				&
				$-1.209\,[-2.288,-0.128]$
				&
				$-1.731\,[-2.394,-1.067]$
				&
				$-1.592\,[-2.711,-0.473]$
				\\
				\bottomrule
			\end{tabular}%
		}
	\end{table*}
	
	Here, Edge-Served denotes the proportion of generated UAV tasks successfully completed through edge execution. As shown in Table~\ref{tab:app_safeguard_decomposition}, Pure DDQN retains the current PT agreements in most STUs, whereas both safeguard-based variants trigger substantially more frequent reconstruction. This difference is reflected directly in agreement-management overhead: Pure DDQN requires considerably fewer renewals and interactions than either DDQN+Guard or Guard-only.
	
	Relative to DDQN+Guard, Pure DDQN reduces the renewal frequency by $62.03\%$ and NI by $60.22\%$. The additional safeguard-triggered renewals yield a $1.457\%$ increase in SW, while the paired differences in UoR, TRLC, and Edge-Served are not statistically distinguishable from zero. Thus, in this comparison, the safeguard primarily increases reconstruction frequency and interaction overhead, with only a modest improvement in welfare.
	
	Guard-only exhibits a similar but more pronounced tradeoff. Compared with Pure DDQN, it achieves higher SW, UoR, TRLC, and Edge-Served, with the corresponding paired confidence intervals excluding zero, but requires about $2.4$ times as many renewals and $2.5$ times as many interactions. These results indicate that more aggressive rule-based reconstruction can improve several execution outcomes, but at substantially higher agreement-management cost.
	
	The three variants therefore represent different operating points between execution performance and agreement-management overhead. Pure DDQN relies solely on the learned state-dependent policy and avoids most of the additional reconstruction induced by the safeguard. We consequently use Pure DDQN in the main evaluation so that the reported AdaptAO results reflect the learned six-action policy without external rule-based intervention.

	\subsection{Discussion of the DDQN Controller}
	\label{app:ddqn_claim_scope}
	
	The controller comparisons reveal different operating points in the tradeoff between execution performance and agreement-management overhead. UCB, Threshold, Random, and the safeguard-based variants generally renew the PT agreements more frequently, which improves some execution metrics but incurs substantially higher interaction and reconstruction overhead. SA represents the opposite case, with infrequent renewal and high TRLC but lower SW and UoR. DDQN operates between these two extremes, maintaining comparatively high welfare and residual-resource utilization while avoiding most of the renewal overhead associated with more aggressive reconstruction.
	
	This tradeoff is consistent with the sequential decision structure of AdaptAO. Each retain-or-renew action determines the agreement configuration and overbooking rate used in the ensuing execution interval, which in turn affects agreement applicability, resource utilization, shortage exposure, and the state observed at subsequent decision instants. By conditioning its decisions on the multidimensional execution state and discounted return, DDQN accounts for these carry-over effects when coordinating agreement retention, renewal, and overbooking-rate selection. The reported results therefore characterize DDQN as a state-dependent controller that balances execution performance and agreement-management overhead under the considered evaluation settings.	}

{
	\section{Overbooking Rate Discretization and Action-Grid Design in AdaptAO}
	\label{app:tau_action_grid}
	
	This appendix examines how the overbooking rate is incorporated into the discrete action space of AdaptAO and evaluates the effect of the candidate-rate grid. We first isolate the effect of the overbooking rate under a common agreement-renewal rule to examine its influence on social welfare, residual-resource utilization, and agreement fulfillment. We then compare coarse, default, and fine candidate-rate grids using independently trained controllers to evaluate how the grid resolution affects the resulting AdaptAO policy.
	
	\subsection{Discrete Overbooking Rate Selection}
	\label{app:tau_discrete_action}
	
	AdaptAO selects the overbooking rate from a finite candidate set when agreement renewal is performed. Its action space is $\mathbb{A}^{\mathsf{(n)}}
	=
	\left\{
	a_{\mathsf{keep}}
	\right\}
	\cup
	\left\{
	a_{\mathsf{renew}}(\tau_k)
	\,\middle|\,
	\tau_k\in\mathcal{T}_{\tau}
	\right\}$, where the default candidate set is
	$\mathcal{T}_{\tau}
	=
	\left\{
	0,0.1,0.2,0.3,0.4
	\right\}$.
	Accordingly, the implemented DDQN contains one agreement-retention action and five agreement-renewal actions. Action $a_{\mathsf{keep}}$ retains the current PT agreements and active overbooking rate without invoking PilotAO, whereas $a_{\mathsf{renew}}(\tau_k)$ invokes PilotAO to reconstruct the PT agreements using the selected overbooking rate $\tau_k$. Thus, the overbooking rate is selected only when agreement renewal is performed.
	
	The selected overbooking rate is a market-level scalar shared across the ESs. For ES $s_j$ with nominal capacity $G_j$ and predicted mean inherent workload $\kappa_j^{\mathsf{(n)}}$, the maximum number of task-level commitments permitted by the signing-capacity constraint is
	\begin{equation}
		\widehat G_j^{\mathsf{(n)}}(\tau)
		=
		\max
		\left\{
		0,\,
		\left\lfloor
		(1+\tau)G_j
		-
		\kappa_j^{\mathsf{(n)}}
		\right\rfloor
		\right\}.
		\label{eq:app_tau_signing_quota}
	\end{equation}
	Hence, $\widehat G_j^{\mathsf{(n)}}(\tau)$ specifies the maximum number of UAV tasks that ES $s_j$ can retain during agreement construction under the selected overbooking rate. The actual number of selected tasks can be smaller because each candidate task--ES contract must also satisfy the corresponding accessibility, price, utility, and risk-feasibility conditions.
	
	The effect of $\tau$ follows from the difference between advance commitments and realized executable demand. A positive overbooking rate allows an ES to sign additional task commitments in anticipation that some contracted tasks may not become executable because of uncertain task generation or UAV mobility. This can improve the utilization of residual ES resources. However, a larger $\tau$ also increases the possibility that executable contracted demand exceeds the realized residual capacity, which can increase ES-side shortage and reduce agreement fulfillment. Therefore, the overbooking rate determines the tradeoff between resource utilization and fulfillment reliability, and its preferred value may vary with the current demand, mobility, and resource conditions.
	
	\subsection{Sensitivity to Fixed Overbooking Rates}
	\label{app:fixed_tau_sensitivity}
	
	We first isolate the effect of the overbooking rate under a common agreement-renewal rule. All configurations use the rule-based renewal trigger of the Guard-only variant in Appx.~\ref{app:safeguard_decomposition}; whenever renewal is triggered, PilotAO reconstructs the PT agreements using a fixed overbooking rate $\tau\in\{0,0.2,0.5\}$. The compared configurations share the same renewal instants, PilotAO procedure, environmental realizations, and matched evaluation seeds, so that they differ only in the overbooking rate used at renewal. Here, $\tau=0$ disables overbooking, $\tau=0.2$ corresponds to the default fixed rate of the Guard-only configuration, and $\tau=0.5$ represents a higher-overbooking setting beyond the default AdaptAO action grid. Each configuration is evaluated with 60 UAVs, 10 ESs, 30 matched evaluation seeds, and 100 STUs per seed.
	
	\begin{table}[t!]
		\centering
		\caption{Sensitivity to fixed overbooking rates, averaged over 30 matched evaluation seeds with 100 STUs per seed.}
		\label{tab:app_fixed_tau_sensitivity}
		\footnotesize
		\renewcommand{\arraystretch}{1.12}
		\setlength{\tabcolsep}{4pt}
		{%
			\begin{tabular}{c|ccccc}
				\toprule
				$\boldsymbol{\tau}$
				&
				\textbf{SW}
				&
				\textbf{UoR (\%)}
				&
				\textbf{TRLC (\%)}
				&
				\textbf{\shortstack{ES Shortage\\(\%)}}
				&
				\textbf{\shortstack{Signed\\/STU}}
				\\
				\midrule
				$0$
				& 147.30
				& 74.18
				& 90.57
				& 9.43
				& 58.66
				\\
				$0.2$
				& 161.69
				& 82.69
				& 83.47
				& 16.53
				& 73.46
				\\
				$0.5$
				& 158.08
				& 79.87
				& 78.26
				& 21.74
				& 76.51
				\\
				\bottomrule
			\end{tabular}%
		}
	\end{table}
	
	Table~\ref{tab:app_fixed_tau_paired} reports the paired differences over the same 30 matched evaluation seeds. Differences in UoR, TRLC, and ES shortage are expressed in percentage points.
	\begin{table*}[t!]
		\centering
		\caption{Paired differences among the fixed overbooking rate settings. Each entry reports the paired mean and its 95\% confidence interval.}
		\label{tab:app_fixed_tau_paired}
		\footnotesize
		\renewcommand{\arraystretch}{1.15}
		\setlength{\tabcolsep}{5.5pt}
		\resizebox{\textwidth}{!}{%
			\begin{tabular}{l|ccccc}
				\toprule
				\textbf{Comparison}
				&
				$\boldsymbol{\Delta}$\textbf{SW}
				&
				$\boldsymbol{\Delta}$\textbf{UoR}
				&
				$\boldsymbol{\Delta}$\textbf{TRLC}
				&
				$\boldsymbol{\Delta}$\textbf{ES Shortage}
				&
				$\boldsymbol{\Delta}$\textbf{Signed/STU}
				\\
				\midrule
				$\tau=0.2$ minus $\tau=0$
				&
				$+14.40\,[13.06,15.73]$
				&
				$+8.52\,[7.77,9.26]$
				&
				$-7.10\,[-7.39,-6.82]$
				&
				$+7.10\,[6.82,7.39]$
				&
				$+14.80\,[14.04,15.56]$
				\\
				$\tau=0.5$ minus $\tau=0$
				&
				$+10.78\,[9.41,12.16]$
				&
				$+5.69\,[4.97,6.41]$
				&
				$-12.31\,[-12.86,-11.76]$
				&
				$+12.31\,[11.76,12.86]$
				&
				$+17.85\,[16.80,18.90]$
				\\
				$\tau=0.5$ minus $\tau=0.2$
				&
				$-3.61\,[-4.64,-2.59]$
				&
				$-2.82\,[-3.42,-2.22]$
				&
				$-5.21\,[-5.65,-4.77]$
				&
				$+5.21\,[4.77,5.65]$
				&
				$+3.05\,[2.37,3.72]$
				\\
				\bottomrule
			\end{tabular}%
		}
	\end{table*}
	Moderate overbooking improves agreement participation and residual-resource utilization relative to the no-overbooking case. Increasing $\tau$ from $0$ to $0.2$ raises Signed/STU by $14.80$, SW by $14.40$, and UoR by $8.52$ percentage points, while reducing TRLC by $7.10$ percentage points. This pattern reflects the intended role of overbooking: additional advance commitments can compensate for contracted tasks that do not become executable, but they also increase competition for the realized residual ES capacity.
	
	Further increasing the rate to $\tau=0.5$ enlarges the signed agreement set but does not yield additional gains in SW or UoR. Relative to $\tau=0.2$, it signs $3.05$ more task commitments per STU, while SW, UoR, and TRLC decrease by $3.61$, $2.82$, and $5.21$ percentage points, respectively, and ES shortage increases accordingly. Thus, beyond the moderate setting, the additional commitments increasingly translate into capacity contention rather than higher effective resource utilization.
	
	These results characterize the utilization--fulfillment tradeoff induced by the overbooking rate. The moderate setting $\tau=0.2$ improves welfare and residual-resource utilization relative to no overbooking, whereas the higher rate admits more agreements at the cost of greater shortage exposure and lower fulfillment. This dependence on the operating point motivates AdaptAO to select the overbooking rate according to the observed execution state rather than using a single fixed rate.
	
	\subsection{Action-Grid Resolution and AdaptAO Performance}
	\label{app:tau_grid_granularity}
	
	We next examine the effect of the candidate overbooking-rate grid on AdaptAO. Three action grids are considered
	$\mathcal{T}_{\tau}^{\mathsf{coarse}}
	=
	\{0,0.2,0.4\} $, $\mathcal{T}_{\tau}^{\mathsf{default}}
	=
	\{0,0.1,0.2,0.3,0.4\} $, and $\mathcal{T}_{\tau}^{\mathsf{fine}}
	=
	\{0,0.05,0.10,\ldots,0.40\} $.
	Including the agreement-retention action, the coarse, default, and fine configurations contain 4, 6, and 10 DDQN output actions, respectively. For each grid, three models are independently trained using seeds $7000$, $8009$, and $9018$ for 220 episodes with 100 STUs per episode, and are subsequently evaluated without further parameter updates over the same 30 evaluation seeds under the 60-UAV/10-ES setting. Tail Reward and Tail SD are calculated over the final 20 training episodes, while Reward AUC denotes the area under the episode-reward curve over the complete training horizon. Because only three independently trained models are available for each grid, their arithmetic means are used only to summarize the observed results.
	
	\begin{table*}[t!]
		\centering
		\caption{Training and evaluation results under different overbooking rate grids. Values within each cell follow the training-seed order $7000/8009/9018$. Tail Reward and Tail SD are calculated over the final 20 training episodes, and each trained model is evaluated over the same 30 evaluation seeds.}
		\label{tab:app_tau_grid_training}
		\footnotesize
		\renewcommand{\arraystretch}{1.15}
		\setlength{\tabcolsep}{3.2pt}
		\resizebox{\textwidth}{!}{%
			\begin{tabular}{l|ccccccc}
				\toprule
				\textbf{Grid}
				&
				\textbf{Tail Reward}
				&
				\textbf{Tail SD}
				&
				\textbf{Evaluation SW}
				&
				\textbf{UoR (\%)}
				&
				\textbf{TRLC (\%)}
				&
				\textbf{\shortstack{Renewals\\/100 STUs}}
				&
				\textbf{Train (min)}
				\\
				\midrule
				\shortstack{Coarse\\(4 actions)}
				&
				4.792/4.792/4.901
				&
				0.616/0.591/0.563
				&
				161.94/162.28/159.05
				&
				84.26/85.04/82.53
				&
				82.03/80.14/81.52
				&
				10.70/1.27/6.43
				&
				10.34/8.95/9.99
				\\
				\shortstack{Default\\(6 actions)}
				&
				4.709/4.711/4.922
				&
				0.653/0.596/0.579
				&
				158.60/161.73/163.83
				&
				81.69/83.27/85.84
				&
				82.03/82.85/80.76
				&
				14.93/22.00/4.97
				&
				12.84/13.71/14.27
				\\
				\shortstack{Fine\\(10 actions)}
				&
				4.732/4.659/4.847
				&
				0.664/0.585/0.584
				&
				164.40/163.93/158.73
				&
				85.38/84.87/81.94
				&
				79.57/81.83/83.08
				&
				16.73/22.33/14.60
				&
				14.44/14.20/14.84
				\\
				\bottomrule
			\end{tabular}%
		}
	\end{table*}
	
	Table~\ref{tab:app_tau_grid_summary} summarizes the corresponding arithmetic means across the three independently trained models for each grid.
	\begin{table*}[t!]
		\centering
		\caption{Arithmetic means across the three independently trained models for each action-grid configuration. Each trained model is evaluated over the same 30 evaluation seeds.}
		\label{tab:app_tau_grid_summary}
		\footnotesize
		\renewcommand{\arraystretch}{1.15}
		\setlength{\tabcolsep}{4.5pt}
		{%
			\begin{tabular}{l|rrrrrrrr}
				\toprule
				\textbf{Grid}
				&
				\textbf{Tail Reward}
				&
				\textbf{Reward AUC}
				&
				\textbf{Evaluation SW}
				&
				\textbf{UoR (\%)}
				&
				\textbf{TRLC (\%)}
				&
				\textbf{\shortstack{Renewals\\/100 STUs}}
				&
				\textbf{NI/STU}
				&
				\textbf{Train (min)}
				\\
				\midrule
				Coarse
				& 4.828
				& 1043.03
				& 161.09
				& 83.95
				& 81.23
				& 6.13
				& 297.55
				& 9.76
				\\
				Default
				& 4.781
				& 1035.63
				& 161.39
				& 83.60
				& 81.88
				& 13.97
				& 611.06
				& 13.61
				\\
				Fine
				& 4.746
				& 1032.17
				& 162.35
				& 84.06
				& 81.49
				& 17.89
				& 759.74
				& 14.49
				\\
				\bottomrule
			\end{tabular}%
		}
	\end{table*}
	The three grids yield comparable training and execution outcomes. The coarse grid has the highest mean Tail Reward and Reward AUC and the shortest training time, but the results across the three independent runs do not show a consistent improvement in training behavior as the grid becomes finer. Likewise, the differences in the principal execution metrics are limited: the spread across the three grids is $1.26$ in SW, $0.46$ percentage points in UoR, and $0.65$ percentage points in TRLC. The fine grid attains the highest mean SW and UoR, whereas the default grid attains the highest mean TRLC, indicating no consistent execution-performance gain from increasing the grid resolution.
	
	The more pronounced difference lies in agreement-management overhead. Moving from the coarse to the default and fine grids increases the mean renewal frequency from $6.13$ to $13.97$ and $17.89$ renewals per 100 STUs, respectively, with corresponding increases in NI and training time. The coarse grid reduces this overhead but excludes the intermediate rates $\tau=0.1$ and $\tau=0.3$, whereas the fine grid introduces additional renewal actions without a consistent improvement in the principal execution metrics. We therefore adopt the default grid as an intermediate action resolution that preserves additional rate-selection flexibility without the larger action space and overhead of the fine configuration.
	
}

{
	\section{AdaptAO Reward Formulation and Sensitivity to Agreement-Renewal Costs}
	\label{app:reward_analysis}
	
	This appendix details the reward formulation of AdaptAO and examines the effects of the two agreement-renewal cost terms on its retain-or-renew decisions. The reward jointly reflects realized service performance, agreement applicability, prediction accuracy, and the overhead associated with agreement renewal. For notational clarity, the STU superscript $\mathsf{(n)}$ is omitted throughout this appendix; quantities indexed by $t$ refer to the considered STU $n$ of practical BTU $t$.
	
	\subsection{Reward Components and Normalization}
	\label{app:reward_specification}
	
	The reward in~\eqref{27} combines realized system performance, the applicability of the current PT agreements, prediction error, and agreement-management overhead. The realized social welfare is defined as
	\begin{equation}
		SW_t
		=
		\sum_{u_i\in\bm{\mathcal U}}
		U_{i,t}^{\mathsf U}
		+
		\sum_{s_j\in\bm{\mathcal S}}
		U_{j,t}^{\mathsf S}.
		\label{eq:app_reward_sw}
	\end{equation}
	Unlike the expected utilities used by PilotAO during agreement construction, the utilities in~\eqref{eq:app_reward_sw} are evaluated from the realized execution outcomes. The UAV-side non-fulfillment transfer $q^{\mathsf U}$ and the ES-side shortage compensation $q^{\mathsf E}$ enter the UAV and ES utilities with opposite signs and therefore cancel when the utilities of both trading sides are aggregated into social welfare.
	
	Let $N_t^{\rm srv}$ denote the number of executable contracted tasks that are successfully served, and let $N_t^{\mathsf E\text{-}\rm sh}$ denote the number of executable contracted tasks that cannot be served because of insufficient residual ES capacity. The total number of executable contracted tasks is
	\begin{equation}
		N_t^{\rm exe}
		=
		N_t^{\rm srv}
		+
		N_t^{\mathsf E\text{-}\rm sh}.
		\label{eq:app_active_contracts}
	\end{equation}
	Accordingly, the task completion rate under PT agreements is defined as
	\begin{equation}
		\mathrm{TRLC}_t
		=
		\begin{cases}
			\dfrac{N_t^{\rm srv}}{N_t^{\rm exe}},
			&
			N_t^{\rm exe}>0,\\[2mm]
			0,
			&
			N_t^{\rm exe}=0.
		\end{cases}
		\label{eq:app_reward_trlc}
	\end{equation}
	Thus, $\mathrm{TRLC}_t$ measures the fraction of executable contracted tasks that are successfully served, while every ES-side shortage event is included once in the denominator.
	
	The realized ES capacity available for UAV services is
	\begin{equation}
		G_t^{\rm avail}
		=
		\sum_{s_j\in\mathcal S}
		\max
		\left\{
		0,
		G_j-\varepsilon_{j,t}
		\right\},
		\label{eq:app_available_capacity}
	\end{equation}
	where $G_j$ denotes the nominal task-level capacity of ES $s_j$ and $\varepsilon_{j,t}$ denotes its realized inherent workload. Since each UAV task consumes one ES resource unit, the residual-resource utilization ratio is
	\begin{equation}
		\mathrm{UoR}_t
		=
		\frac{
			N_t^{\rm srv}
		}{
			\max
			\left\{
			1,
			G_t^{\rm avail}
			\right\}
		}.
		\label{eq:app_reward_uor}
	\end{equation}
	Hence, $\mathrm{TRLC}_t$ characterizes fulfillment among executable contracted tasks, whereas $\mathrm{UoR}_t$ characterizes the utilization of the realized residual ES capacity.
	
	The applicability of the current PT agreements is represented by
	\begin{equation}
		A_t
		=
		\frac{
			N_t^{\rm app}
		}{
			\max
			\left\{
			1,
			N_t^{\rm ctr}
			\right\}
		},
		\label{eq:app_reward_applicability}
	\end{equation}
	where $N_t^{\rm ctr}$ denotes the number of currently contracted task--ES pairs and $N_t^{\rm app}$ denotes the number that remain applicable under the current execution state. A contracted task--ES pair is classified as applicable when it remains accessible and satisfies the corresponding execution-feasibility conditions, its estimated successful-service probability is no smaller than $0.56$, its realized service valuation is no smaller than the corresponding service cost plus the minimum margin of $0.02$, and the realized service valuation covers the signed payment. When $N_t^{\rm ctr}=0$, $A_t$ is set to zero.
	
	The aggregate non-fulfillment ratio is defined as
	\begin{equation}
		D_t
		=
		\frac{
			N_t^{\mathsf U\text{-}\rm def}
			+
			N_t^{\mathsf E\text{-}\rm sh}
		}{
			\max
			\left\{
			1,
			N_t^{\rm ctr}
			\right\}
		},
		\label{eq:app_reward_default}
	\end{equation}
	where $N_t^{\mathsf U\text{-}\rm def}$ denotes the number of contracted tasks that do not become executable because of UAV-side non-fulfillment. Thus, $D_t$ accounts for both UAV-side non-fulfillment and ES-side capacity shortage while preserving their distinct transfer responsibilities.
	
	The prediction error between the estimated and realized execution states is measured by
	\begin{equation}
		E_t
		=
		\operatorname{clip}_{[0,2]}
		\left(
		\frac{1}{3}
		\sum_{z\in\{d,m,r\}}
		\frac{
			\left|
			z_t^{\rm actual}
			-
			z_t^{\rm pred}
			\right|
		}{
			\max
			\left\{
			0.1,
			z_t^{\rm actual}
			\right\}
		}
		\right),
		\label{eq:app_reward_belief_error}
	\end{equation}
	where $d$, $m$, and $r$ denote the task-demand, UAV-mobility, and residual-resource states, respectively, and
	\begin{equation}
		\operatorname{clip}_{[0,2]}(x)
		=
		\min
		\left\{
		2,
		\max
		\left\{
		0,
		x
		\right\}
		\right\}.
		\label{eq:app_reward_clipping}
	\end{equation}
	The clipping operation prevents an unusually large prediction error from dominating the remaining reward components, while the final scalar reward itself is not clipped.
	
	Let $\mathrm{Age}_t^{-}$ denote the elapsed retention duration of the current PT-agreement set immediately before action $a_t$ is applied. The agreement-renewal cost is
	\begin{equation}
		C_t^{\rm upd}
		=
		\mathbbm{1}_
		{\left\{
			a_t\in\mathcal A_{\mathsf{renew}}
			\right\}}
		\left[
		c_{\mathsf{fix}}
		+
		c_{\mathsf{early}}
		\max
		\left\{
		0,
		H_{\mathsf{age}}
		-
		\mathrm{Age}_t^{-}
		\right\}
		\right].
		\label{eq:app_reward_update_cost}
	\end{equation}
	Here, $c_{\mathsf{fix}}$ denotes the fixed cost of reconstructing the PT agreements, while $c_{\mathsf{early}}$ controls the additional renewal cost when the current agreements have been retained for less than the reference retention horizon $H_{\mathsf{age}}$. The additional term becomes zero once $\mathrm{Age}_t^{-}\geq H_{\mathsf{age}}$.
	
	The interaction term $\mathrm{NI}_t$ counts the proposal, acceptance, rejection, price-update, and final agreement-confirmation interactions generated when PilotAO is invoked for agreement renewal. It is zero when the current PT agreements are retained.
	
	Table~\ref{tab:app_reward_parameters} summarizes the reward parameters used in the default AdaptAO configuration. The normalization scale $s_{\mathsf{sw}}$ brings the social-welfare term to a numerical range comparable with the ratio-based reward components, while the remaining coefficients determine the relative contributions of fulfillment, utilization, applicability, non-fulfillment, prediction error, and agreement-management overhead.
	
	\begin{table}[t!]
		\centering
		\caption{Default parameters of the AdaptAO reward formulation.}
		\label{tab:app_reward_parameters}
		\scriptsize
		\renewcommand{\arraystretch}{1.10}
		\setlength{\tabcolsep}{2.5pt}
		\begin{tabular}{
				@{}
				>{\centering\arraybackslash}p{1.45cm}
				>{\centering\arraybackslash}p{1.1cm}
				>{\raggedright\arraybackslash}p{5.35cm}
				@{}
			}
			\toprule
			\textbf{Parameter}
			&
			\textbf{Value}
			&
			\textbf{Definition}
			\\
			\midrule
			$s_{\mathsf{sw}}$
			&
			$24$
			&
			Normalization scale for realized social welfare
			\\
			$\lambda_{\mathsf{sw}}$
			&
			$1$
			&
			Weight of normalized social welfare
			\\
			$\lambda_{\mathsf{ful}}$
			&
			$1.2$
			&
			Weight of agreement fulfillment
			\\
			$\lambda_{\mathsf{util}}$
			&
			$0.65$
			&
			Weight of residual-resource utilization
			\\
			$\lambda_{\mathsf{app}}$
			&
			$0.70$
			&
			Weight of agreement applicability
			\\
			$\lambda_{\mathsf{def}}$
			&
			$1.4$
			&
			Penalty weight for aggregate non-fulfillment
			\\
			$\lambda_{\mathsf{err}}$
			&
			$0.80$
			&
			Penalty weight for prediction error
			\\
			$c_{\mathsf{fix}}$
			&
			$0.06$
			&
			Fixed cost of reconstructing the PT agreements
			\\
			$c_{\mathsf{early}}$
			&
			$0.18$
			&
			Additional renewal cost before the reference retention horizon
			\\
			$H_{\mathsf{age}}$
			&
			$4$
			&
			Reference retention horizon
			\\
			$\lambda_{\mathsf{ni}}$
			&
			$5\times10^{-5}$
			&
			Penalty weight for agreement-management interactions
			\\
			\bottomrule
		\end{tabular}
	\end{table}
	
	Substituting the values in Table~\ref{tab:app_reward_parameters} into~\eqref{27} gives the numerical reward used in the default AdaptAO configuration:
	\begin{equation}
		\begin{aligned}
			r_t
			={}&
			\frac{SW_t}{24}
			+
			1.2\,\mathrm{TRLC}_t
			+
			0.65\,\mathrm{UoR}_t
			+
			0.70\,A_t
			\\
			&
			-
			1.4\,D_t
			-
			0.80\,E_t
			-
			5\times10^{-5}\mathrm{NI}_t
			\\
			&
			-
			\mathbbm{1}_
			{\left\{
				a_t\in\mathcal A_{\mathsf{renew}}
				\right\}}
			\left[
			0.06
			+
			0.18
			\max
			\left\{
			0,
			4-\mathrm{Age}_t^{-}
			\right\}
			\right].
		\end{aligned}
		\label{eq:app_exact_reward}
	\end{equation}
	Here, the factor $24$ is used only to normalize the scale of realized social welfare.
	
	\subsection{Sensitivity to Agreement-Renewal Costs}
	\label{app:reward_cost_sensitivity}
	
	The two renewal-cost coefficients penalize different aspects of PT-agreement reconstruction. The fixed cost $c_{\mathsf{fix}}$ applies to every renewal action, whereas $c_{\mathsf{early}}$ adds an age-dependent penalty when the current agreements have been retained for less than the reference horizon $H_{\mathsf{age}}$. Because changing either coefficient modifies the reward function, the DDQN controller is independently trained for each tested configuration.
	
	The default setting is
	$
	\left(
	c_{\mathsf{fix}},
	c_{\mathsf{early}}
	\right)
	=
	(0.06,0.18).
	$
	We additionally consider $c_{\mathsf{fix}}\in\{0.03,0.12\}$ with $c_{\mathsf{early}}=0.18$, and $c_{\mathsf{early}}\in\{0.09,0.36\}$ with $c_{\mathsf{fix}}=0.06$. For each configuration, three models are independently trained using seeds $7000$, $8009$, and $9018$ for 220 episodes with 100 STUs per episode, and are subsequently evaluated without further parameter updates over the same 30 evaluation seeds under the 60-UAV/10-ES setting. Table~\ref{tab:app_retrained_update_cost} reports the individual results in the training-seed order $7000/8009/9018$. Mean $\tau$ is calculated only over STUs in which agreement renewal is selected. Final-20 Reward denotes the mean episode reward over the final 20 training episodes; because the reward function changes across cost settings, this quantity is interpreted only within each configuration. Arithmetic means across the three trained models are used only to summarize the observed trends.
	
	\begin{table*}[t!]
		\centering
		\caption{Sensitivity to agreement-renewal costs after independent DDQN training. Slash-separated values follow the training-seed order $7000/8009/9018$. Each trained model is evaluated over the same 30 evaluation seeds. Mean $\tau$ is calculated only over STUs in which agreement renewal is selected.}
		\label{tab:app_retrained_update_cost}
		\footnotesize
		\renewcommand{\arraystretch}{1.15}
		\setlength{\tabcolsep}{3.2pt}
		\resizebox{\textwidth}{!}{%
			\begin{tabular}{l|ccccccc}
				\toprule
				\textbf{Setting}
				&
				\textbf{SW}
				&
				\textbf{UoR (\%)}
				&
				\textbf{TRLC (\%)}
				&
				\textbf{\shortstack{Renewals\\/100 STUs}}
				&
				\textbf{NI/STU}
				&
				\textbf{\shortstack{Mean $\boldsymbol{\tau}$\\Given Renewal}}
				&
				\textbf{Final-20 Reward}
				\\
				\midrule
				Default
				&
				158.60/161.73/163.83
				&
				81.69/83.27/85.84
				&
				82.03/82.85/80.76
				&
				14.93/22.00/4.97
				&
				640.51/940.14/252.54
				&
				0.237/0.185/0.135
				&
				4.709/4.711/4.922
				\\
				$c_{\mathsf{fix}}=0.03$
				&
				160.85/162.16/163.52
				&
				82.96/84.21/85.53
				&
				81.86/82.23/80.74
				&
				18.31/26.13/6.23
				&
				773.54/1095.88/290.15
				&
				0.236/0.195/0.148
				&
				4.743/4.748/4.947
				\\
				$c_{\mathsf{fix}}=0.12$
				&
				158.92/161.16/163.19
				&
				81.83/83.22/85.35
				&
				82.23/82.77/80.91
				&
				11.67/16.57/3.47
				&
				491.56/692.97/189.44
				&
				0.240/0.186/0.144
				&
				4.670/4.686/4.885
				\\
				$c_{\mathsf{early}}=0.09$
				&
				161.10/160.35/167.12
				&
				83.45/82.27/86.77
				&
				81.23/83.72/80.35
				&
				13.83/23.87/33.30
				&
				587.01/983.02/1586.62
				&
				0.184/0.173/0.213
				&
				4.779/4.773/4.922
				\\
				$c_{\mathsf{early}}=0.36$
				&
				161.89/163.29/162.16
				&
				84.29/85.44/84.63
				&
				79.72/80.05/81.83
				&
				8.20/3.10/2.97
				&
				378.60/174.38/163.05
				&
				0.264/0.175/0.134
				&
				4.707/4.680/4.865
				\\
				\bottomrule
			\end{tabular}%
		}
	\end{table*}
	
	The fixed renewal cost primarily affects how frequently AdaptAO reconstructs the PT agreements. Increasing $c_{\mathsf{fix}}$ from $0.03$ to $0.06$ and $0.12$ reduces the mean renewal frequency from $16.89$ to $13.97$ and $10.57$ renewals per 100 STUs, respectively, with a corresponding reduction in NI. In contrast, SW, UoR, and TRLC vary over substantially narrower ranges. This behavior is consistent with $c_{\mathsf{fix}}$ imposing a direct penalty on every invocation of PilotAO.
	
	The early-renewal cost has a stronger effect on reconstruction before the reference retention horizon. Increasing $c_{\mathsf{early}}$ to $0.36$ reduces the mean renewal frequency to $4.76$ renewals per 100 STUs, whereas reducing it to $0.09$ increases the mean to $23.67$. The associated NI changes in the same direction, while the principal execution metrics remain comparatively stable. Thus, $c_{\mathsf{early}}$ mainly discourages repeated reconstruction while the current agreements are still relatively recent.
	
	Across the tested configurations, the mean SW, UoR, and TRLC remain within ranges of $161.09$--$162.86$, $83.47\%$--$84.79\%$, and $80.53\%$--$81.97\%$, respectively, whereas renewal frequency and NI vary much more substantially. Moreover, the mean overbooking rate conditioned on renewal remains close to $0.19$ across the configurations. These results indicate that the two renewal-cost coefficients primarily regulate \emph{when} AdaptAO reconstructs the PT agreements and the resulting agreement-management overhead, rather than systematically changing the overbooking rate selected upon renewal. We therefore retain $(c_{\mathsf{fix}},c_{\mathsf{early}})=(0.06,0.18)$ as the default setting within the tested range.

	{
		\section{Component-Level Ablations and Utility-Preference Sensitivity of PAST}
		\label{app:component_sensitivity}
		
		This appendix examines the respective contributions of PilotAO agreement formation, AdaptAO selective renewal, and agreement reuse, and further studies how different preferences for latency and energy savings affect the resulting PAST outcomes. The overbooking rate, controller alternatives, probability estimator, PilotAO parameters, action-grid resolution, and renewal-cost coefficients are analyzed separately in Appxs.~\ref{app:fixed_tau_sensitivity}, \ref{app:ddqn_controller_analysis}, \ref{app:probability_estimation}, \ref{app:pilot_parameter_sensitivity}, \ref{app:tau_action_grid}, and \ref{app:reward_analysis}, respectively. Unless otherwise specified, the following component-level comparisons use 60 UAVs, 10 ESs, 30 matched evaluation seeds, and 100 STUs per seed.
		
		\subsection{Contribution of the PilotAO Agreement-Formation Procedure}
		\label{app:pilotao_matcher_ablation}
		
		To examine the contribution of the agreement-formation procedure in PilotAO, we compare PilotAO with GreedyNoPrice while keeping the environmental realizations, AdaptAO actions, selected overbooking rates, renewal instants, and agreement-execution rules unchanged. GreedyNoPrice ranks candidate task--ES contracts according to their bilateral expected surplus at the initial payments and then performs agreement selection without the iterative proposal, rejection, and payment-adjustment process of PilotAO. The same feasibility and risk conditions are retained in both settings. Hence, the comparison changes the agreement-formation mechanism while preserving the subsequent execution conditions.
		
		\begin{table}[t!]
			\centering
			
			\caption{Ablation of the PilotAO agreement-formation procedure over 30 matched evaluation seeds with 100 STUs per seed.}
			\label{tab:app_pilotao_matcher_ablation}
			\footnotesize
			\renewcommand{\arraystretch}{1.15}
			\setlength{\tabcolsep}{4pt}
			{%
				\begin{tabular}{l|cccc}
					\toprule
					\textbf{Matcher}
					&
					\textbf{SW}
					&
					\textbf{UoR (\%)}
					&
					\textbf{TRLC (\%)}
					&
					\textbf{Signed/STU}
					\\
					\midrule
					PilotAO
					& 158.60
					& 81.69
					& 82.03
					& 76.28
					\\
					GreedyNoPrice
					& 146.59
					& 75.04
					& 78.75
					& 72.86
					\\
					\bottomrule
				\end{tabular}%
			}
		\end{table}
		
		As shown in Table~\ref{tab:app_pilotao_matcher_ablation}, PilotAO increases SW by $12.01$, UoR by $6.65$ percentage points, TRLC by $3.28$ percentage points, and Signed/STU by $3.42$ compared with GreedyNoPrice. The simultaneous improvements in agreement volume, resource utilization, fulfillment, and social welfare indicate that the iterative agreement-formation process enables additional mutually acceptable task--ES contracts and produces more effective assignments than selection based only on the initial-price surplus ranking.
		
		These gains are accompanied by additional proposal and payment-adjustment operations in PilotAO. Their interaction overhead and expected-valuation preprocessing complexity are analyzed separately in Appxs.~\ref{app:pilot_parameter_sensitivity} and~\ref{app:mathematical_derivations}, respectively. Overall, this ablation shows that the complete PilotAO agreement-formation procedure improves the resulting trading and execution outcomes under the same AdaptAO decisions and environmental realizations.
		
		\subsection{Roles of AdaptAO and Agreement Reuse}
		\label{app:adapt_reuse_component_ablation}
		
		We next examine the roles of AdaptAO and PT-agreement reuse through three agreement-management strategies. PAST-DDQN selectively retains or renews the current PT agreements according to AdaptAO. PAST without AdaptAO constructs the initial PT agreements using PilotAO and retains them throughout the evaluation. PAST-PerSTU disables the retention action and reconstructs the PT agreements in every STU, while otherwise using the same trained DDQN, probability estimator, PilotAO procedure, candidate overbooking-rate set, and execution model as PAST-DDQN. Upon each reconstruction, the overbooking rate is determined by the highest-valued action among the five renewal actions of the same trained DDQN. The three configurations therefore represent permanent retention, selective renewal, and per-STU reconstruction, respectively.
		
		\begin{table}[t!]
			\centering
			\caption{Ablation of AdaptAO and PT-agreement reuse under the default setting over 30 matched evaluation seeds with 100 STUs per seed.}
			\label{tab:app_adapt_reuse_component_ablation}
			\scriptsize
			\renewcommand{\arraystretch}{1.15}
			\resizebox{\columnwidth}{!}{%
				\begin{tabular}{l|ccccc}
					\toprule
					\textbf{Method}
					&
					\textbf{SW}
					&
					\textbf{UoR (\%)}
					&
					\textbf{TRLC (\%)}
					&
					\textbf{Edge-Served (\%)}
					&
					\textbf{NI/STU}
					\\
					\midrule
					PAST-DDQN
					& 158.60
					& 81.69
					& 82.03
					& 45.73
					& 640.51
					\\
					PAST without AdaptAO
					& 152.06
					& 80.75
					& 77.26
					& 47.21
					& 47.57
					\\
					PAST-PerSTU
					& 175.72
					& 85.70
					& 79.68
					& 49.78
					& 3780.16
					\\
					\bottomrule
				\end{tabular}%
			}
		\end{table}
		
		Relative to permanent agreement retention, PAST-DDQN improves SW, UoR, and TRLC by $6.54$, $0.94$ percentage points, and $4.77$ percentage points, respectively, although its Edge-Served ratio is $1.48$ percentage points lower. This comparison shows that selective renewal can improve the quality and fulfillment of the retained PT agreements as execution conditions evolve, while requiring additional agreement-management interactions.
		
		Per-STU reconstruction exhibits the opposite operating point. Compared with PAST-DDQN, PAST-PerSTU achieves higher SW, UoR, and Edge-Served, but its TRLC is $2.35$ percentage points lower and its NI increases from $640.51$ to $3780.16$ per STU. Reconstructing the agreements in every STU therefore provides greater responsiveness to changing demand and resource conditions at substantially higher interaction cost.
		
		The three configurations illustrate the tradeoff between agreement reuse and reconstruction. Permanent retention minimizes repeated agreement-management overhead but does not adapt the initial agreements to subsequent execution changes, whereas per-STU reconstruction maximizes update frequency at a much higher interaction cost. AdaptAO operates between these two extremes by selectively reconstructing the PT agreements according to the observed execution state.

		\subsection{Sensitivity to the Latency--Energy Preference}
		\label{app:utility_preference_sensitivity}
		
		The UAV service valuation combines latency and energy savings through the weighting coefficients $\omega_4$ and $\omega_5$. To vary their relative emphasis while keeping the total coefficient scale unchanged, we define the latency-preference ratio as
		\begin{equation}
			\eta_{\mathrm{lat}}
			=
			\frac{\omega_4}{\omega_4+\omega_5},
			\label{eq:app_latency_preference}
		\end{equation}
		and keep $\omega_4+\omega_5$ fixed. A larger $\eta_{\mathrm{lat}}$ places greater emphasis on latency reduction, whereas a smaller value places greater emphasis on energy saving. For each tested value of $\eta_{\mathrm{lat}}$, DDQN is independently trained using seeds $7000$, $8009$, and $9018$, and each trained model is subsequently evaluated over the same 30 evaluation seeds. The default weighting coefficients correspond to $\eta_{\mathrm{lat}}\approx0.696$.
		
		\begin{table*}[t!]
			\centering
			\caption{Sensitivity to the latency--energy preference after independent DDQN training. Slash-separated values correspond to training seeds $7000/8009/9018$, and each trained model is evaluated over the same 30 evaluation seeds.}
			\label{tab:app_latency_energy_preference}
			\footnotesize
			\renewcommand{\arraystretch}{1.15}
			{%
				\begin{tabular}{c|ccccc}
					\toprule
					$\boldsymbol{\eta_{\mathrm{lat}}}$
					&
					\textbf{SW}
					&
					\textbf{UoR (\%)}
					&
					\textbf{TRLC (\%)}
					&
					\textbf{Edge-Served (\%)}
					&
					\textbf{\shortstack{Renewals\\/100 STUs}}
					\\
					\midrule
					0.2
					&
					133.79/136.24/132.12
					&
					84.90/86.65/83.70
					&
					81.86/81.06/82.44
					&
					47.48/48.56/46.84
					&
					9.93/14.17/9.23
					\\
					0.5
					&
					151.59/153.56/151.51
					&
					85.11/86.16/85.36
					&
					81.12/78.32/81.03
					&
					47.54/48.28/47.64
					&
					9.50/12.37/3.83
					\\
					\textbf{0.696 (default)}
					&
					\textbf{158.60/161.73/163.83}
					&
					\textbf{81.69/83.27/85.84}
					&
					\textbf{82.03/82.85/80.76}
					&
					\textbf{45.73/46.75/47.89}
					&
					\textbf{14.93/22.00/4.97}
					\\
					0.8
					&
					168.78/168.63/170.01
					&
					84.25/84.38/85.24
					&
					80.66/79.67/81.39
					&
					47.09/47.24/47.66
					&
					14.27/6.80/8.47
					\\
					\bottomrule
				\end{tabular}%
			}
		\end{table*}
		
		The descriptive mean SW increases from $134.05$ at $\eta_{\mathrm{lat}}=0.2$ to $152.22$, $161.39$, and $169.14$ as $\eta_{\mathrm{lat}}$ increases to $0.5$, $0.696$, and $0.8$, respectively. Since changing $\eta_{\mathrm{lat}}$ changes the valuation assigned to latency and energy savings, these SW values correspond to different utility preferences and should not be interpreted as performance measurements under an identical welfare objective. Instead, the trend reflects the change in aggregate valuation induced by placing greater emphasis on latency reduction.
		
		The remaining execution outcomes do not exhibit a monotonic dependence on $\eta_{\mathrm{lat}}$. Across the four settings, the descriptive mean UoR ranges from $83.60\%$ to $85.54\%$, TRLC from $80.16\%$ to $81.88\%$, Edge-Served from $46.79\%$ to $47.82\%$, and renewal frequency from $8.57$ to $13.97$ renewals per 100 STUs. The default setting achieves the highest mean TRLC, whereas $\eta_{\mathrm{lat}}=0.5$ achieves the highest mean UoR and Edge-Served ratio. Hence, changing the latency--energy preference modifies the resulting agreement and execution behavior without producing a single setting that dominates all execution metrics.
		
		The influence of $\eta_{\mathrm{lat}}$ propagates through both stages of PAST. Changing the weighting coefficients modifies the service valuations and preference rankings used during PilotAO agreement construction. The resulting PT agreements subsequently affect the execution states observed by AdaptAO and can therefore alter its renewal decisions. The preferred latency--energy weighting can thus be selected according to the service requirements of different UAV applications, depending on whether latency reduction or onboard-energy saving is emphasized.
		
		\section{Behavior of PAST under Recoverable and Nonrecoverable Dynamic Shifts}
		\label{app:dynamic_stress}
		
		This appendix evaluates PAST when the execution environment changes after the initial PT agreements have been established. We distinguish two types of changes. The first creates an agreement mismatch that can potentially be mitigated by reconstructing the PT agreements, for example when residual capacity remains available at alternative ESs or when the current agreements no longer match the realized demand, mobility, or channel conditions. The second causes a system-wide reduction in physical service capacity, which cannot be compensated for by changing the agreement assignment alone. This distinction allows us to examine both the adaptive capability and the operating limits of agreement renewal.
		
		\subsection{Post-Shift Evaluation Protocol}
		\label{app:dynamic_protocol}
		
		All experiments use 60 UAVs, 10 ESs, 30 matched evaluation seeds, and a horizon of 100 STUs. PAST-Full and PAST-w/o-AdaptAO begin with the same PT agreements constructed by PilotAO under the baseline conditions and use this common agreement set during STUs~1--30. Immediately before STU~31, a controlled environmental shift is introduced and remains active through STU~100. During this post-shift period, PAST-Full either retains the current agreements or reconstructs them using an overbooking rate selected from $\{0,0.1,0.2,0.3,0.4\}$ by the trained DDQN, whereas PAST-w/o-AdaptAO retains the initial agreements throughout the remaining horizon.
		
		The environmental shift does not invalidate the existing PT agreements. Retained agreements continue to be executed according to the realized task generation, UAV--ES accessibility, and residual ES capacity. For each evaluation seed, the two methods share the same task, mobility, channel, and ES-load realizations. SW, UoR, and TRLC are computed for each STU and averaged over the 70 post-shift STUs, whereas Edge-Served and ES Shortage are computed from the corresponding aggregated event counts over the same period. Consequently, the paired differences in TRLC and ES Shortage need not be exact complements because the two metrics use different aggregation procedures.
		
		\begin{table*}[t!]
			\centering
			\caption{Post-shift endpoint results under the dynamic-shift evaluation protocol. Each $\Delta$ denotes the paired PAST-Full-minus-PAST-w/o-AdaptAO mean difference and its two-sided Student-$t$ 95\% confidence interval over 30 matched evaluation seeds, each containing 70 post-shift STUs. TRLC is averaged at the STU level, whereas ES shortage is calculated from aggregated event counts.}
			\label{tab:app_dynamic_stress}
			\footnotesize
			\renewcommand{\arraystretch}{1.15}
			\setlength{\tabcolsep}{3.8pt}
			\resizebox{\textwidth}{!}{%
				\begin{tabular}{l|c|c|c|c|c|c|c}
					\toprule
					\textbf{Stress Endpoint}
					&
					\textbf{\shortstack{Direct Metric\\at Endpoint}}
					&
					\textbf{\shortstack{PAST-Full Endpoint\\SW / Edge / UoR / TRLC}}
					&
					$\boldsymbol{\Delta}$\textbf{SW}
					&
					$\boldsymbol{\Delta}$\textbf{Edge}
					&
					$\boldsymbol{\Delta}$\textbf{UoR}
					&
					$\boldsymbol{\Delta}$\textbf{TRLC}
					&
					$\boldsymbol{\Delta}$\textbf{Shortage}
					\\
					\midrule
					Location error: 100 m
					&
					RMSE: 145.358 m
					&
					\shortstack{$171.051/50.567/$\\$81.406/87.579$}
					&
					$+9.060$ $[5.300,12.820]$
					&
					$-0.241$ $[-1.646,1.163]$
					&
					$+2.541$ $[0.163,4.919]$
					&
					$+4.998$ $[3.027,6.969]$
					&
					$-4.605$ $[-6.424,-2.787]$
					\\
					Demand multiplier: 1.5
					&
					Arrivals: 110.031/STU
					&
					\shortstack{$178.141/50.601/$\\$84.428/85.562$}
					&
					$+11.702$ $[7.819,15.586]$
					&
					$+0.351$ $[-0.962,1.665]$
					&
					$+3.576$ $[1.443,5.709]$
					&
					$+3.872$ $[1.850,5.894]$
					&
					$-3.573$ $[-5.484,-1.661]$
					\\
					Uniform ES load: $\times2$
					&
					\shortstack{Load: 101.903/STU\\Capacity: 24.297/STU}
					&
					\shortstack{$54.462/13.256/$\\$60.694/54.660$}
					&
					$-11.612$ $[-15.263,-7.962]$
					&
					$-7.658$ $[-9.001,-6.316]$
					&
					$-30.673$ $[-35.426,-25.921]$
					&
					$+20.896$ $[16.273,25.518]$
					&
					$-17.842$ $[-21.348,-14.337]$
					\\
					Redistributive ES load: $\times2$
					&
					\shortstack{Expected total-load ratio: 1.000\\Capacity: 71.757/STU}
					&
					\shortstack{$177.505/52.596/$\\$81.218/88.658$}
					&
					$+44.167$ $[38.459,49.874]$
					&
					$+11.438$ $[9.463,13.412]$
					&
					$+20.758$ $[18.318,23.198]$
					&
					$+21.895$ $[19.275,24.514]$
					&
					$-21.042$ $[-23.408,-18.676]$
					\\
					Channel attenuation: 8 dB
					&
					Rate: 7.057 Mbps
					&
					\shortstack{$170.279/51.632/$\\$83.020/86.436$}
					&
					$+13.442$ $[9.218,17.666]$
					&
					$+0.823$ $[-0.700,2.347]$
					&
					$+4.156$ $[1.752,6.560]$
					&
					$+3.855$ $[2.058,5.653]$
					&
					$-3.578$ $[-5.289,-1.867]$
					\\
					\bottomrule
				\end{tabular}%
			}
		\end{table*}
		
		For each evaluation seed, the reported difference is defined as PAST-Full minus PAST-w/o-AdaptAO. The corresponding two-sided paired Student-$t$ 95\% confidence interval is
		\begin{equation}
			\overline{\Delta}
			\pm
			t_{0.975,29}
			\frac{s_{\Delta}}{\sqrt{30}},
			\label{eq:app_dynamic_paired_ci}
		\end{equation}
		where $\overline{\Delta}$ and $s_{\Delta}$ denote the mean and standard deviation of the 30 paired seed-level differences, respectively. The trained DDQN parameters remain fixed throughout the post-shift evaluations.
		
		Five controlled shifts are considered. For the \textbf{location-prediction shift}, zero-mean two-dimensional Gaussian errors are added to the UAV positions used for subsequent agreement reconstruction, with per-coordinate standard deviation $\sigma_{\mathrm{loc}}\in\{0,25,50,75,100\}$ m, while the realized UAV trajectories remain unchanged. This shift increases the mismatch between the mobility information available for agreement reconstruction and the realized UAV locations.
		
		For the \textbf{demand-activity shift}, the task-generation propensity from STU~31 onward is adjusted as
		\begin{equation}
			p_{i,m,\gamma}^{\mathsf{arr,(n)}}
			=
			\min
			\left\{
			1,\,
			\gamma_d
			p_{i,m}^{\mathsf{arr,(n)}}
			\right\},
			~
			\gamma_d\in\{0.64,0.8,1.0,1.2,1.5\}.
		\end{equation}
		The multiplier therefore changes the task-generation probability rather than directly scaling the realized task count.
		
		For the \textbf{uniform ES-load shift}, the inherent workload of every ES is doubled from STU~31 onward, reducing residual capacity across the network. This setting represents a system-wide capacity contraction that cannot be removed through agreement reassignment alone.
		
		For the \textbf{redistributive ES-load shift}, ESs are ranked within each evaluation seed according to the number of task commitments in the common initial agreement set. The top $40\%$ of ESs, which hold $53.356\%$ of the initial task commitments on average, are assigned a two-fold inherent-load multiplier. The loads of the remaining ESs are reduced using a seed-specific multiplier with mean $0.338$ so that the total expected inherent workload is preserved. This shift changes the spatial distribution of residual ES capacity without changing the aggregate expected inherent workload.
		
		For the \textbf{channel-attenuation shift}, the channel-gain scale from STU~31 onward is multiplied by $10^{-a/10}$, where $a\in\{0,2,4,6,8\}$ dB. The resulting attenuation reduces the transmission rate and consequently changes the service valuation used during subsequent agreement reconstruction.

		\subsection{Post-Shift Endpoint Results}
		\label{app:dynamic_endpoint_results}
		
		Table~\ref{tab:app_dynamic_stress} reports the most severe tested endpoint for each dynamic shift. The PAST-Full endpoint column lists SW, Edge-Served, UoR, and TRLC in this order, and differences in percentage-valued metrics are expressed in percentage points.
		
		Under the location-prediction, demand-activity, and channel-attenuation shifts, selective renewal consistently improves SW, UoR, and TRLC while reducing ES-side shortage, with the corresponding confidence intervals excluding zero. The Edge-Served differences are not statistically distinguishable from zero in these three cases. Thus, when the environmental change primarily introduces a mismatch between the retained agreements and the realized UAV--edge conditions, agreement reconstruction improves the utilization and fulfillment of executable contracts without materially changing the overall fraction of tasks served at the edge.
		
		The two ES-load shifts further distinguish redistributable mismatch from system-wide capacity loss. Under the uniform-load shift, PAST-Full increases TRLC by $20.896$ percentage points and reduces ES shortage by $17.842$ percentage points, but SW, Edge-Served, and UoR decrease by $11.612$, $7.658$, and $30.673$ percentage points, respectively. Agreement renewal can therefore improve the composition and fulfillment of the remaining executable contracts, but cannot restore residual capacity that has been lost across all ESs. By contrast, under the redistributive-load shift, where the aggregate expected inherent workload is preserved, PAST-Full improves all four execution metrics and reduces ES shortage; in particular, SW and UoR increase by $44.167$ and $20.758$ percentage points. These gains arise from reallocating task commitments toward the changed spatial distribution of residual ES resources.
		
		These endpoint results show that the benefit of selective renewal depends on whether the post-shift mismatch can be mitigated through agreement reconstruction. When sufficient service capacity remains available but its correspondence with the existing agreements changes, AdaptAO can improve the alignment between task commitments and the realized execution conditions. When physical capacity contracts system-wide, renewal remains useful for improving fulfillment among the remaining executable contracts, but cannot compensate for the lost service resources.
		
		\subsection{Verification under Predefined Renewal Actions}
		\label{app:dynamic_fixed_action}
		
		To isolate the effect of agreement reconstruction under the redistributive ES-load shift, we apply one predefined renewal immediately before STU~31 for each $\tau\in\{0,0.1,0.2,0.3,0.4\}$. For a given $\tau$, the same renewal action is applied across all evaluation seeds. Compared with permanent agreement retention, all tested renewal settings yield positive paired differences in SW, UoR, and TRLC, with the corresponding 95\% confidence intervals excluding zero.
		
		For instance, renewal with $\tau=0.4$ increases SW by $37.736$ with a 95\% confidence interval of $[31.460,44.012]$, UoR by $20.036$ percentage points with an interval of $[16.635,23.436]$, and TRLC by $16.954$ percentage points with an interval of $[13.516,20.392]$. These results show that, following the redistributive shift, agreement reconstruction can exploit the residual ES resources available under the updated load distribution and improve both resource utilization and agreement fulfillment.

	}

\end{document}